\documentclass[traditabstract]{aa}
\pdfoutput=1
\usepackage{amsmath}
\usepackage{txfonts}
\usepackage{natbib,hyperref,ifthen}
\usepackage{graphicx}
\usepackage{multirow}
\usepackage{array}
\usepackage{color}

\bibpunct{(}{)}{;}{a}{}{,}

\title{Three-dimensional neutrino-driven supernovae:
Neutron star kicks, spins, and asymmetric
ejection of nucleosynthesis products}
\abstract
{We present three-dimensional (3D) simulations of supernova explosions
  of nonrotating stars, triggered by the delayed neutrino-heating
  mechanism with a suitable choice of the core-neutrino
  luminosity. Our results show that asymmetric mass ejection caused by
  hydrodynamic instabilities can accelerate the neutron star (NS) up
  to recoil velocities of more than 700\,km\,s$^{-1}$ by the
  ``gravitational tug-boat mechanism'', which is sufficient to explain
  most observed pulsar space velocities.  The associated NS spin
  periods for our nonrotating progenitors are 
  about 100\,ms to 8000\,ms without any obvious correlation between
  spin and kick magnitudes or directions. This suggests that faster
  spins and a possible spin-kick alignment might require angular
  momentum in the progenitor core prior to collapse. Our simulations
  for the first time demonstrate a clear correlation between the size
  of the NS kick and anisotropic production and distribution of heavy
  elements created by explosive burning behind the shock. In the case
  of large pulsar kicks, the explosion is significantly stronger
  opposite to the kick vector. Therefore the bulk of the explosively
  fused iron-group elements, in particular nickel,
  are ejected mostly
  in large clumps against the kick direction. This contrasts with the
  case of low recoil velocity, where the nickel-rich lumps are more
  isotropically distributed.  Explosively produced intermediate-mass
  nuclei heavier than $^{28}$Si (like $^{40}$Ca and $^{44}$Ti) also
  exhibit significant enhancement in the hemisphere opposite to the
  direction of fast NS motion, while the distribution of $^{12}$C,
  $^{16}$O, and $^{20}$Ne is not affected, and that of $^{24}$Mg only
  marginally. Mapping the spatial distribution of the heavy elements
  in supernova remnants with identified pulsar motion may offer an
  important diagnostic test of the kick
  mechanism. Unlike kick
  scenarios based on anisotropic neutrino emission, our hydrodynamical
  acceleration model predicts enhanced ejection of iron-group elements
  and of their nuclear precursors in the opposite direction to the NS
  recoil.}
\author{A. Wongwathanarat \and
        H.-Th. Janka \and
        E. M\"uller}
\institute{Max-Planck Institut f\"{u}r Astrophysik,
  Karl-Schwarzschild-Stra\ss e 1, 85748 Garching, Germany}
\authorrunning{Wongwathanarat et al.}
\titlerunning{Neutron star kicks and spins and supernova nucleosynthesis}


\begin{document}
\maketitle

\section{Introduction}

Young pulsars have been measured to possess high
space velocities with an average value of several 100\,km\,s$^{-1}$
\citep[e.g.,][]{Arzoumanianetal02,Hobbsetal05,FaucherKaspi06}. Some
of them move with an estimated transverse velocity even higher 
than 500\,km\,s$^{-1}$, such as the compact stellar
remnant RX\,J0822-4300 in Puppis~A with a velocity of
$672\pm115$\,km\,s$^{-1}$ \citep{Beckeretal12}, and PSR B1508+55 with
a velocity of $1083^{+103}_{-90}$\,km\,s$^{-1}$
\citep{Chatterjeeetal05}. Such velocities are much too high to 
be explained by the disruption of close binary systems in the
supernova (SN) event that gave birth to the neutron star (NS).
Further evidence of natal kicks and associated NS
spin-up can be deduced from orbital and spin parameters and kinematic
information of double NS systems
\citep[e.g.,][]{Wongetal10,Farretal11,Lai01}. 
 
Numerous models for producing the kicks have been proposed, either
attempting to explain the NS recoil by anisotropic neutrino emission
during the Kelvin-Helmholtz cooling of the nascent remnant or by
asymmetric mass ejection in the SN explosion \citep[cf.][for a
  review]{Laietal01}. Both scenarios constitute fundamental
possibilities that exploit momentum conservation in the transition
from the progenitor star, on the one hand, to the relic compact object,
expelled gas, and $\sim$0.2$\,M_\odot$ of mass-equivalent neutrino
loss, on the other. The challenge, however, is connected to
explaining the exact physical reason of the
necessary asymmetry. 

Although only one percent anisotropy of the several $10^{53}$\,erg of
gravitational binding energy released through neutrinos in the NS
formation event would be sufficient to account for
an NS recoil 
velocity of about 300\,km\,s$^{-1}$, it turns out that an asymmetry of
this size or greater is extremely difficult to
obtain. Superstrong 
($\ga 10^{16}$\,G) internal, ordered magnetic fields, i.e., fields
with a very strong dipolar component, would be needed in the NS. Such
fields could impose a direction dependence on the matter-interactions
of active neutrinos in the dense NS core
\citep[e.g.,][]{Bisnovatyi-Kogan96,LaiQian98,SagertSchaffner08}, or
they could create an asymmetry of resonant Mikheev-Smirnov-Wolfenstein
flavor transformations between active neutrinos and sterile neutrinos
with masses of several keV
\citep[e.g.,][]{FryerKusenko06,Kusenko09,Kishimoto11}.  Also a
radiatively driven magnetoacoustic instability, termed neutrino-bubble
instability, has been proposed as a possible cause of sizable neutrino
emission anisotropies \citep{Socratesetal05}.  However, all these
scenarios require the favorable combination of several unestablished
ingredients.

In contrast, NS kick theories based on anisotropic mass ejection
during the explosion can be motivated by the observational fact that
core-collapse SNe and their gaseous remnants exhibit large-scale
asphericity essentially in all observed cases
\citep{Leonardetal06,WheelerAkiyama10,Vink12}. While
the fundamental requirement of the ``hydrodynamical'' recoil mechanism
is thus well consolidated, the exact origin and size of the mass
ejection asymmetry needs to be worked out and requires hydrodynamical
modeling for quantitative answers. \citet{JankaMueller94}, using
numerical simulations and analytic estimates, showed that the
interaction of the proto-neutron star (PNS) with the violent mass
flows in the neutrino-heated postshock layer prior to and around the
onset of the SN explosion cannot explain NS velocities of many
100\,km\,s$^{-1}$. This conclusion is supported by more recent
three-dimensional (3D) simulations by
\citet{FryerYoung07}. These works relied on
asymmetries that developed only after core 
bounce by neutrino-driven convection and the standing-accretion shock
instability \citep[SASI;][]{Blondinetal03,Foglizzo02}.  These
hydrodynamic instabilities can grow from random seed perturbations,
which can be expected to be present with small amplitudes in any
convectively stirred stellar environment. In contrast,
\citet{BurrowsHayes96} assumed the existence of a large, dipolar
asymmetry already in the precollapse stellar core as
the relic of 
convective shell-burning during the late evolution
stages of the 
progenitor star.  With the density accordingly reduced in a wide polar
funnel, they obtained a pronounced pole-to-pole asymmetry of the
explosion, because the ejecta could accelerate more readily to higher
velocities on the side of the low-density funnel. As a consequence,
the NS received a kick of more than 500\,km\,s$^{-1}$ opposite to the
direction of the stronger explosion. Although the assumed density
reduction might be motivated by the observation of vigorous dynamical
interaction between different burning shells in combination with large
(low-order mode) deviations from spherical symmetry in two-dimensional
(2D) stellar calculations for a stage close to core collapse
\citep{ArnettMeakin11}, the amplitude and structure of such density
and velocity perturbations in precollapse stellar
cores are still 
highly uncertain.  The determination of these perturbations requires
full-sphere 3D simulations over much longer periods of
precollapse evolution than the existing 2D models.\\
\\
\indent \citet{Schecketal04,Schecketal06} therefore also started with
minimal and least predetermining assumptions about inhomogeneities in
the progenitor before collapse. Like \citet{JankaMueller94} they
applied small random perturbations (typically 0.1\%--1\% of the
density and/or velocity) to trigger the development of large-scale,
nonradial flows in the postshock accretion layer, but in contrast to
the previous work they could follow the asymmetric, neutrino-powered
SN blast wave well beyond the onset of the explosion for evolution
times of at least one second (corresponding to shock radii of
$\sim$10,000\,km and more). Their results for a large set of 2D
simulations demonstrated that due to the action of neutrino-driven
convection and low-mode SASI activity, the postshock shell can be
expelled with sufficiently large asymmetry to accelerate the NS over a
timescale of seconds to velocities up to $\sim$1000\,km\,s$^{-1}$ and
more.  The NS recoil develops in the direction opposite to the side of
the strongest explosion.  It is mediated mainly by the gravitational
forces between the compact remnant and the most slowly moving, massive
ejecta ``clumps'', whose radial propagation lags behind the rest of
the expelled postshock material. The long-distance coupling by
gravitational forces can still act on the NS long after hydrodynamical
interactions have ceased. The influence of long-range gravity is
therefore much more efficient in accelerating the NS into a well
defined direction than the random kicks exerted by convective
downdrafts during the accretion phase. The impacts of accretion flows
buffet the accretor in varying directions. Such a random-walk like
process is ineffective in producing high recoil velocities. Recent 2D
results of \citet{Nordhausetal10,Nordhausetal12} lend support to the
``gravitational tug-boat mechanism''\footnote{We adopt this term
  following a suggestion by Jeremiah Murphy.}, and \citet{kickletter}
presented a first, small set of 3D simulations which showed that NS
recoil velocities of at least $\sim$500\,km\,s$^{-1}$ can also be
obtained in the absence of the artificial contraint to axisymmetry
associated with 2D modeling. 

In this paper we discuss results of a significantly extended set of 3D
simulations for nonrotating 15\,$M_\odot$ and 20\,$M_\odot$ progenitor
stars, some of which were continued for nearly 3.5 seconds beyond
bounce to follow the saturation of the NS kick velocity. As in the
previous works by \citet{Schecketal04,Schecketal06} and
\citet{kickletter}, we initiate the neutrino-driven explosions
artificially by imposing suitably chosen neutrino luminosities at the
inner grid boundary.  The latter is placed well inside the
neutrinosphere (at neutrino optically depths between $\sim$10 and some
100), and neutrino effects on the computational grid are treated by an
approximate description of the neutrino transport.  We also present
estimates of the spins that are acquired by the NS. Moreover, we
argue and demonstrate by results that the explosive production of
heavy elements during the first seconds of the SN blast becomes highly
anisotropic in the case of large NS kicks. In particular nuclei
between silicon and the iron group are ejected in considerably larger
amounts in the hemisphere pointing away from the NS velocity.  We
therefore predict that the bulk of such nucleosynthetic products
should possess a momentum opposite to that of a high-velocity
NS. Confirming such a hemispheric momentum asymmetry in SN remnants
with determined NS velocities would provide extremely valuable
information about the underlying NS kick mechanism. In particular, it
might allow one to discriminate our gravitational tug-boat mechanism
from scenarios like the neutrino-driven kicks discussed by
\citet{FryerKusenko06}, who expect the stronger explosion (and thus
enhanced explosive nucleosynthesis) in the direction of the NS motion.
High-resolution spectral investigations of young SN remnants like
Puppis~A \citep{Katsudaetal08,Katsudaetal10}, G11.2-0.3
\citep{Moonetal09}, and Cassiopeia~A \citep{Isenseeetal10,
Delaneyetal10, Restetal11, HwangLaming12} in different wavebands
may offer a promising perspective.

The outline of our paper is as follows.  In Sect.~\ref{sec:numerics}
we describe our numerical setup, employed methods, and the computed
models.  In Sect.~\ref{sec:kicks} we present our results for NS kicks,
discuss their connection to blast-wave asymmetries
(Sect.~\ref{sec:modeanalysis}), give a detailed description of the
physics of the NS acceleration mechanism (Sect.~\ref{sec:theory}),
provide analytic estimates of the achievable NS velocities
(Sect.~\ref{sec:analytics}), and address possible progenitor
dependences (Sect.~\ref{sec:progkick}) and the contribution from
anisotropic neutrino emission (Sect.~\ref{sec:neutrinokick}).  In
Sect.~\ref{sec:nsspins} we evaluate our models concerning the spin-up
of the nascent NSs, and in Sect.~\ref{sec:heavyelements} we discuss
our expectations for the connection between NS kicks and observable
asymmetries of the explosive nucleosynthesis of heavy
elements. Finally we summarize and draw conclusions in
Sect.~\ref{sec:summary}.


%
\begin{figure}
\centering
\includegraphics[width=0.75\hsize]{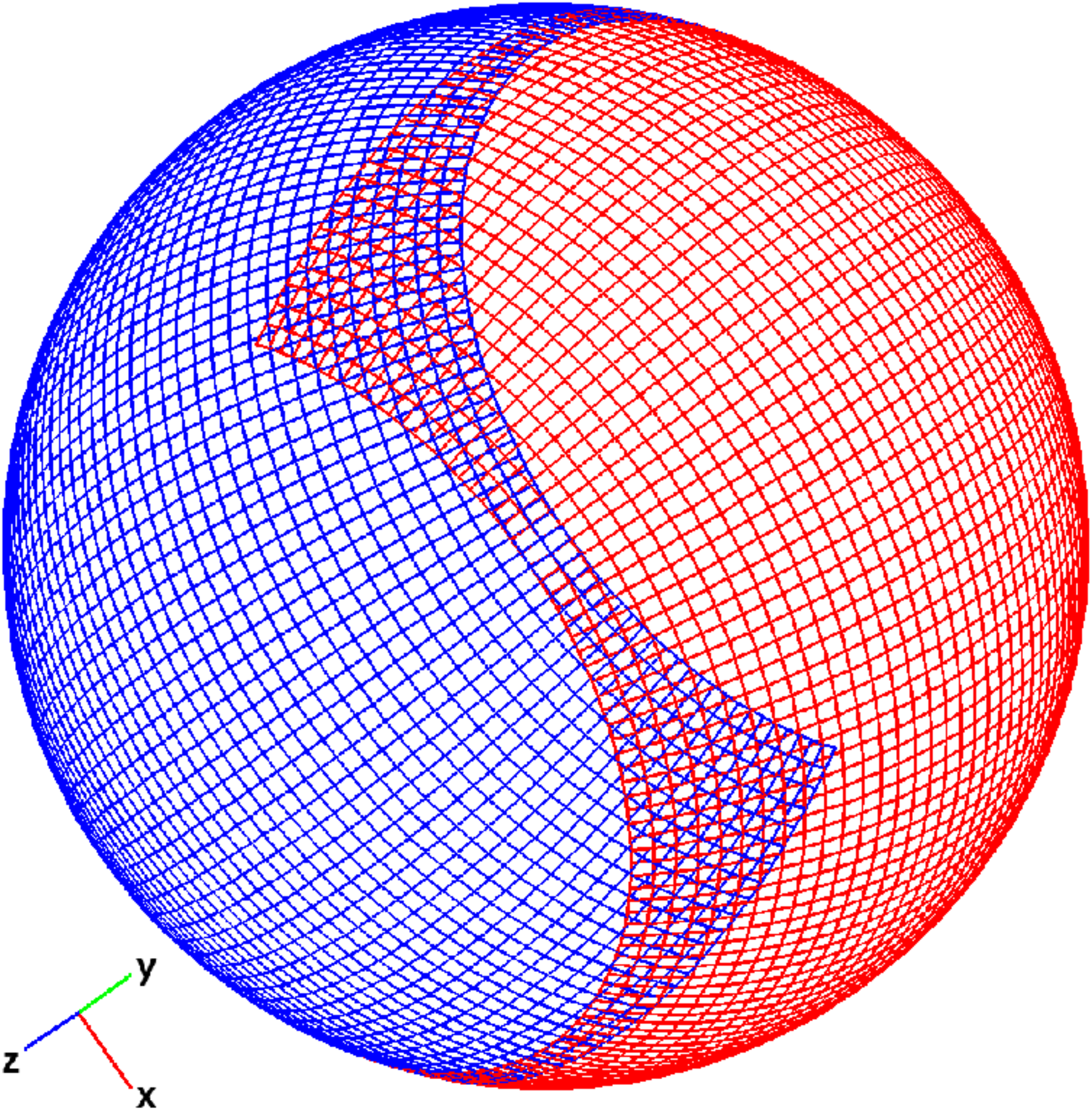}
\caption{Axis-free Yin-Yang grid plotted on a spherical surface
  demonstrating symmetric geometry of the Yin (red) and Yang (blue)
  regions. Both grids are the low-latitude part of a spherical polar
  grid, and therefore contain grid cells which are almost equidistant
  in angular directions.}
\label{fig:yygrid}
\end{figure}
%

\begin{table}
\setlength{\tabcolsep}{1.2mm}
\caption{Chosen values of parameters for all simulated models (see 
Sect.~\ref{sec:models} for definitions).}
\centering
\begin{tabular}{lccccccc}
\hline
\hline
\multirow{2}{*}{Model} & $L_{\nu_e,\mathrm{ib}}$ &
$L_{\bar{\nu_e},\mathrm{ib}}$ & $R_\mathrm{ib}$ & 
$R_\mathrm{ib}^f$ & $t_\mathrm{ib}$ & $R_\mathrm{ob}$ & $R_\mathrm{c}$ \\ 
 & [B/s] & [B/s] & [km] & [km] & [s] & [km] & [km] \\
\hline
W15-1    & 23.8 & 14.3 & 65.4 & 15 & 1 & 18000 & 80 \\
W15-2    & 23.8 & 14.3 & 65.4 & 15 & 1 & 18000 & 80 \\
W15-3    & 23.8 & 14.3 & 65.4 & 15 & 1 & 18000 & 80 \\
W15-4    & 23.8 & 14.3 & 65.4 & 15 & 1 & 20000 & 80 \\
W15-5-lr & 23.8 & 14.3 & 65.4 & 15 & 1 & 18000 & 80 \\
W15-6    & 23.8 & 14.3 & 65.4 & 15 & 1 & 18000 & 80 \\
W15-7    & 25.3 & 15.2 & 65.4 & 15 & 1 & 18000 & 80 \\
W15-8    & 22.3 & 13.4 & 65.4 & 15 & 1 & 18000 & 80 \\
\hline
L15-1    & 25.3 & 15.2 & 81.8 & 25 & 1 & 18000 & 115 \\
L15-2    & 28.3 & 17.0 & 81.8 & 25 & 1 & 18000 & 115 \\
L15-3    & 25.3 & 15.2 & 81.8 & 25 & 1 & 20000 & 115 \\
L15-4-lr & 25.3 & 15.2 & 81.8 & 25 & 1 & 18000 & 115 \\
L15-5    & 23.8 & 14.3 & 81.8 & 25 & 1 & 20000 & 115 \\
\hline
N20-1-lr & 32.8 & 19.7 & 65.4 & 15 & 1 & 18000 & 80 \\
N20-2    & 35.8 & 21.5 & 65.4 & 15 & 1 & 20000 & 80 \\
N20-3    & 32.8 & 19.7 & 65.4 & 15 & 1 & 20000 & 80 \\
N20-4    & 29.8 & 17.9 & 65.4 & 15 & 1 & 20000 & 80 \\
\hline
B15-1    & 26.8 & 16.1 & 54.5 & 15 & 1 & 22000 & 80 \\
B15-2    & 26.8 & 16.1 & 54.5 & 15 & 1 & 22000 & 80 \\
B15-3    & 25.3 & 15.2 & 54.5 & 15 & 1 & 22000 & 80 \\
\hline
\end{tabular}
\label{tab:parameters}
\end{table}

\section{Numerical methods and setup}
\label{sec:numerics}

\subsection{Code, grid, and boundary conditions}

Our numerical code is based on the explicit finite-volume, Eulerian,
multi-fluid hydrodynamics code {\sc Prometheus}
\citep{PROMET1,PROMET3,PROMET2}. It integrates the multi-dimensional
hydrodynamic equations using dimensional splitting following
\citet{Strang68}, piecewise parabolic reconstruction
\citep[PPM;][]{PPM}, and a Riemann solver for real gases
\citep{CollelaGlaz85}. Inside grid cells with strong grid-aligned
shocks, fluxes computed from the Riemann solver are replaced by the
AUSM+ fluxes of \citet{Liou96} in order to prevent odd-even decoupling
\citep{Quirk94}. The code treats advection of nuclear species by
employing the Consistent Multi-fluid Advection (CMA) scheme of
\citet{CMA}.

The code employs an axis-free overlapping ``Yin-Yang'' grid
\citep{YinYang} in spherical polar coordinates, which was recently
implemented into {\sc Prometheus}, for spatial discretization
\citep{YYmethod}. Figure~\ref{fig:yygrid} shows the Yin-Yang grid
plotted on a spherical surface. The Yin-Yang grid helps relaxing the
restrictive CFL-timestep condition and avoiding numerical artifacts
near the polar axis. 
Each grid patch is simply a part of the usual
  spherical polar grid, and is geometrically identical. This allows us
  to still make use of the ``ray-by-ray'' neutrino transport scheme
  already available in our code. On the other hand, it also means that
  the Yin-Yang grid still retains the problem of the singularity at
  the coordinate origin that is present in a spherical polar
  grid. This, however, is irrelevant in our simulations from the
  technical point of view because, as in \citet{Schecketal06}, we
  excise the central region of the computational domain interior to a
  chosen value of the time-dependent inner grid boundary
  R$_\mathrm{ib}$.
Our standard grid configuration consists of
$400(r)\times47(\theta)\times137(\phi)\times2$ grid cells, 
corresponding to an angular resolution of $2^\circ$ and covering the
full $4\pi$ solid angle. The grid configuration with reduced angular
resolution (distinguishable from the $2^\circ$
  configuration by the suffix ``lr'' in the model name) consists of
$400(r)\times20(\theta)\times56(\phi)\times2$ zones, corresponding to
an angular resolution of $5^\circ$. The radial 
grid has a constant spacing of 0.3\,km from the inner grid boundary
$R_\mathrm{ib}$ up to a radius $R_\mathrm{c}$ of
80\,km for the progenitors W15, 
N20, and B15, and 115\,km for the progenitor L15 (see
Sect.~\ref{sec:models}). Beyond this radius the radial grid is
logarithmically spaced. The outer grid boundary $R_\mathrm{ob}$ is at
18000--22000\,km, which is sufficient to prevent the
supernova (SN) 
shock from leaving the computational domain during the simulated
time. Hydrostatic equilibrium is assumed at the inner grid boundary
$R_\mathrm{ib}$, while a free outflow boundary condition is employed
at the outer one.

\subsection{Input physics}

Self-gravity is taken into account by solving Poisson's equation in
its integral form, using an expansion into spherical harmonics as in
\citet{MuellerSteinmetz95}.  The monopole term of the potential is
corrected for general relativistic effects as described in
\citet{Schecketal06} and \citet{Arconesetal07}.  The inner core of the
PNS with densities well above those of the neutrinospheric layer is
excised and replaced by a point mass pinned to the
coordinate origin. Thus the PNS core is not allowed to move. The
implication of this fact will be discussed in Sect.~\ref{sec:theory}. The 
cooling of the PNS is then described by neutrino emission properties
(luminosities and mean spectral energies) that are prescribed as
boundary conditions at $R_\mathrm{ib}$. The boundary
  neutrino luminosities are assumed to be constant during the first
  second of the evolution and decline with time as $t^{-3/2}$ afterwards
  \citep[cf.][]{Schecketal06,Arconesetal07}. The mean
  energies are prescribed as functions of the temperature in the first
  radial cell of the computational grid as in
  \citet{Uglianoetal12}. The contraction of the PNS is 
mimicked by a radial grid movement with defined velocity according to
\citet{Schecketal06}; detailed information on the chosen parameters
can be found in Table~\ref{tab:parameters}. 
There is no mass flow through the inner grid
boundary as long as we follow the evolution of the PNS surface layers.

``Ray-by-ray'' neutrino transport and neutrino-matter interactions are
approximated as in \citet{Schecketal06} by radial integration of the
one-dimensional (spherical), grey transport equation for $\nu_e$,
$\bar\nu_e$, and heavy-lepton neutrinos and all angular grid
directions ($\theta$,\,$\phi$) independently. This approach allows us
to take into account angular variations in the
neutrino fluxes 
produced by the matter accreted onto the newly forming NS. The
accretion luminosity adds to the neutrino fluxes imposed spherically
symmetrically at the inner grid boundary so that
  the outgoing luminosities are considerably different from the values
  assumed at the inner grid boundary \citep[see,
    e.g.,][]{Schecketal06,Arconesetal07}. The neutrino spectra are
assumed to have Fermi-Dirac shape with chemical potentials that are
equal to the equilibrium values in optically thick regions and
constant, limiting values in the free streaming regime. The spectral
temperatures are computed from inverting the ratio of neutrino energy
and number fluxes, whose transport is solved simultaneously. Details
can be found in \citet{Schecketal06}. The tabulated equation of state
(EoS) of \citet{JankaMueller96} is used to describe the stellar
fluid. It includes arbitrarily degenerate and arbitrarily relativistic
electrons and positrons, photons, and four predefined nuclear species
(n, p, $\alpha$, and a representative Fe-group nucleus) in nuclear
statistical equilibrium.
 
\subsection{Nucleosynthesis}
\label{sec:nucleosynthesis}
 
In order to follow approximately the explosive nucleosynthesis, we
solve a small $\alpha$-chain reaction network, similar to the network
described in \citet{Kifonidisetal03}. For our simulations we choose a
subset of nine $\alpha$ nuclei, $^4$He, $^{12}$C, $^{16}$O, $^{20}$Ne,
$^{24}$Mg, $^{28}$Si, $^{40}$Ca, $^{44}$Ti, $^{56}$Ni, and an
additional ``tracer nucleus''. We omit $^{32}$S, $^{36}$Ar, $^{48}$Cr,
and $^{52}$Fe from the complete $\alpha$-series in order to reduce the
network size and thus to save computing time.  The two gaps between
$^{28}$Si --$^{40}$Ca and $^{44}$Ti --$^{56}$Ni are bridged by combining
the intermediate reaction steps and estimating the new effective
reaction rates by taking the slowest rate of the intermediate
reactions.  The tracer nucleus is produced via the reaction
$^{44}$Ti(3$\alpha$,$\gamma$)$^{56}$Ni within grid cells whose
electron fraction $Y_e$ is below 0.49. Such conditions are found in
the neutrino-heated ejecta of our models. The tracer nucleus
represents iron-group species that are formed under conditions of
neutron excess. It thus allows us to keep track of nucleosynthesis in
neutron-rich regions. However, it should be noted that some fraction
of the tracer material may actually be $^{56}$Ni, because our
approximations in the neutrino transport tend to underestimate $Y_e$
in the neutrino-heated ejecta, while more sophisticated
energy-dependent neutrino transport yields (slightly) proton-rich
conditions in the neutrino-processed material expelled during the
early explosion
\citep[e.g.,][]{Pruetetal06,Froehlichetal06,Fischeretal10,BMuelleretal12}. 
The network is solved in grid cells whose temperature is within the
range of $10^8$\,K and $8\times10^9$\,K. We neglect the feedback from
the network composition to the EoS and the effect of the nuclear
energy release on the hydrodynamic flow. The energy release by nuclear
reactions is of minor relevance for the dynamics because the
production of 0.1\,M$_\odot$ of $^{56}$Ni means a contribution of only
$\sim10^{50}$\,erg to the explosion energy. It is important to note
that our model parameters are chosen to give energetic explosions
already by neutrino energy input. Not using an exact nuclear
composition in the EoS is an acceptable approximation because at
neutrino-heated ejecta conditions the contributions of nuclei to
pressure, energy density, and entropy are dwarfed by those of
electrons, positrons, and photons. To prevent too small nuclear
burning timesteps we do not perform any network calculations above
$8\times10^9$\,K. When freeze-out from nuclear statistical equilibrium
(NSE) is followed in ejecta cooling down from above $8\times10^9$\,K,
we start from pure $\alpha$-composition, which is compatible with the
high-temperature NSE composition in the absence of free neutrons and
protons, in particular since $Y_e=n_e/n_B$ has values close to 0.5 in
all of the ejecta. As the hot ejecta expand and their temperature
decreases these $\alpha$-particles recombine to produce heavier
$\alpha$-nuclei considered in our $\alpha$-network, among them
$^{56}$Ni. Once the temperature drops below $10^8$\,K, all nuclear
reactions are switched off because they become too slow to change the
nuclear composition on the explosion timescale. Nuclear burning in
shock-heated matter is also described by our $\alpha$-network.  

\subsection{Long-time simulations}
\label{sec:long-time}
 
To follow the propagation of the SN shock wave until very late times,
we map our results after 1.1--1.4\,s onto a new computational grid,
whose inner and outer radial boundaries are placed at 500\,km and
$3.3\times10^8$\,km, respectively. The latter value is near the
stellar surface. At the inner grid boundary we assume the inflow of a
spherically symmetric neutrino-driven wind, which is considered to be
the consequence of ongoing neutrino emission and corresponding energy
deposition near the PNS.  We do not directly simulate neutrino-matter
interactions during our long-time runs, because details of the
neutrino physics become less important during the late evolution
phases when the explosion conditions are already determined. By
including a wind inflow, however, we continue to account for the
consequences of neutrino heating at the PNS surface. Since the imposed
wind has spherical symmetry it will not affect the anisotropy of the
leading ejecta that we intend to investigate. The hydrodynamic inflow
quantities are obtained as the angular averages of the flow properties
at 500\,km at the time of the mapping and kept constant for another
2\,s of evolution.  

The wind is replaced by a free outflow boundary condition
afterwards. At the outer boundary we apply a free outflow condition at
all times. Since we keep the angular grid resolution unchanged but
increase the number of radial zones, our computational grid grows to
$1200(r)\times47(\theta)\times137(\phi)\times2$ zones. The radial
resolution is approximately $\Delta r/r \approx 0.01$. We successively
discard the innermost radial zone and thus move the inner grid
boundary outward to a new location whenever its radius becomes less
than 2\% of the minimum shock radius. Moving the inner grid boundary
to a larger radius helps relaxing the CFL timestep and allows us to
follow the evolution to very late times with acceptable computational
costs. 

Also during the long-time runs we consider nucleosynthesis processes
and account for the self-gravity of the stellar gas by a Newtonian
description. We switch to the tabulated EoS of \citet{TimmesSwesty00},
appropriate for the lower density and temperature values present in
the outer stellar layers \citep[not covered by the EoS table
  of][]{JankaMueller96}. Moreover, besides considering the 
contributions from arbitrarily degenerate and relativistic electrons
and positrons and from a photon gas, it treats the nucleonic gas
components by a mix of ideal Boltzmann gases with 11 nuclear species:
p, $^4$He, $^{12}$C, $^{16}$O, $^{20}$Ne, $^{24}$Mg, $^{28}$Si,
$^{40}$Ca, $^{44}$Ti, $^{56}$Ni, and $X$.  This is sufficient for a
reasonably good representation of the composition in the outer core,
mantle, and envelope of the progenitor.

\subsection{Computed models}
\label{sec:models}

We investigate three 15\,$M_\odot$ models and one 20\,$M_\odot$
progenitor denoted as W15, L15, B15, and N20, respectively. W15 is
based on the nonrotating 15\,$M_\odot$ model s15s7b2 of
\citet{WoosleyWeaver95}, L15 corresponds to a star evolved by
\citet{Limongi2000}, B15 is a blue supergiant (SN~1987A) progenitor of
\citet{Woosleyetal88}, and N20 was computed up to core collapse by
\citet{ShigeyamaNomoto90}. 
  The 15\,$M_\odot$ progenitors W and L were
  followed through collapse to 15\,ms after bounce with the {\sc
  Prometheus-Vertex} code \citep{RamppJanka02} in
  one dimension (A.~Marek and R.~Buras, private communication). The {\sc
  Prometheus-Vertex} code is an Eulerian multi-D hydro-code
  (based on a finite volume method) coupled to a two-moment closure
  scheme with variable Eddington factor for neutrino transport. Using
  the same code, the progenitor N20 was evolved until 11\,ms after
  bounce (A.~Marek, private communication), whereas the model B15 was
  collapsed in a 1D simulation \citep{Bruenn93} using a Lagrangian
  hydro-code with   multi-group flux-limited neutrino diffusion
  \citep{Bruenn85}. Both simulation codes are Newtonian, but {\sc
    Prometheus-Vertex} employs correction of the gravitational
  potential due to general relativistic effects \citep{Mareketal06}. 
To break spherical symmetry, random seed perturbations with an
amplitude of 0.1\% (and 
cell-to-cell variations) are imposed on the radial velocity ($v_r$)
field\footnote{Unfortunately, 3D progenitor models with self-consistent
  information about the asymmetries present in the stellar core before
  collapse are not available yet.}. 
Explosions with chosen energy are initiated by neutrino heating at a
rate that depends on suitable values of the neutrino luminosities
assumed at the lower boundary.

We computed in total 20 models varying the neutrino luminosities
imposed at the inner grid boundary and thus the resulting explosion
energies (Table~\ref{tab:results}). The choice of parameters for each
model is listed in Table~\ref{tab:parameters}. These are
$L_{\nu_e,\mathrm{ib}}$ and $L_{\bar{\nu_e},\mathrm{ib}}$, the
luminosities of electron neutrinos and electron antineutrinos at the
inner grid boundary, respectively\footnote{We repeat, however, that we
  do not use a light-bulb approximation, but in our simulations the
  outgoing neutrino luminosities are considerably modified by
  accretion contributions to the values imposed at the inner grid
  boundary and listed in Table~\ref{tab:parameters}}, $R_\mathrm{ib}$,
the initial radius of the inner grid boundary, $R_\mathrm{ib}^f$, the
asymptotic radius of the inner grid boundary as $t\rightarrow\infty$,
$t_\mathrm{ib}$, the timescale for the boundary contraction,
$R_\mathrm{ob}$, the radius of the outer grid boundary, and
$R_\mathrm{c}$, the radius up to where the radial grid resolution is
kept constant at 0.3\,km. Some of our models (W15-1 and W15-2; B15-1
and B15-2) differ only by the initial seed perturbations.  The
evolution including our treatment of neutrino transport is followed
until 1.3\,s after bounce for the W15 and N20 models, while the L15
simulations are continued until 1.4\,s postbounce and the B15 runs
until about 1.1\,s after bounce. The time at which we stop our
calculations differs from progenitor to progenitor depending on how
fast the explosion sets in.  Explosions in the L15 models are the most
delayed ones, while the B15 models show the fastest explosion. The
simulations with detailed neutrino treatment are stopped when the
maximum SN shock's radius is close to $R_{\mathrm{ob}}$. Models which
are simulated with the reduced angular grid resolution of 
$5^\circ$ are indicated by the suffix ``lr'' appended to their model
names. In addition, a subset of 9 models is continued into the
long-time evolution as described in Sect.~\ref{sec:long-time} in order
to study the further acceleration of the PNS and the 
explosive nucleosynthesis behind the outgoing shock.

\begin{table*}
\caption{Explosion and NS properties for all models (see 
Sect.~\ref{sec:simresults} for definitions).}
\centering
\begin{tabular}{lcccccccccccc}
\hline
\hline
\multirow{2}{*}{Model} &
$M_\mathrm{ns}$ & $t_\mathrm{exp}$ & $E_\mathrm{exp}$ &
$v_\mathrm{ns}$ & $a_\mathrm{ns}$ & $v_{\mathrm{ns},\nu}$ &
$\alpha_{\mathrm{k}\nu}$ & $v_\mathrm{ns}^{\mathrm{long}}$ &
$a_\mathrm{ns}^{\mathrm{long}}$ & $J_{\mathrm{ns},46}$ &
$\alpha_\mathrm{sk}$ & $T_\mathrm{spin}$\\
      & [$M_\odot$] & [ms] & [B] &
[km/s] & [km/s$^2$] & [km/s] & [$^\circ$] & [km/s] & [km/s$^2$] &
[10$^{46}$\,g\,cm$^2$/s] & [$^\circ$] & [ms]\\
\hline
W15-1
& 1.37 & 246 & 1.12 & 331 & 167 & 2 & 151 & 524 &  44 & 1.51 & 117 & 652\\
W15-2
& 1.37 & 248 & 1.13 & 405 & 133 & 1 & 126 & 575 &  49 & 1.56 & 58  & 632\\
W15-3
& 1.36 & 250 & 1.11 & 267 & 102 & 1 & 160 & -   &  -  & 1.13 & 105 & 864\\
W15-4
& 1.38 & 272 & 0.94 & 262 & 111 & 4 & 162 & -   &  -  & 1.27 & 43  & 785\\
W15-5-lr
& 1.41 & 289 & 0.83 & 373 & 165 & 2 & 129 & -   &  -  & 1.63 & 28  & 625\\
W15-6
& 1.39 & 272 & 0.90 & 437 & 222 & 2 & 136 & 704 &  71 & 0.97 & 127 & 1028\\
W15-7
& 1.37 & 258 & 1.07 & 215 & 85  & 1 &  81 & -   &  -  & 0.45 & 48  & 2189\\
W15-8
& 1.41 & 289 & 0.72 & 336 & 168 & 3 & 160 & -   &  -  & 4.33 & 104 & 235\\
\hline
L15-1
& 1.58 & 422 & 1.13 & 161 & 69  & 5 & 135 & 227 &  16 & 1.89 & 148 & 604\\
L15-2
& 1.51 & 382 & 1.74 & 78  & 14  & 1 & 150 & 95  &  4  & 1.04 & 62  & 1041\\
L15-3
& 1.62 & 478 & 0.84 & 31  & 27  & 1 &  51 & -   &  -  & 1.55 & 123 & 750\\
L15-4-lr
& 1.64 & 502 & 0.75 & 199 & 123 & 4 & 120 & -   &  -  & 1.39 & 93  & 846\\
L15-5
& 1.66 & 516 & 0.62 & 267 & 209 & 3 & 147 & 542 & 106 & 1.72 & 65  & 695\\
\hline
N20-1-lr
& 1.40 & 311 & 1.93 & 157 & 42  & 7 & 118 & -   &  -  & 5.30 & 122 & 190\\
N20-2
& 1.28 & 276 & 3.12 & 101 & 12  & 4 & 159 & -   &  -  & 7.26 & 43  & 127\\
N20-3
& 1.38 & 299 & 1.98 & 125 & 15  & 5 & 138 & -   &  -  & 4.42 & 54  & 225\\
N20-4
& 1.45 & 334 & 1.35 & 98  & 18  & 1 & 98  & 125 &  9  & 2.04 & 45  & 512\\
\hline
B15-1
& 1.24 & 164 & 1.25 & 92  & 16  & 1 & 97  & 102 &  1  & 1.03 & 155 & 866\\
B15-2
& 1.24 & 162 & 1.25 & 143 & 37  & 1 & 140 & -   &  -  & 0.12 & 162 & 7753\\
B15-3
& 1.26 & 175 & 1.04 & 85  & 19  & 1 & 24  & 99  &  3  & 0.44 & 148 & 2050\\
\hline
\end{tabular}
\label{tab:results}
\end{table*}

%
\begin{figure*}
\centering
\includegraphics[width=0.35\hsize]{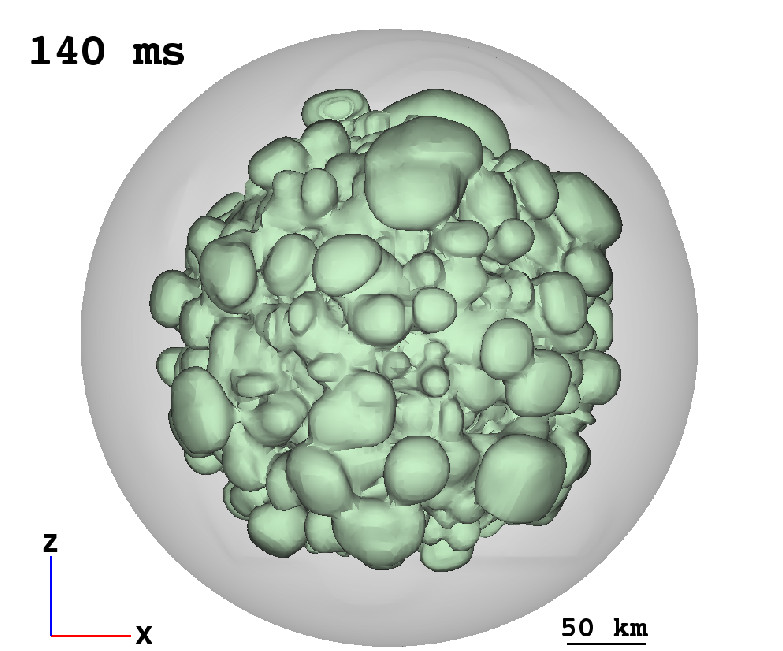}\hspace{1cm}
\includegraphics[width=0.35\hsize]{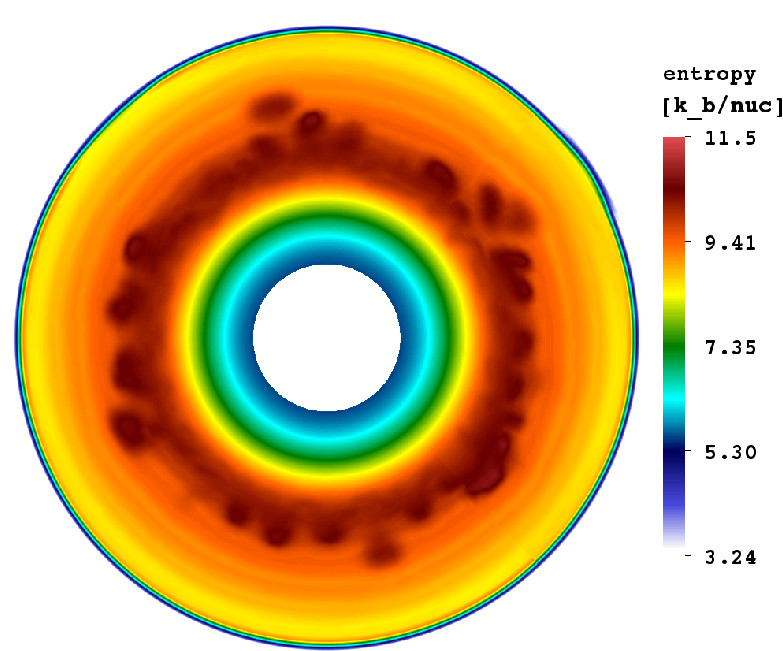}\\
\includegraphics[width=0.35\hsize]{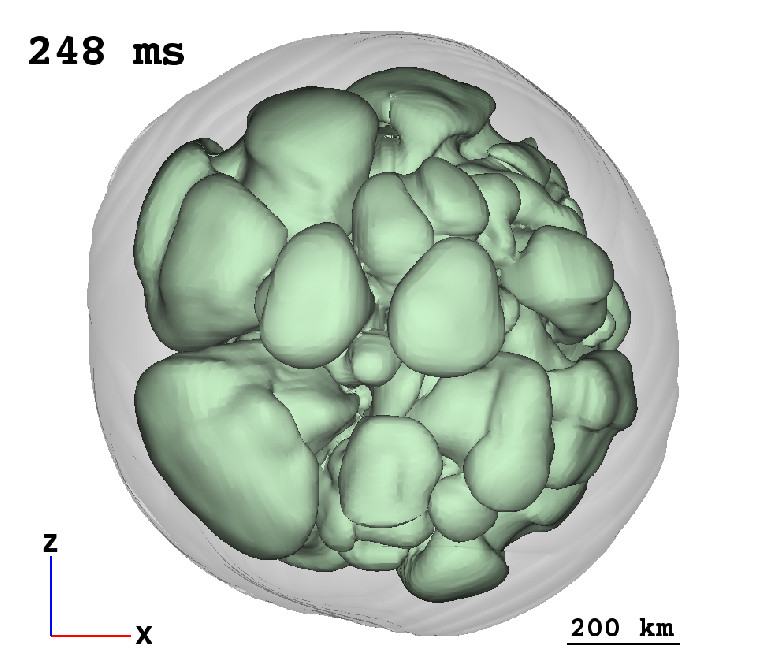}\hspace{1cm}
\includegraphics[width=0.35\hsize]{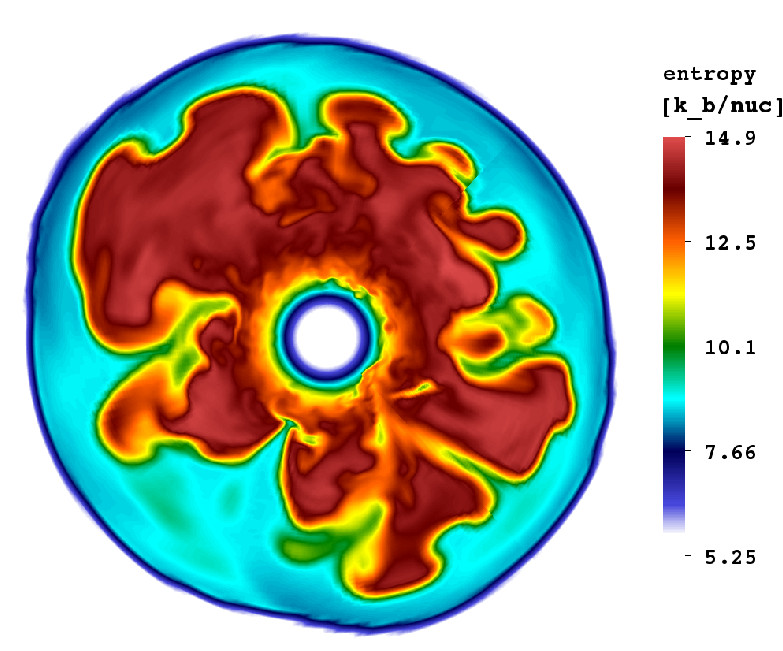}\\
\includegraphics[width=0.35\hsize]{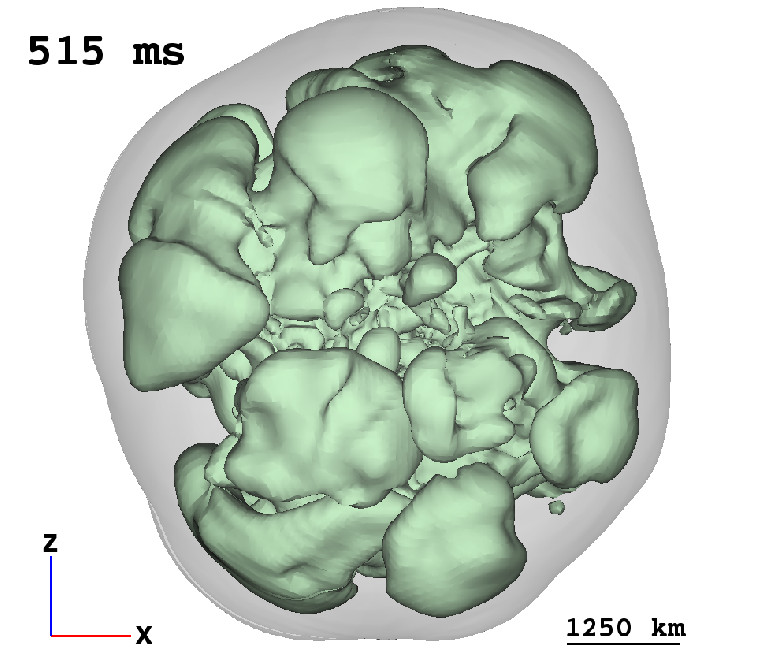}\hspace{1cm}
\includegraphics[width=0.35\hsize]{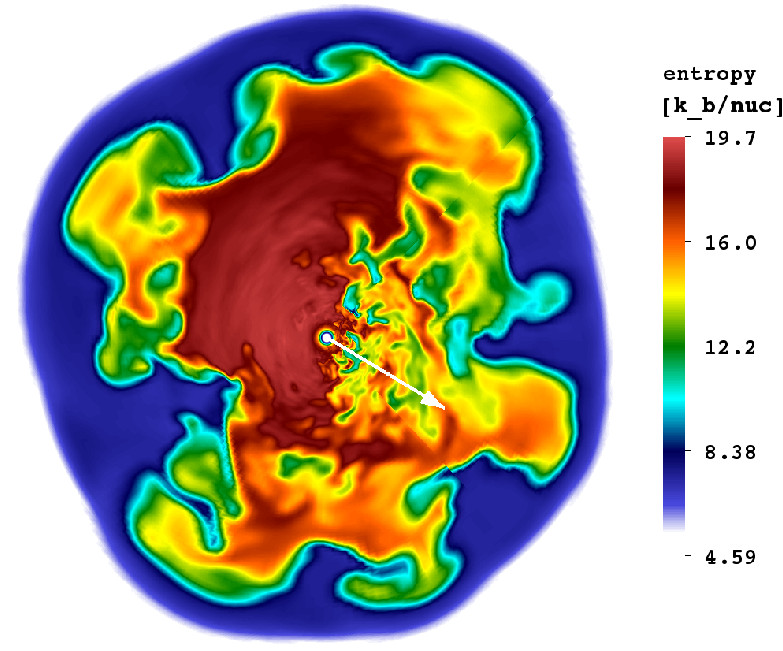}\\
\includegraphics[width=0.35\hsize]{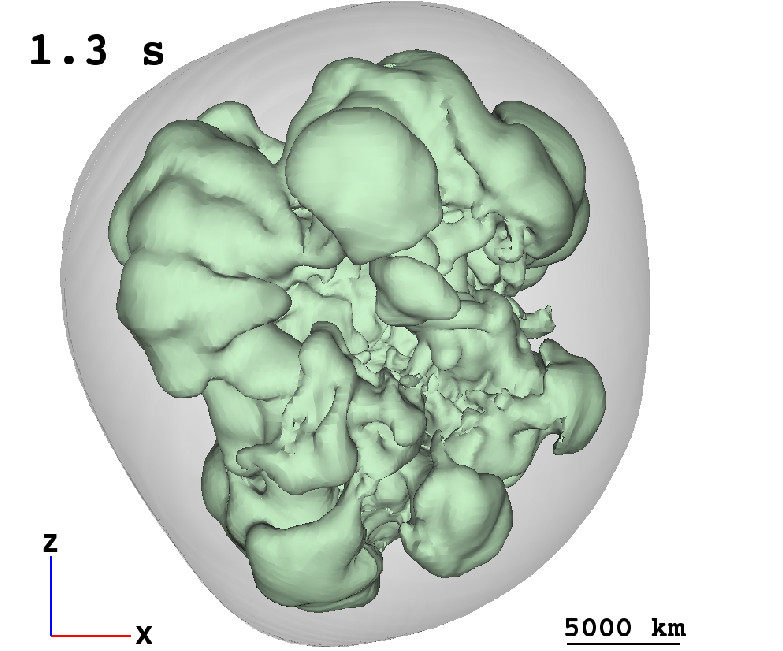}\hspace{1cm}
\includegraphics[width=0.35\hsize]{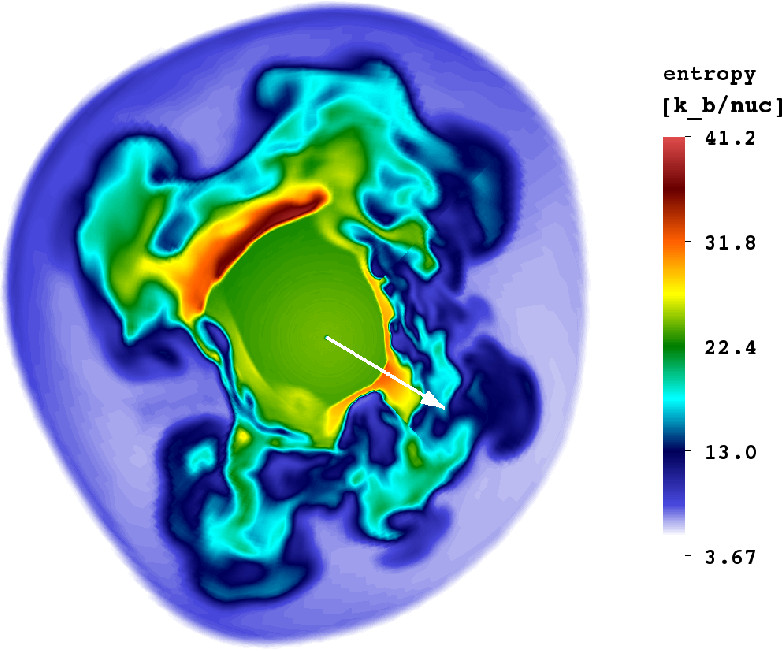}\\
\caption{Entropy-isosurfaces ({\em left}) of the SN shock (grey) and
  of the high-entropy bubbles (green), and entropy distribution in a
  cross-sectional plane ({\em right}) at $t=140,248,515$\,ms, and
  $1.3$\,s for model W15-2. The viewing direction is normal to the
  $xz$-plane. Length scales are given by the yardsticks below each
  image. Small Rayleigh-Taylor mushrooms start growing at
  $\sim100$\,ms ({\em top panels}). The rising high-entropy bubbles
  merge and rearrange to form larger bubbles by the time the explosion
  sets in ({\em second from top}). The NS starts accelerating due to
  the asymmetric distribution of the ejecta, the acceleration reaching
  its maximum at $\sim$500\,ms ({\em third from top}).  At this epoch,
  the ejecta show a clear dipolar distribution with more dense,
  low-entropy material concentrated in the hemisphere containing the
  kick direction.  The final NS kick direction has already found its
  orientation, and its {\em projection} onto the $xz$-plane is shown
  by the white arrows pointing to the lower right direction. The {\em
    bottom panels} show the entropy structure when the essentially
  spherically symmetric neutrino-driven wind has developed, visible as
  the green central region enclosed by the highly aspherical
  wind-termination shock (lower right panel).}
\label{fig:W15-2sequence}
\end{figure*}
%

%
\begin{figure*}
\centering
\includegraphics[width=0.33\hsize]{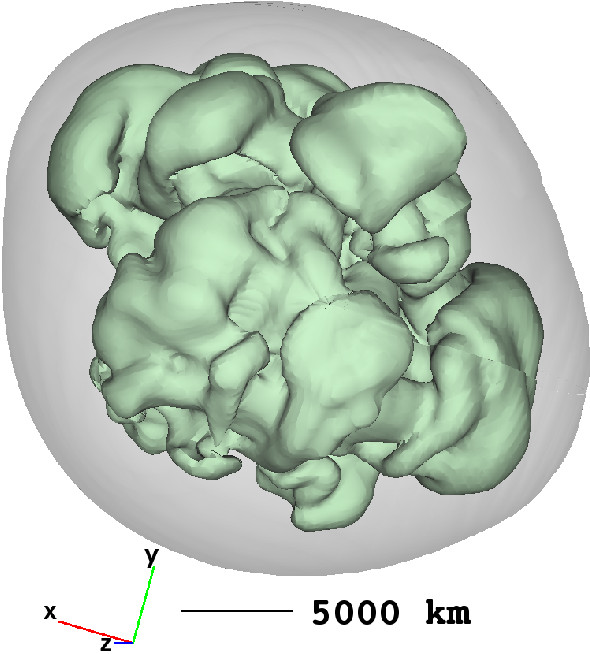}
\includegraphics[width=0.33\hsize]{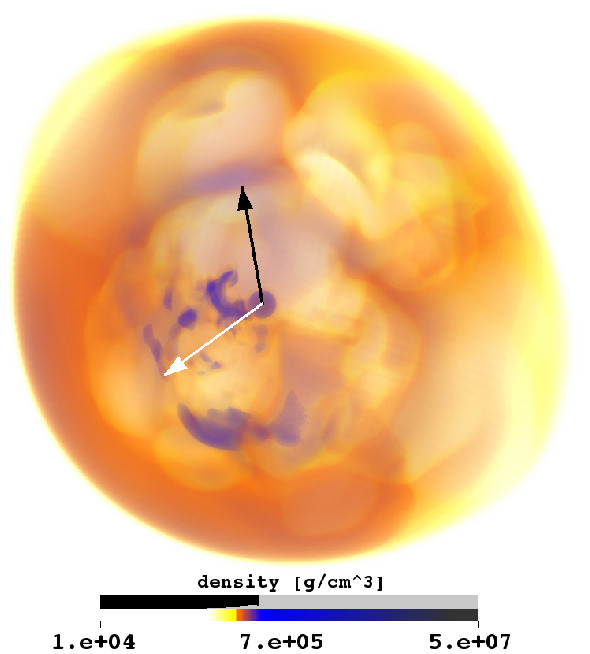}
\includegraphics[width=0.33\hsize]{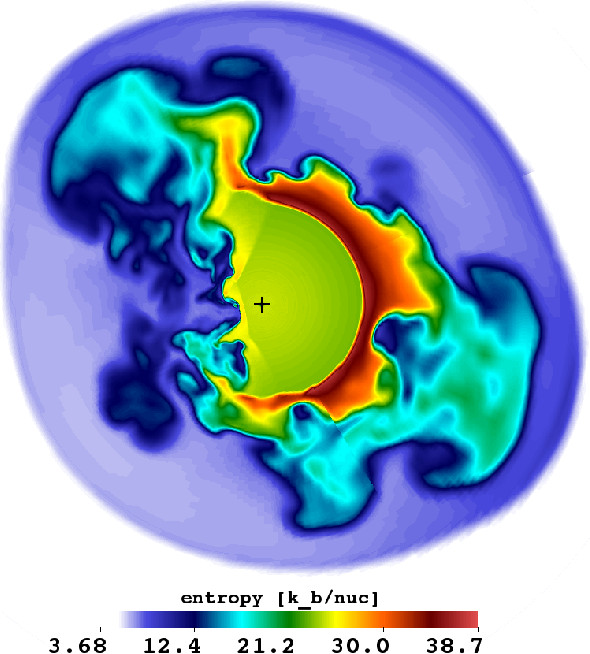}\\
\includegraphics[width=0.33\hsize]{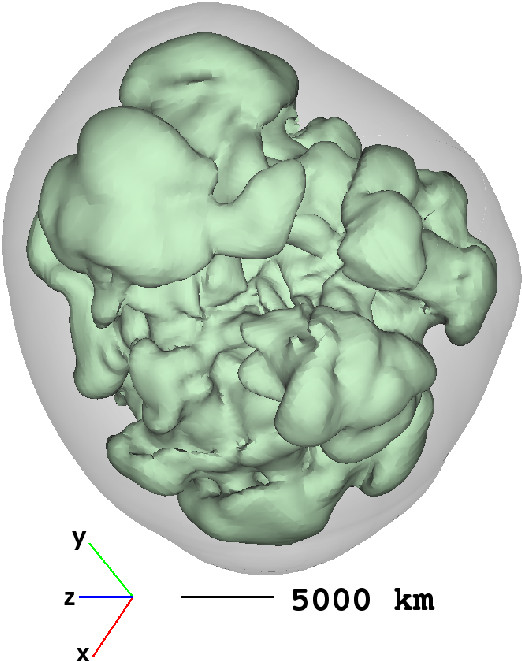}
\includegraphics[width=0.33\hsize]{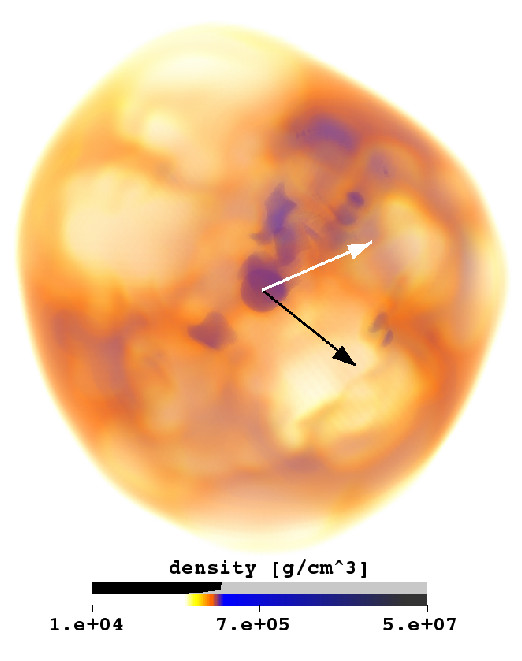}
\includegraphics[width=0.33\hsize]{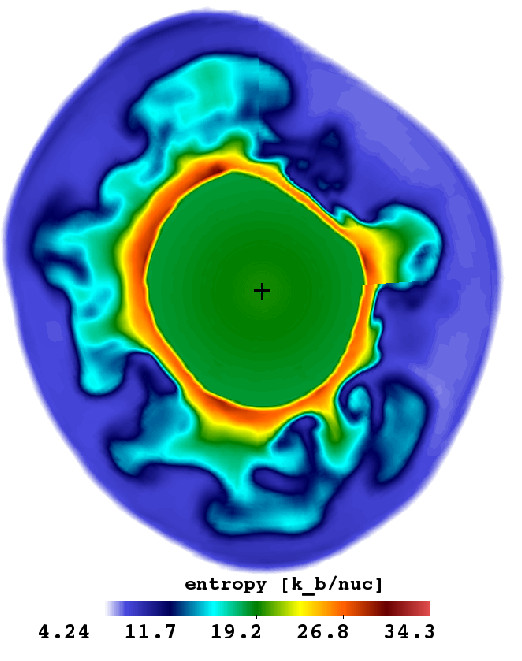}\\
\caption{Entropy-isosurfaces ({\em left}) of the SN shock (grey) and
  of the high-entropy bubbles (green), ray-casting images of the
  density ({\em middle}), and entropy distribution in a
  cross-sectional plane ({\em right}) for models W15-1 ({\em top}) and
  L15-2 ({\em bottom}) at about 1.3\,s and 1.4\,s after core bounce.
  The viewing directions are normal to the plane spanned by the NS
  kick and spin vectors of each model, which also defines the plane
  for the entropy slices. The SN shock has an average radius of
  13,000\,km and 14,000\,km for models W15-1 and L15-2, respectively
  (a length of 5000\,km is given by yardsticks below the left images).
  The kick and spin directions are indicated by the white and black
  arrows, respectively, in the {\em middle figures}. The NS locations
  are marked by black crosses in the {\em right plots}.  The images in
  the {\em middle} correspond roughly to the projections of the
  density distribution on the observational planes. Dilute bubble
  regions are light-colored in white and yellow, while dense clumps
  appear more intense in reddish and bluish hues. The purple circular
  areas around the NS represent the dense inner region of the
  essentially spherically symmetric neutrino-driven wind. The wind is
  visible in green in the {\em right images} and is bounded by the
  aspherical wind-termination shock. The NS is clearly displaced from
  the geometrical center of the expanding shock in the direction of
  the kick vector pointing to the lower left for model W15-1. The NS
  in model L15-2 has a much smaller kick velocity than in model W15-1,
  and does not show any clear displacement but remains located roughly
  at the center of the expanding ejecta shell.}
\label{fig:kick}
\end{figure*}
%

\section{Neutron star kicks}
\label{sec:kicks}

\citet{kickletter} presented results of NS kicks for a small set of 3D
computations conducted until 1.3--1.4\,s after bounce. Here, we
describe the evolution of these models for a much longer time as well
as more results of a significantly enlarged model set that includes
runs for three 15\,$M_\odot$ progenitors and the 20\,$M_\odot$
star. Moreover, we analyze our numerical results in the light of
theoretical considerations and simple analytical estimates (extending
the brief discussion in \citealt{kickletter} and sharpening arguments
made by \citealt{Schecketal06} and \citealt{Nordhausetal10}). We also
evaluate our simulations for recoil velocities caused by anisotropic
neutrino emission in comparison to the far dominant kicks by
asymmetric mass ejection.

\subsection{Simulation results}
\label{sec:simresults}

Figure~\ref{fig:W15-2sequence} presents snapshots of model W15-2 at
four different epochs: 0.140, 0.248, 0.515, and 1.30\,s after
bounce. Violent buoyancy and mass overturn is triggered by neutrino
heating around the NS.  At roughly 100\,ms, Rayleigh-Taylor mushrooms
begin to grow from the imposed seed perturbations in the convectively
unstable layer that builds up between the gain radius (that separates
neutrino-cooling and -heating regions) and the stalled SN shock.
These high-entropy bubbles evolve in a highly non-stationary manner,
start rising, growing, merging, partially collapse again and emerge
once more to inflate to larger sizes. Along with the slowly expanding
shock their angular diameter increases, adjusting roughly to the
radial extension of the convective layer. Aided by convective overturn
and global shock motions, possibly associated with activity due to the
SASI, the delayed, neutrino-driven explosion sets in at about 250\,ms
postbounce.

To identify the explosion epoch, we perform a time-dependent
evaluation of the explosion energy $E_{\mathrm{exp}}$ defined as the
sum of the total (i.e., internal plus kinetic plus gravitational)
energy of all grid cells where this energy is positive. We define the
explosion time $t_{\mathrm{exp}}$ as the moment when the computed
explosion energy exceeds $10^{48}$\,erg. This roughly coincides with
the time when the average shock radius exceeds 400--500\,km.

This is also the time when the NS develops a growing kick velocity
along a well defined direction. While the kick direction was randomly
fluctuating before, the acceleration vector now settles and the NS
velocity grows \citep[see][]{kickletter}. A cross-sectional view of
the entropy distribution clearly shows a dipolar asymmetry with dense,
low-entropy material concentrated at small radii primarily in the
hemisphere to which the kick vector is pointing
(Fig.~\ref{fig:W15-2sequence}, third row and also Fig.~\ref{fig:kick},
upper row).  At a later stage, the pattern of the nonradial structures
freezes and the ejecta continue to expand in a nearly self-similar
manner. An entropy slice shows the emergence of an essentially
spherically symmetric neutrino-driven wind enclosed by an aspherical
wind-termination shock. On the side where the explosion is stronger,
which manifests itself by faster shock expansion and higher entropy
behind the SN shock (see the 10--11 o'clock position in the bottom
right panel of Fig.~\ref{fig:W15-2sequence} and the upper right panel
of Fig.~\ref{fig:kick}), the wind is shocked to higher entropies
because it passes the termination shock at larger radii and thus with
higher velocity (see also \citealt{Arconesetal07} and
\citealt{ArconesJanka11}).

Table~\ref{tab:results} summarizes explosion and NS properties for all
computed models: $M_\mathrm{ns}$, $t_\mathrm{exp}$, $E_\mathrm{exp}$,
$v_\mathrm{ns}$, $a_\mathrm{ns}$, $v_{\mathrm{ns},\nu}$,
$\alpha_{\mathrm{k}\nu}$, $J_{\mathrm{ns}}$, $\alpha_\mathrm{sk}$, and
$T_\mathrm{spin}$ denote the NS mass, time of the onset of the SN
explosion, explosion energy, NS kick velocity by the gravitational
tug-boat mechanism, corresponding NS acceleration, additional
neutrino-induced kick velocity, angle between gravitational and
neutrino-induced NS kick, NS angular momentum, angle between NS spin
and kick directions, and estimated final NS spin period.  While
results for all models are provided after 1.1--1.4\,s of post-bounce
evolution, $v_\mathrm{ns}^{\mathrm{long}}$ and
$a_\mathrm{ns}^{\mathrm{long}}$ give the NS velocity and acceleration
at 3.1--3.4\,s after bounce for the subset of the simulations
that was continued until this later time.

The NS mass $M_\mathrm{ns}$ is defined as the baryonic mass enclosed
by the NS radius $R_\mathrm{ns}$, which is considered as the angular
dependent radial location where the density is
$10^{11}$\,g\,cm$^{-3}$. In estimating the NS recoil velocity,
we assume conservation of the linear momentum of the whole progenitor
star.  Since the high-density core of the PNS is excised in our
simulations, the NS cannot move when it exchanges momentum with the
surrounding matter. Thus it behaves like an object with infinite
inertial mass, absorbing momentum in analogy to a wall that reflects a
bouncing ball.  Nevertheless, following \citet{Schecketal06}, we can
estimate the NS velocity $\mathbf{v}_\mathrm{ns}$ from the negative of
the total linear momentum of the ejecta gas as
\begin{equation}
\mathbf{v}_\mathrm{ns}(t)=-\mathbf{P}_\mathrm{gas}(t)/M_\mathrm{ns}(t)
\, ,
\label{eq:vns}
\end{equation}
where $\mathbf{P}_\mathrm{gas}=\int_{R_\mathrm{ns}}^{R_\mathrm{ob}}
\mathrm{d}V\,\rho\mathbf{v}$ is the integral of the gas momentum over
the computational volume between PNS radius and outer grid
boundary\footnote{It is important to note that there is no momentum
  flux across the outer grid boundary.}. 

In our simulations we find NS kick velocities ranging from about
30\,km\,s$^{-1}$ (L15-3) to roughly 440\,km\,s$^{-1}$ (W15-6) at
1.1--1.4\,s after bounce. The NS acceleration $a_\mathrm{ns}$ is
obtained by numerically differentiating the NS velocity calculated
from Eq.~(\ref{eq:vns}) and listed in Table~\ref{tab:results}. Because
of the nonvanishing acceleration, the recoil velocity at 1.1--1.4\,s
p.b.\ is usually not the final NS velocity. Only models ending with a
low kick (like L15-2 and L15-3) may have reached nearly the asymptotic
value already after this period of time. In the three long-time runs
W15-1, W15-2, and W15-6, we find a significant increase
in the recoil 
velocity between 1.3 and 3.3\,s, and even then the acceleration
continues, though at a considerably smaller rate (compare the values
of $v_\mathrm{ns}$ and $a_\mathrm{ns}$ with and without the
superscript ``long'' in Table~\ref{tab:results}).  Although we evolved
several of the long-time models to much later times we do not compute
the NS velocities beyond 3.4\,s, because the subsequent shift of the
inner grid boundary and the application of a free outflow condition,
can lead to anisotropic momentum flow across the boundary can
occur. The latter would have to be accounted for with high precision
if the argument of momentum conservation were to be used for
estimating the NS recoil also later on.

%
\begin{figure}
\includegraphics[width=\hsize]{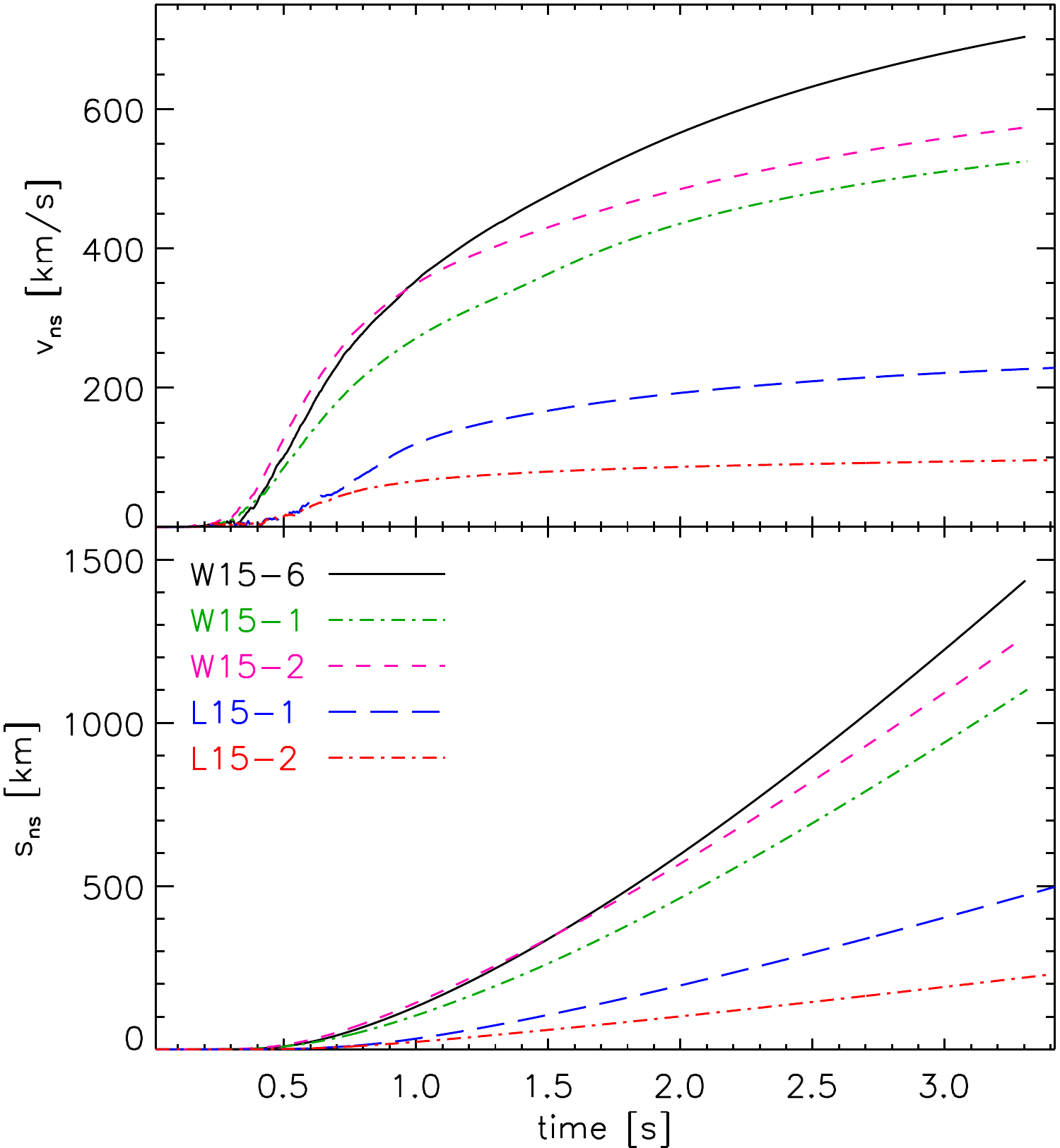}\\
\caption{Long-time evolution of the NS velocity ({\em top}) and of the
  corresponding NS displacement ({\em bottom}) for models W15-6,
  W15-1, W15-2, L15-1, and L15-2. The displacement of the kicked NS
  from the coordinate center, although not taken into account in our
  simulations, can be estimated by time integration of the NS
  velocity.  Within 3.3 seconds the NS in model W15-6 exceeds a
  velocity of 700\,km\,s$^{-1}$ and continues to be accelerated at a
  modest rate. Note also that the NS velocities in models L15-1 and
  L15-2 have nearly reached their terminal values after 3 seconds. In
  all cases the NS shift relative to the coordinate center is
  negligibly small before the explosion takes off, and even in model
  W15-6 with its most extreme NS kick it is always dwarfed by the
  diameter of the shock wave and postshock ejecta, which expand with
  typically $\sim$10,000\,km\,s$^{-1}$.  }
\label{fig:longvns}
\end{figure}
%

In Fig.~\ref{fig:kick} the asymmetry of the SN shock, the structure of
the high-entropy bubbles, and the inhomogeneous, dense ejecta shell
behind the expanding shock are compared for two models at
1.3--1.4\,s. One model, W15-1 (top), has a large kick, the other
model, L15-2 (bottom), a small one. The analogous combination of plots
was provided for model W15-2, our second highest kick case, in Fig.~3
of \citet{kickletter}.  The three panels display different aspects of
the anisotropic mass distribution. While in the left image the
geometry is visible in overview by three-dimensional
entropy-isosurfaces, the middle panel gives an impression of the
asymmetry of the gas (column) density visualized by a ray-casting
technique. The viewing direction is perpendicular to the plane defined
by the NS kick (white) and spin (black) vectors. The right plot shows
the entropy distribution in that plane.  The ray-casting images
visualize the density projected along the lines of sight. They show
that the kick vector points to the region with the highest
concentration of dense clumps. Moreover, the NS in model W15-1 is
visibly displaced from the geometrical center of the ejecta and of the
SN shock in the recoil direction; this effect is even more pronounced
for W15-2. In contrast, in the low-kick case of L15-2 the high-entropy
bubbles are of relatively similar sizes on all sides. The expansion of
the SN shock is therefore nearly equally strong in all directions,
i.e. the shock has no strong dipolar deformation (though a quadrupolar
asymmetry mode is visible), and the NS is close to the center of the
ejecta shell.  Correspondingly, the wind-termination shock is
relatively spherical in this model. While there
are denser, clumpy structures near the one o'clock position of model
L15-2, there are also similarly massive, though less concentrated
filaments in the opposite hemisphere. The explosion asymmetry has a
dipolar component that is not large enough to produce a great pulsar
kick. 

Figure~\ref{fig:longvns} displays the growth of the NS velocities in
five of our nine long-time runs over the whole period of 3.3--3.4\,s.
In the most extreme cases, models W15-6 and W15-2, the NS velocity
reaches an asymptotic value over 700\,km\,s$^{-1}$ and near
600\,km\,s$^{-1}$, respectively.  Several of the W15-models therefore
develop kicks clearly higher than the peak of the measured velocity
distribution of young pulsars. Only more simulations for a wider
variety of progenitors, however, will be able to provide answers to
the questions whether velocities of even 1000\,km\,s$^{-1}$ and more
can be obtained by the discussed gravitational tug-boat mechanism, and
whether and how such extreme recoils might be connected to particular
features of the progenitor stars or especially favorable conditions of
the explosion itself. Analytic estimates in Sect.~\ref{sec:analytics}
will show that the mechanism is likely to be viable of producing kicks
well in excess of 1000\,km\,s$^{-1}$ for reasonable assumptions about
the explosion asymmetry. Yet it is not clear that the magnitude of the
kick is merely a stochastic result of the chaotic growth of
asymmetries by hydrodynamic instabilities in the SN core. Some trends
of our results seem to indicate that certain properties of the
progenitor and the explosion could be supportive for large kicks (see
also Sect.~\ref{sec:progkick}).

\subsection{Mode analysis}
\label{sec:modeanalysis}

The causal connection between the explosion asymmetries, especially
the density asymmetries in the ejecta shell, and the natal NS kicks,
which is visually suggested by Figs.~\ref{fig:W15-2sequence} and
\ref{fig:kick} as well as by Fig.~3 of \citet{kickletter}, can be
quantitatively supported by an expansion of variables that
characterize the mass distribution in the postshock layer into
spherical harmonics.

\begin{figure*}
\includegraphics[width=0.48\hsize]{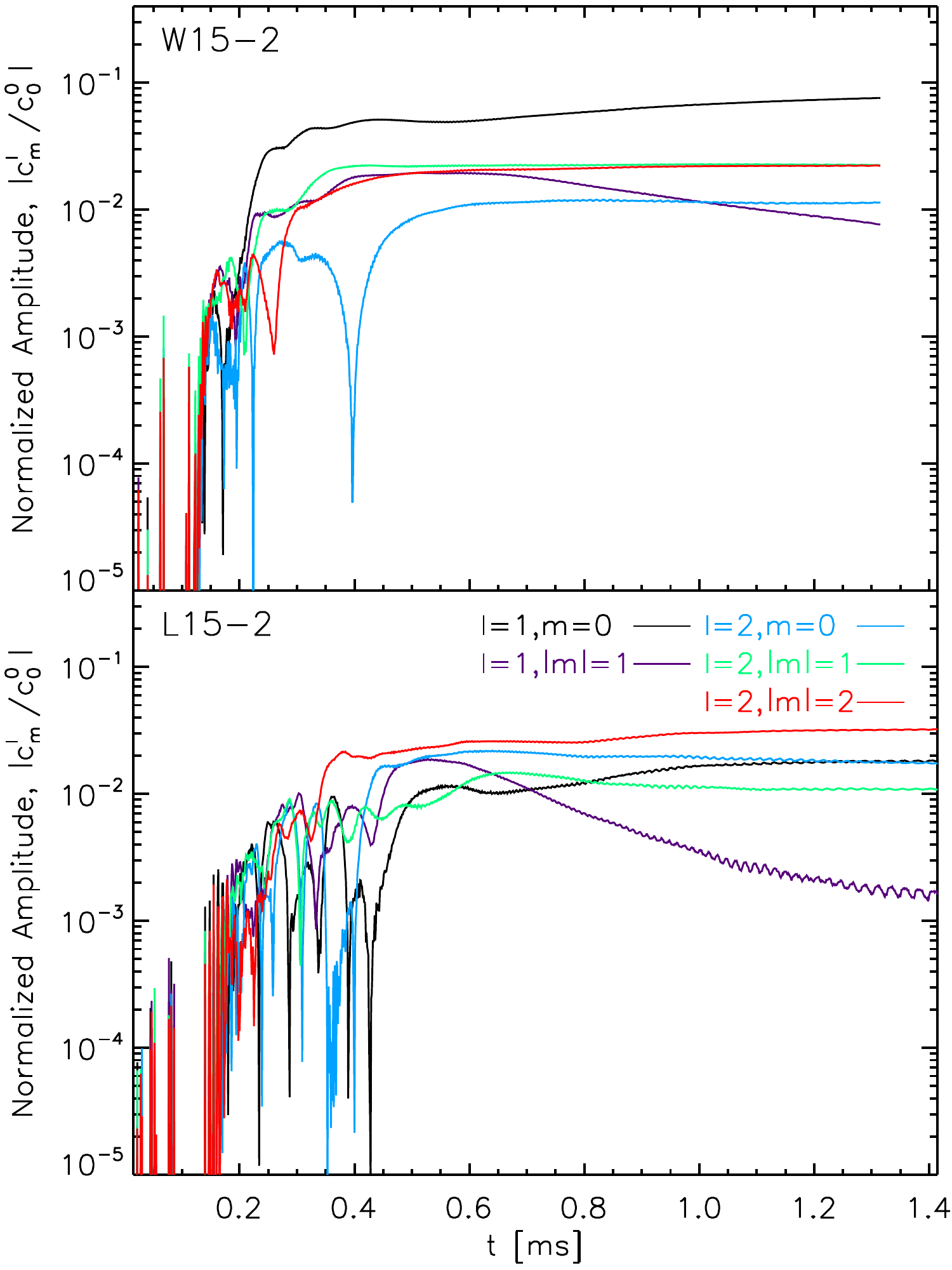}
\hspace{0.03\hsize}
\includegraphics[width=0.48\hsize]{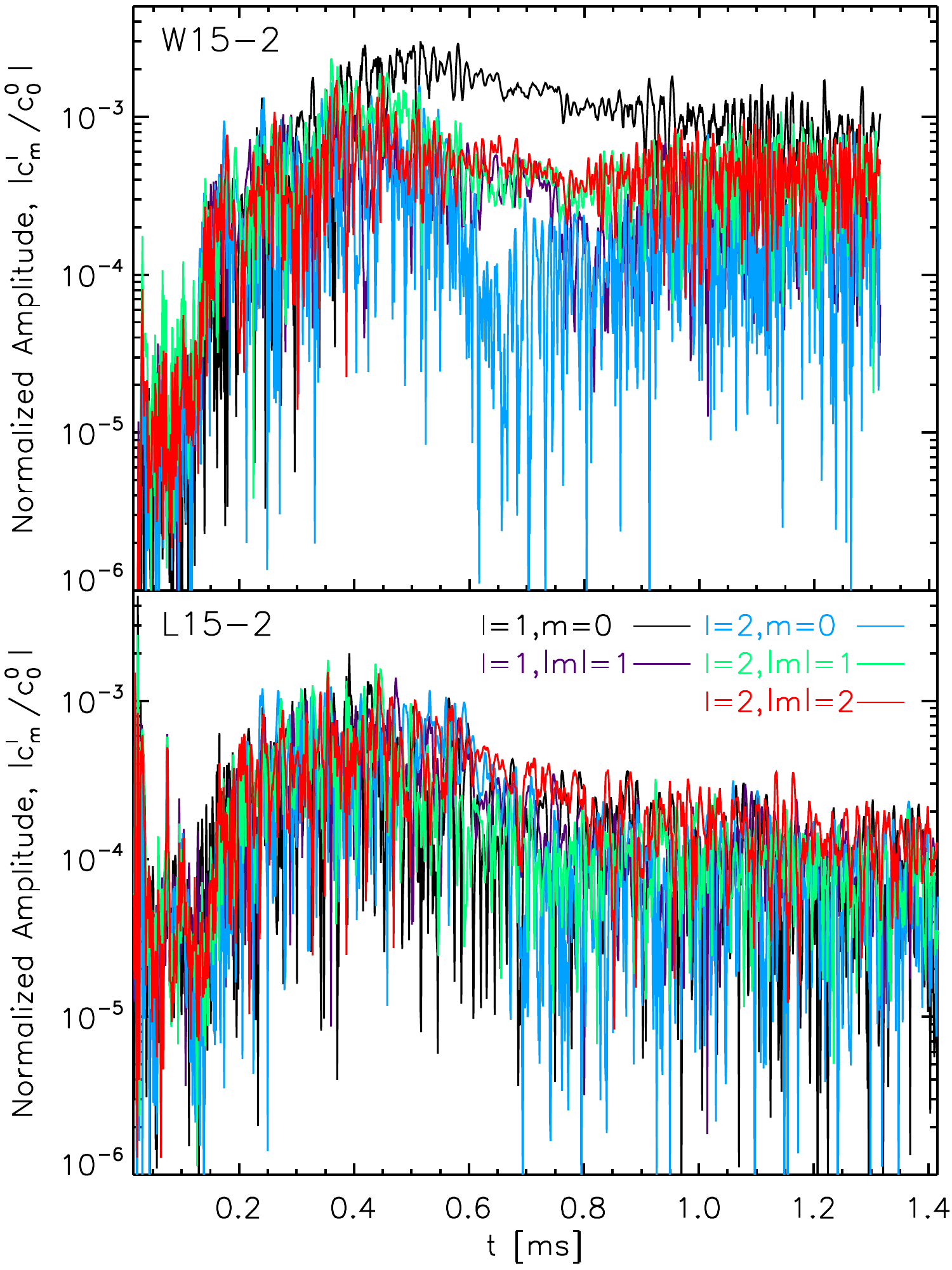}
\caption{Time evolution of the normalized spherical harmonic expansion
  coefficients for different $(\ell,m)$ modes computed for the shock
  surface $R_\mathrm{s}(\theta,\phi)$ ({\em left}) and for the surface
  mass density $D(\theta,\phi)$ (Eq.~\ref{eq:surfmass}; {\em right})
  of models W15-2 (top) and L15-2 (bottom). The polar axis of the
  coordinate system chosen for the expansion is aligned with the final
  NS kick direction. Note the dominance of the $(\ell,m)=(1,0)$ mode
  for model W15-2 with the highest kick velocity, while this dominance
  is absent in the case of model L15-2.}
\label{fig:mode_rotk}
\end{figure*}

Testing different possibilities, we ascertained that the most
suitable quantity for this purpose is the ``surface mass density''
\begin{equation}
D(\theta,\phi) \equiv 
\int_{R_\mathrm{ns}}^{R_\mathrm{s}}\mathrm{d}r\,\rho(\mathbf{r})\, ,
\label{eq:surfmass}
\end{equation}
where $R_{\mathrm{s}}(\theta,\phi)$ is the direction-dependent radius
of the SN shock. The integrand of Eq.~(\ref{eq:surfmass}) means that
mass elements $\Delta m$ at different distances from the NS are
weighted according to $\Delta m/r^2$, reflecting the importance of
their contribution to the NS acceleration by gravitational effects. In
contrast, the shock radius $R_\mathrm{s}(\theta,\phi)$ or the
integrated mass per solid angle,
\begin{equation}
\Sigma(\theta,\phi) \equiv
\int_{R_\mathrm{ns}}^{R_\mathrm{s}}\mathrm{d}r\,r^2\rho(\mathbf{r}) 
\, ,
\label{eq:masspersa}
\end{equation}
put too much weight on the region just behind the shock, which is not
necessarily the most relevant one for accelerating the NS. The NS kick
is mainly determined by the density inhomogeneities in the deeper
ejecta regions. Although a large kick velocity is typically also
linked to a pronounced $\ell = 1$ displacement of the shock surface
relative to the grid center (in our most extreme cases the dipole
amplitude can reach nearly 10\% of the monopole one, see
Fig.~\ref{fig:mode_rotk}, left panel), we find that the inverse
conclusion does not always hold: A dominant $\ell = 1$ shock
asphericity does not necessarily imply a strong NS recoil. In fact, we
observe the $\ell = 1$ spherical harmonics of the shock surface to be
larger than all higher-order multipole amplitudes at late postbounce
times for the majority of our models (except for the few cases like 
L15-2, L15-3 or N20-2, all of which are among our runs with the
lowest NS kicks).  For a discriminating diagnostic quantity we
therefore prefer to evaluate $D(\theta,\phi)$, whose mode amplitudes
are more conclusive about ejecta asymmetries that are causal for large
recoil velocities of the compact remnant. Large kicks correlate with
an $\ell = 1$ mode of $D(\theta,\phi)$ that clearly dominates the
higher modes during (at least) a longer period of the postbounce
evolution, whereas for cases with small kicks the $\ell = 1$ mode is
not much stronger at any time than the other modes.

%
%

\begin{figure*}
\includegraphics[width=\hsize]{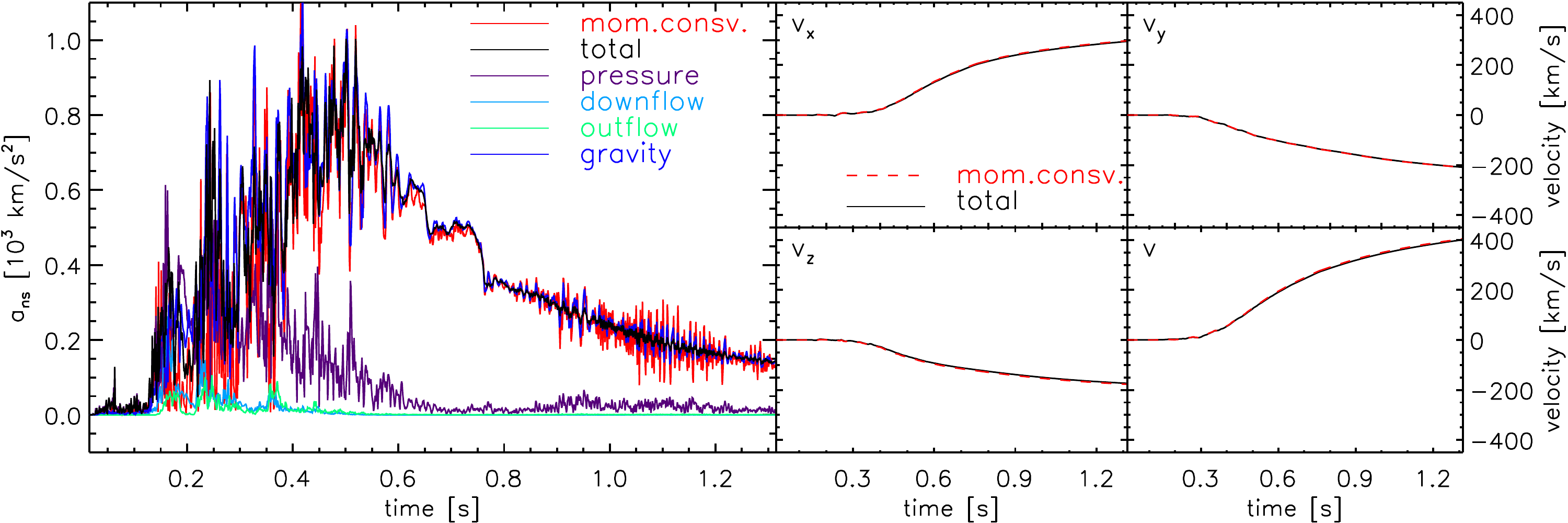}
\caption{{\em Left:} Time evolution of the NS acceleration in model
  W15-2. The acceleration computed by the assumption of momentum
  conservation, i.e., by the negative derivative of the total linear
  momentum of the gas surrounding the NS (red; denoted by
  ``mom.consv.''), is compared with the value obtained by the
  integration of all hydrodynamical and gravitational forces that act
  on a sphere of radius $1.3R_\mathrm{ns}$ around the PNS (black;
  denoted as ``total'').  Also shown are the absolute values of the
  individual contributions of all terms in Eq.~(\ref{eq:pnsdot}),
  namely the momentum transfers by downflows (light blue), outflows
  (green), anisotropic pressure forces (purple), and the gravitational
  attraction by the asymmetric ejecta (blue). The last contribution
  clearly dominate the evolution over long periods.  {\em Right:} NS
  velocity components in all coordinate directions and total value of
  the NS velocity. The two lines in each panel, which 
  correspond to our two methods of computing the NS recoil
  acceleration in the left panel, can hardly be distinguished from
  each other.}
\label{fig:pnsdot}
\end{figure*}

The decomposition of $D(\theta,\phi)$ in spherical harmonics is
written as
\begin{equation}
\label{eq:ylmexpansion}
D(\theta,\phi)=\sum_{\ell=0}^\infty\sum_{m=-\ell}^\ell
c_\ell^mY_\ell^m(\theta,\phi)\, ,
\end{equation}
where $c_\ell^m$ are the expansion coefficients and the spherical 
harmonics $Y_\ell^m$ are functions of
the associated Legendre polynomials $P_\ell^m$ as
\begin{equation}
\label{eq:ylm}
Y_\ell^m(\theta,\phi)=K_\ell^mP_\ell^m(\cos{\theta})e^{im\phi}\, ,
\end{equation}
with
\begin{equation}
\label{eq:klm}
K_\ell^m=\sqrt{\frac{2\ell+1}{4\pi}\frac{(\ell-m)!}{(\ell+m)!}}\, .
\end{equation}
Multiplying Eq.~(\ref{eq:ylmexpansion}) by the complex conjugate of the
spherical harmonic, $Y_\ell^{m*}$, and integrating over the solid angle,
the expansion coefficient is found to be
\begin{equation}
\label{eq:clm}
c_\ell^m=\int_0^{2\pi}d\phi\int_0^\pi
d\theta\;\sin{\theta}\,D(\theta,\phi)Y_\ell^{m*}(\theta,\phi)\,.
\end{equation}
We choose the coordinate system for the expansion such that the polar
axis is aligned with the final NS kick direction. Before performing
the integration of Eq.~(\ref{eq:clm}), we interpolate, using bi-linear
interpolation, the surface mass density computed on the Yin-Yang grid
onto the new coordinate system.  Figure~\ref{fig:mode_rotk}, right
panel displays the resulting expansion coefficients of different
$(\ell,m)$ modes versus time for models W15-2 and L15-2. In the
former, high-kick model the amplitude of the $(\ell,m)=(1,0)$ mode
reaches its maximum at about 400\,ms after bounce, i.e., approximately
at the same time when the NS acceleration becomes maximal
(Fig.~\ref{fig:pnsdot}).  After this time, at $t\ga 0.4$\,s
postbounce, the $(\ell,m)=(1,0)$ mode clearly dominates. Such a
behavior is absent in the low-kick model L15-2.

Consistent with what is visible in Figs.~\ref{fig:W15-2sequence},
\ref{fig:kick} and Fig.~3 of \citet{kickletter}, the mode analysis
shows that a dominant $\ell = 1$ (dipolar) asymmetry of the mass
distribution in the ejecta correlates with a high pulsar kick along
the dipole axis. In model L15-2 all of the $\ell = 1$ and $\ell = 2$
modes displayed in the lower right panel of Fig.~\ref{fig:mode_rotk}
exhibit similar amplitudes and thus the gravitational pull of the
ejecta does not develop a preferred direction. The acceleration of the
NS remains correspondingly low.  This result provides strong support
to the described gravitational tug-boat mechanism for NS kicks, which
is linked to the anisotropic ejection of matter in the first second of
the supernova explosion and which is mediated to the NS on
significantly longer timescales by the asymmetric gravitational pull
of the expanding ejecta gas. In the next section we will describe in
more detail how the asymmetries are established and how the
acceleration mechanism works. Moreover, we will present simple toy
models to approximately estimate its possible strength in dependence
of the properties of the supernova ejecta.

%
%

\subsection{Kick mechanism: theory and tests}
\label{sec:theory}

The NS acceleration that is associated with the asymmetric ejecta
distribution developing during the onset of the SN explosion is
achieved mainly by gravitational forces in combination with
hydrodynamical forces.  This will be detailed below. The acceleration
proceeds in three steps:
\begin{itemize}
\item[(1)] When violent convective mass flows and possibly SASI
  sloshing motions conspire to stir the postshock layer, an anisotropy
  of the mass-energy distribution around the PNS is
  created. Convective downdrafts, channelling accretion flows from the
  stalled shock through the neutrino-heating region to the vicinity of
  the gain radius, get deflected to feed an asymmetric pattern of
  high-entropy bubbles. The bubbles form, grow, contract again, partly
  collapse, and reappear in a quasi-chaotic way to become smaller or
  larger, absorbing less or more neutrino energy. This stochastic
  bubble formation, however, does not cause an appreciable recoil of
  the NS \citep[see Fig.~\ref{fig:pnsdot} and][]{kickletter}.
\item[(2)] When the explosion sets in, the shock and postshock gas
  begin to expand aspherically, driven by the asymmetric inflation of
  the bubbles. The ejecta gas therefore gains radial momentum and its
  c.o.m.\ moves away from the coordinate origin
  (Figs.~\ref{fig:W15-2sequence}, \ref{fig:kick}, right panels): The
  ejecta shell acquires a net linear momentum because of different
  strengths of the explosion in different directions.  The initial
  {\em energy and mass asymmetry} is thus transformed into a {\em
    momentum asymmetry} by the conversion of thermal to kinetic energy
  through hydrodynamical forces. When the expansion timescale becomes
  shorter than the timescale of lateral mixing, the asymmetric ejecta
  structures essentially freeze out.
\item[(3)] Because of linear momentum conservation, the NS must
  receive the negative of the total momentum of the anisotropically
  expanding ejecta mass. Hydrodynamic pressure forces alone cannot
  achieve the NS acceleration \citep{Schecketal06}. As long as
  accretion downdrafts reach the NS, momentum is transferred by
  asymmetric gas flows. Stronger accretion on the weaker side of the
  blast and more mass loss in the neutrino-driven wind on the other
  side cause a recoil opposite to the main explosion
  direction. However, the largest kick contribution, which continues
  even after accretion has ended and after the wind outflow has become
  spherical (Fig.~\ref{fig:W15-2sequence}, bottom right panel, and
  Fig.~\ref{fig:kick}, right panels), results from the gravitational
  pull of the anisotropic shell of ejecta
  (\citealt{Schecketal06,kickletter} and also
  \citealt{Nordhausetal10}).
\end{itemize}

Why can our models produce kick velocities in excess of
$\sim$500\,km\,s$^{-1}$ by hydrodynamically created explosion
asymmetries although such a mechanism appeared to be highly disfavored
even for the extreme case of dipolar ejecta deformation in an analysis
of 2D explosion models by \citet{JankaMueller94}? It should be noted
that the kick scenario discussed in the latter paper is fundamentally
different from the one described here.  \citet{JankaMueller94} assumed
the kicks to originate from an {\em impulsive event}, happening during
a short period of a few hundred milliseconds around the onset of the
anisotropic explosion. Correspondingly, in their estimates they
considered the momentum asymmetry of the ejecta shell during this
early phase, expecting the kick to be established before the accretion
is over.  Indeed, also in the simulations described in the present
paper the NS velocity at the end of the accretion phase is much lower
than 500\,km\,s$^{-1}$ in all cases.  However, in the scenario
discussed here the NS acceleration is a {\em long-duration}
phenomenon, which continues for several seconds even after accretion
has ended, because it is promoted by the long-range gravitational
forces of the asymmetric, dense ejecta shell that expands away from
the NS behind the outgoing SN shock.

To elaborate on this point and to check our estimates of the NS recoil
velocity on the basis of Eq.~(\ref{eq:vns}), we follow
\citet{Schecketal06} and perform an analysis of the different
contributions to the NS kick by summing up all forces acting on a
central sphere of radius $r_0=1.3R_\mathrm{ns}$ \citep[][adopted a
  similar analysis for their recent 2D
  results]{Nordhausetal10}. According to Eq.~(5) of
\citet{Schecketal06} the time-derivative of the NS momentum can be
deduced from the Euler equations as
\begin{equation}
\label{eq:pnsdot}
\dot{\mathbf{P}}_\mathrm{ns}\approx-\oint_{r=r_0}\mathcal{P}\;
\mathrm{d}\mathbf{S} 
-\oint_{r=r_0}\rho\mathbf{v}v_r\;\mathrm{d}S+
\int_{r>r_0}GM_\mathrm{ns}\frac{\mathbf{r}} 
{r^3}\;\mathrm{d}m\, .
\end{equation}
The corresponding NS acceleration is
$\mathbf{a}_\mathrm{ns}=\dot{\mathbf{P}}_\mathrm{ns}/M_\mathrm{ns}$
and the NS velocity $\mathbf{v}_\mathrm{ns}(t)=\int_0^t {\mathrm
  d}t^\prime\; \mathbf{a}_\mathrm{ns}(t^\prime)$. The first term on
the rhs of Eq.~(\ref{eq:pnsdot}) accounts for the variation
in the
pressure force around the sphere. The second term is the flux of
momentum going through the sphere. This term can be decomposed into
downflow and outflow contributions depending on the sign of the radial
velocity $v_r$ being negative or positive, respectively. The third
term is the contribution from the long-range gravitational forces
exerted on the NS by the anisotropic mass distribution outside the
sphere. We note in passing that a change in the radius of evaluation,
$r_0$, in a reasonable range around $1.3R_\mathrm{ns}$ (i.e., in the
neighborhood of the NS surface) leads to different weights of the
three summed terms \citep[as noticed recently by][]{Nordhausetal10}
but still yields the same result for the integrated total effect.

Figure~\ref{fig:pnsdot} compares $a_\mathrm{ns}$ and $v_\mathrm{ns}$
for model W15-2 as calculated by the two different methods, i.e.\ from
the requirement of total momentum conservation (Eq.~\ref{eq:vns}) and
from the integration of the responsible hydrodynamical and
gravitational forces according to Eq.~(\ref{eq:pnsdot}).  NS
acceleration as well as velocity exhibit excellent agreement of the
results obtained in the two different ways. Between $\sim$130 and
$\sim$600\,ms after bounce the forces by the anisotropic pressure
distribution and by the gravitational pull both contribute to the
total acceleration on significant levels, while downflows and outflows
transfer sizable amounts of momentum mostly during transient phases of
high activity when strong accretion streams impact the nascent NS
anisotropically. At later stages, i.e., at $t \ga 400$\,ms when the
asymmetry of the ejecta has been established, and in particular at $t
\ga 600$\,ms when accretion to the PNS has ceased and the
neutrino-driven wind has become essentially spherically symmetric
(Fig.~\ref{fig:W15-2sequence}), the gravitational force term starts to
dominate and therefore accounts for the long-term increase
in the NS
kick. A major fraction of the final NS velocity is thus caused by the
gravitational attraction of the anisotropic ejecta.

%
\begin{figure*}
\includegraphics[width=0.48\hsize]{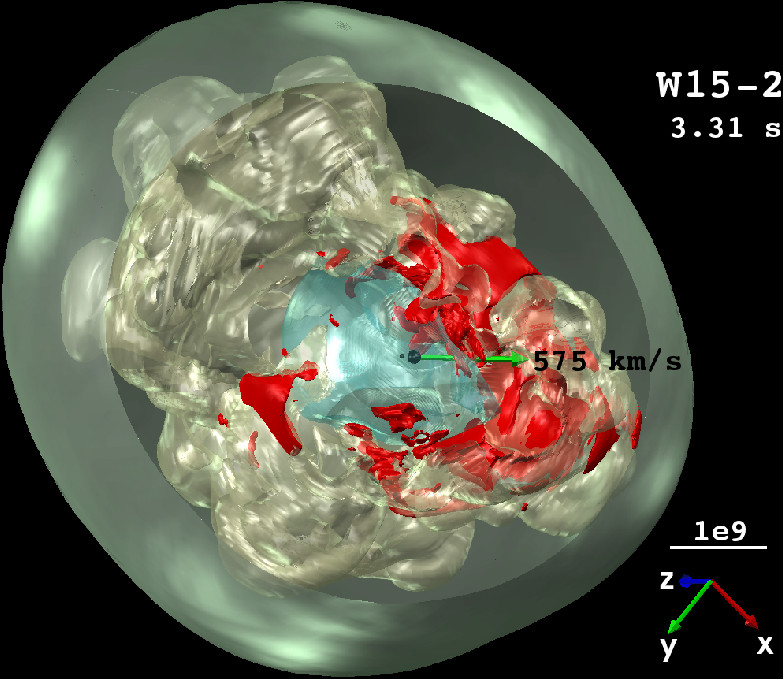}\hspace{0.03\hsize}
\includegraphics[width=0.48\hsize]{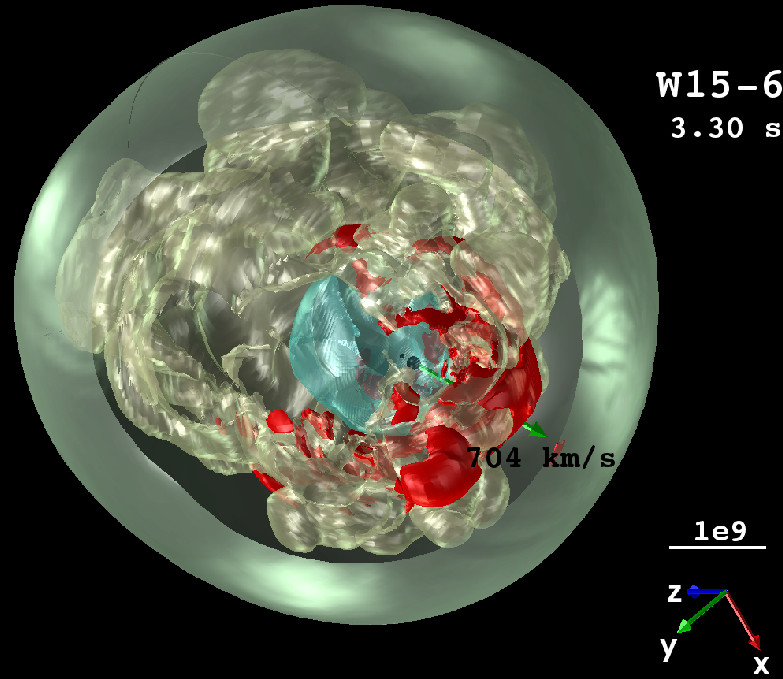}
\caption{Ejecta morphology of the high-kick models W15-2 ({\em left})
  and W15-6 ({\em right}) at approximately 3.3\,s after bounce.  The
  NS (indicated by black spheres, enlarged by a factor of 20 for
  better visibility) are accelerated by the gravitational tug of the
  asymmetrically distributed, dense ejecta to velocities of
  575\,km\,s$^{-1}$ and 704\,km\,s$^{-1}$, respectively (green arrows
  pointing to the right). The ejecta ``clumps'' with the highest
  densities, which are mainly responsible for the gravitational
  attraction, are visualized by the red-colored surfaces of constant
  density of $10^4$\,g\,cm$^{-3}$. The small black dot left of the NS
  in both images marks the coordinate origin. The NS displacement
  relative to this location was computed by time integration of the NS
  velocity. After $\sim$3.3\,s it is only 1260\,km in W15-2 and
  1440\,km in W15-6. The neutrino-heated, asymmetrically distributed
  inner ejecta are bounded by the semi-transparent, beige surface from
  outside and the semi-transparent, light blue surface from
  inside. The former surface is defined by a constant $^{56}$Ni mass
  fraction of 5\%, the latter surface conincides with the termination
  shock of the neutrino-driven wind. The outermost, semi-transparent
  green surface (cut open towards the observer for better visibility
  of the inner volume) corresponds to the SN shock wave. The average
  shock radius is approximately 36,000\,km at the displayed time. A
  yardstick of 10,000\,km along with a tripod indicating the
  coordinate directions are provided in the lower right corner of each
  plot.  }
\label{fig:displacement}
\end{figure*}
%

Our analysis, compared to the 2D results of \citet{Schecketal06},
reveals that in the 3D case the direct hydrodynamical momentum
transfer by gas downflows to and outflows from the NS is much less
important than in 2D. While in 2D the toroidal nature of all
structures favors the formation of large hemispheric differences with
inflow to the NS dominating in one hemisphere and outflow in the other
(corresponding to an $\ell = 1$ geometry), the 3D downflow and outflow
pattern is more often and for longer periods of time characterized by
higher multipoles. The momentum transferred to the NS by inflows
is thus almost balanced by the momentum carried away from the NS by
outflows on smaller angular scales, resulting only in a small
contribution to the NS kick velocity. The gravitational tug by the
asymmetric ejecta distribution is therefore clearly the dominant
mediator of the NS acceleration in a time-integrated sense, although
it is not necessarily the only important effect, in particular not
during the very early phase of the NS recoil and not in 2D.  Sizable
directional differences of the explosion strength can occur and can
lead to large ejecta inhomogeneities and asymmetries also in 3D (most
pronounced in models W15-6 as well as W15-1 and W15-2, see
Figs.~\ref{fig:W15-2sequence} and \ref{fig:kick}).  Large recoil
velocities can therefore be produced despite the fact that the pattern
of convective bubbles and downdrafts seems to be characterized by the
presence of higher-order multipoles.

Allowing the PNS to move (instead of being attached to the 
coordinate center because of the excision of the central core in our
simulations) is unlikely to change fundamental aspects of our
results. This expectation is based on a simple fact visualized in
Figs.~\ref{fig:longvns} (bottom panel) and \ref{fig:displacement}.
Even for the highest computed kick velocities the displacement of the
NS from the coordinate origin during the computed evolution of several
seconds (at most $\sim$1500\,km) is very small compared to the radial
extension of the explosion shock and asymmetric ejecta distribution
(typically several 10,000\,km). It is therefore not astonishing that
our expectation was recently confirmed by
\citet{Nordhausetal10,Nordhausetal12} with 2D simulations, in which
the NS was able to move.  The minor influence of the NS motion had
been tested before by \citet{Schecketal06}, who performed a large set
of 2D simulations in which some of the kicks were similarly strong as
the 3D cases discussed here. For doing the tests, \citet{Schecketal06}
applied a Galilei transformation with velocity $-v_\mathrm{ns}$ to the
gas outside of the PNS (see Sects.~2.3, 7.2, Appendix~B, and Table~A.5
in the cited paper), thus mimicing the effects of an accelerating NS
moving through the surrounding medium. Naturally, because of the
stochastic and chaotic growth of local and global asymmetries, the
simulations developed values of $v_\mathrm{ns}$ that could be
different for a fixed progenitor and the same choices of boundary
conditions (core neutrino luminosities and PNS contraction) and
initial perturbations. Nevertheless, the whole set of models with the
NS motion taken into account did not exhibit any systematic
differences compared to the standard set without the Galilei
transformation applied \citep[cf.\ Tables~A.1 and A.5
  in][]{Schecketal06}.

\subsection{Analytic estimates of the kick magnitude}
\label{sec:analytics}

What are the maximum kick velocities that can be mediated to the NS by
gravitational forces of asymmetrically ejected matter?  The maximum
velocity of more than 700\,km\,s$^{-1}$ obtained in our present set of
3D models is no upper limit, and recoil velocities beyond
1000\,km\,s$^{-1}$ were already found in 2D simulations by
\citet{Schecketal06}. Simple considerations can yield
order-of-magnitude estimates that confirm this possibility.  In the
following we discuss two different, idealized cases that allow one to
integrate the gravitational pull over the whole period of acceleration
(see also \citealt{kickletter} and alternative, simple estimates by
\citealt{Nordhausetal10}).

In the first case we consider a hemispheric asymmetry of the mass
distribution in an expanding, spherical ejecta shell. The shell with
an average radius $r_\mathrm{s}$ is assumed to contain an extra
clump-like mass $\Delta m$ in one hemisphere, while an equal mass
deficit exists in the other hemisphere. A mass asymmetry of this kind
may be established by asymmetric accretion flows during the violently
convective postbounce phase. Downflows may, for example, be stopped
and blown out again before they can reach down to the PNS surface, or
only some part of the downflows is accreted onto the PNS while the
rest is deflected and ejected again after neutrino heating. Let us
assume that the shell with its time-dependent radius $r_\mathrm{s}(t)$
expands with a constant velocity $v_\mathrm{s}$, starting at an
initial radius $r_\mathrm{i}$. If the NS is displaced by a distance
$s$ from the center of the shell ($s = 0$ at $t = 0$), the
gravitational forces lead to a NS velocity $v_\mathrm{ns} = \dot s$
via an acceleration $a_\mathrm{ns} = \ddot s$ given by
\begin{equation}
\ddot s = \frac{\mathrm{d}v_\mathrm{ns}}{\mathrm{d}t} = 
G\Delta m \left\lbrack \frac{1}{(r_\mathrm{s}-s)^2} +
\frac{1}{(r_\mathrm{s}+s)^2} \right\rbrack \,.
\label{eq:accel1}
\end{equation}
Using $r_\mathrm{s} = r_\mathrm{i} + v_\mathrm{s}t$ and assuming
$s \ll r_\mathrm{s}$ at all times, the integration of 
Eq.~(\ref{eq:accel1}) from $t = 0$ to $t = \infty$ yields:
\begin{equation}
v_\mathrm{ns} \approx \frac{2G\Delta m}{r_\mathrm{i}v_\mathrm{s}}
\approx 540\,\left[\frac{\mathrm{km}}{\mathrm{s}}\right]\,
\frac{\Delta m_{-3}}{r_{\mathrm{i},7} \, v_{\mathrm{s},5000}} \,,
\label{eq:kickest1}
\end{equation}
where $\Delta m$ is normalized by 10$^{-3}\,M_\odot$, $r_\mathrm{i}$
by $10^7$\,cm, and $v_\mathrm{s}$ by 5000\,km\,s$^{-1}$.  A
$10^{-3}\,M_\odot$ hemispheric asymmetry of the shell expanding with a
constant velocity of 5000\,km\,s$^{-1}$ can thus pull the NS to a
velocity of 540\,$\mathrm{km\,s}^{-1}$ \citep[this result was already
  mentioned in][]{kickletter}.  Ejecta asymmetries can therefore very
effectively mediate a long-lasting acceleration of the NS, which is
pulled in the direction of the higher mass concentration. According to
Eq.~(\ref{eq:kickest1}) the kick becomes larger for a lower expansion
velocity, because the gravitational tug from the asymmetric ejecta
shell remains strong for a longer time. While a value of
5000\,km\,s$^{-1}$ tends to be on the high side for the postshock
ejecta shell during the relevant times, a clumpy anisotropy of
$10^{-3}\,M_\odot$ means an asymmetry of the mass distribution of only
$\la$1\% in a shell that typically contains $10^{-1}\,M_\odot$ or
more. Asymmetries of several percent appear well within reach and
therefore velocities of more than 1000\,km\,s$^{-1}$ as
well. It is interesting that the first-order terms of $s/r_\mathrm{s}$
cancel in Eq.~(\ref{eq:accel1}).  This means that the NS recoil is
affected by the NS motion and displacement from the explosion center
only on the level of second-order correction terms of
$s/r_\mathrm{s}\ll 1$. 

In the second case we consider a difference of the expansion velocity
of two clumps $\Delta m$ of the ejecta in both hemispheres instead of
a hemispheric mass difference.  This means that we assume that these
clumps have been accelerated differently strongly and propagate away
from the center of the blast according to
$r_{\mathrm{s},1} = r_\mathrm{i} + v_{\mathrm{s},1}t$ and
$r_{\mathrm{s},2} = r_\mathrm{i} + v_{\mathrm{s},2}t$ with
$v_{\mathrm{s},1} < v_{\mathrm{s},2}$. 
Taking $s$ to be again the displacement of the NS from the blast
center, the compact remnant in this situation experiences a
gravitational acceleration
\begin{equation}
\ddot s = \frac{\mathrm{d}v_\mathrm{ns}}{\mathrm{d}t} =
G\Delta m \left\lbrack \frac{1}{(r_{\mathrm{s},1}-s)^2} -
\frac{1}{(r_{\mathrm{s},2}+s)^2} \right\rbrack \,.
\label{eq:accel2}
\end{equation}
To lowest order in $s/r_\mathrm{s} \ll 1$, time integration from
0 to $\infty$ leads to
\begin{eqnarray}
v_\mathrm{ns} \approx \frac{G\Delta m}{r_\mathrm{i}}\,
\frac{v_{\mathrm{s},2}-v_{\mathrm{s},1}}{v_{\mathrm{s},1}\,v_{\mathrm{s},2}}
&=& \frac{G\Delta m}{r_\mathrm{i}}\,
\frac{\Delta v_\mathrm{s}}{{\bar{v}}_\mathrm{s}^2}     \cr
& \approx& 260\,\left[\frac{\mathrm{km}}{\mathrm{s}}\right]\,
\frac{\Delta m_{-3}}{r_{\mathrm{i},7}\,{\bar{v}}_{\mathrm{s},5000}}\,
\frac{\Delta v_\mathrm{s}}{{\bar{v}}_\mathrm{s}} \,,
\label{eq:kickest2}
\end{eqnarray}
where we have introduced the definitions $\Delta v_\mathrm{s} =
v_{\mathrm{s},2}-v_{\mathrm{s},1}$ and ${\bar{v}}_\mathrm{s} =
\sqrt{v_{\mathrm{s},1}v_{\mathrm{s},2}}$.  In the last expression we
have again normalized $\Delta m$ by $10^{-3}\,M_\odot$, $r_\mathrm{i}$
by $10^7$\,cm, and ${\bar{v}}_\mathrm{s}$ by 5000\,km\,s$^{-1}$.  The
acceleration is opposite to the faster expanding hemisphere, i.e.\ it
is against the direction of the stronger explosion, and of the same
order of magnitude as the result of Eq.~(\ref{eq:kickest1}), if we
assume a velocity asymmetry of 100\%, $\Delta
v_\mathrm{s}/{\bar{v}}_\mathrm{s} \sim 1$.  This, however, is on the
very extreme side, and $\Delta v_\mathrm{s}/{\bar{v}}_\mathrm{s} <
0.5$ seems more realistic. According to Eqs.~(\ref{eq:accel2}) and
(\ref{eq:kickest2}) a velocity asymmetry is therefore less efficient
in accelerating the NS than the mass asymmetry in
Eqs.~(\ref{eq:accel1}) and (\ref{eq:kickest1}). The first-order
correction term in $s/r_\mathrm{s} \ll 1$, which we have suppressed
when going from Eq.~(\ref{eq:accel2}) to Eq.~(\ref{eq:kickest2}),
reads $\mathrm{d}v_\mathrm{ns}^{(1)}/\mathrm{d}t = 2G\Delta
m\,s(r_{\mathrm{s},1}^{-3} + r_{\mathrm{s},2}^{-3})$.  It is not
necessarily small compared to the leading term
$\mathrm{d}v_\mathrm{ns}^{(0)}/\mathrm{d}t = G\Delta
m\,s(r_{\mathrm{s},1}^{-2} - r_{\mathrm{s},2}^{-2})$, but the
corresponding acceleration amplifies the effect of the zeroth-order
term, because the displacement by the kick brings the NS closer to the
attracting slower parts of the ejecta.

A closer inspection of Eqs.~(\ref{eq:kickest1}) and
(\ref{eq:kickest2}) clarifies the meaning and implications of the
non-impulsive nature of the described gravitational tug-boat mechanism
for NS acceleration.  The instantaneous linear momentum carried by
asymmetrically distributed clumps with representative masses and
velocities, $P_\mathrm{gas} \sim 2\Delta m\,v_\mathrm{s}\approx
2\times 10^{39}\, \Delta m_{-3}v_{\mathrm{s},5000}$\,g\,cm\,s$^{-1}$,
is one to two orders of magnitude smaller than the momentum associated
with the estimated possible NS kick, $P_\mathrm{ns} =
M_{\mathrm{ns},1.5}\,v_{\mathrm{ns},8} \approx 3\times
10^{41}$\,g\,cm\,s$^{-1}$ with $M_{\mathrm{ns},1.5} =
M_\mathrm{ns}/1.5\,M_\odot$. Efficient NS acceleration can therefore
not be explained by considering just the instantaneous momentum of the
clumps at any time. Exactly this is the problem connected with purely
hydrodynamically mediated kicks where the NS receives its recoil by
hydrodynamical forces during the relatively short (hundreds of
milliseconds) phase of anisotropic mass accretion and
ejection. Correspondingly, \citet{JankaMueller94} showed that even
with the most extreme assumptions for the explosion asymmetry these 
more impulsive hydrodynamical kicks are unlikely to lead to recoil
velocities of more than a few 100\,km\,s$^{-1}$.  Instead, stronger
kicks require a long-lasting accelerating force, which is provided by
the gravitational interaction between the NS and the asymmetric
ejecta. For this gravitational tug-boat mechanism to work, it is
necessary to sustain the momentum of the dense clumps in the ejecta
shell over many seconds, despite the gravitational transfer of clump
momentum to the NS. This requires a continuous clump acceleration
which can happen by hydrodynamical forces within the inhomogeneous
ejecta, whereby internal energy of the ejecta is converted into
kinetic energy provided the ejecta do not expand ballistically
yet. The latter requirement limits the radial scales and time
intervals that can contribute to the pulsar kick. The
clumps need to be hydrodynamically coupled to the energetic
environment they are embedded in. This guarantees that they are
continuously accelerated while they transfer momentum and energy to
the NS through their long-range gravitational attraction. The energy
thus pumped into the motion of the NS can become quite significant,
$E_\mathrm{k,ns} = \frac{1}{2} M_\mathrm{ns}v_\mathrm{ns}^2 \sim 10^{49}
M_{\mathrm{ns},1.5}v_{\mathrm{ns},8}^2$\,erg.  This, however, is still
a small fraction of the $\sim$10$^{51}$\,erg of a normal SN explosion.

%
\begin{figure}[!]
\includegraphics[width=\hsize]{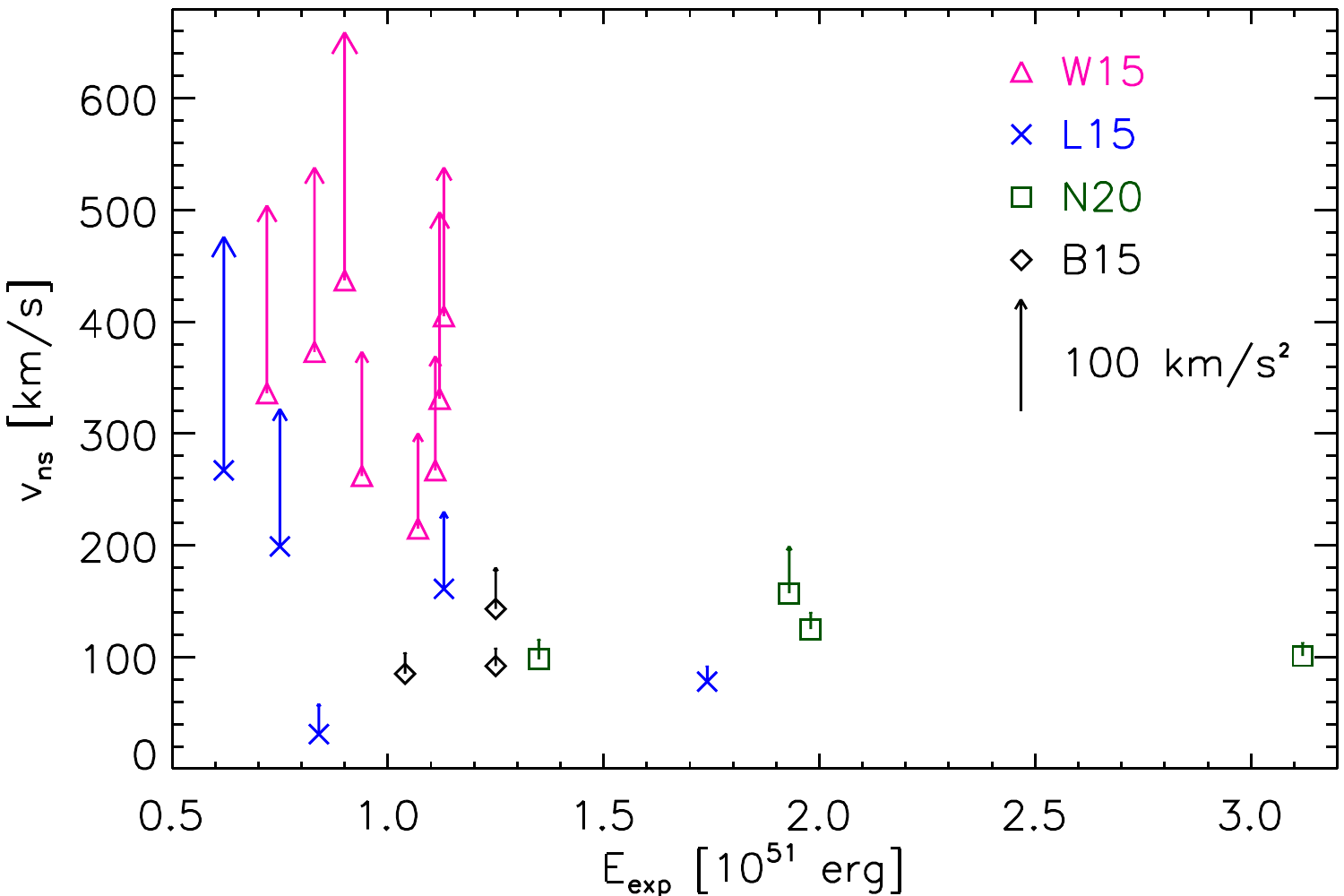}
\caption{NS kick velocities and acceleration values (indicated by the
  length of the arrows attached to the symbols) versus explosion
  energies for all computed models after 1.1--1.4\,s of simulated
  postbounce evolution. Models corresponding to different progenitors
  are shown by different symbols as defined in the figure.}
\label{fig:vns-Eexp}
\end{figure}
%

\subsection{Progenitor dependence of neutron-star kicks}
\label{sec:progkick}

The data of Table~\ref{tab:results} seem to suggest an interesting
trend in dependence of the progenitor models: All computed cases for
the W15-progenitor yield NS recoil velocities that clearly exceed
200\,km/s, and in all cases the acceleration is still relatively high
at the end of the simulation. Also two of the L15 models (the
low-resolution run L15-4-lr and L15-5) develop kicks of
$\sim$200\,km/s and higher with still large acceleration at the end of
the simulated evolution, but all models of the B and N progenitors
produce clearly slower NSs.

Both model subsets of strong and weak kicks exhibit a range of
explosion energies without any clear correlations between energy and
kick velocity (Fig.~\ref{fig:vns-Eexp}). It is important to note that
for each investigated progenitor the number of models is still limited
and includes either a wider range of kick velocities for high as well
as low explosion energies (W15, L15) or little variation in the kick 
velocities for all tested energies (B15, N20). We refrain from
claiming a lack of high-kick cases for strong explosions
($E_\mathrm{exp}\ga 1.2\times10^{51}$\,erg) because no W15 simulation
was done in this energy range and a trend could be connected with
progenitor dependencies. In general, our present results are
compatible with conclusions drawn by \citet{Schecketal06} on the basis
of a large set of more than 
70 simulations in two dimensions and for three different progenitor
models with and without rotation. \citet{Schecketal06} observed kick
velocities on the low and high sides of the velocity distribution for
all investigated progenitors and for less as well as more energetic
explosions.

Presently we have no clue which model conditions, i.e., which
progenitor or explosion or NS properties, could be favorable or
disfavorable for high kick velocities.  Neither are W15-models with
their typically higher kicks distinct concerning explosion energy or
assumed NS contraction, nor is their progenitor structure clearly
special in any aspect compared to the other stars. For example,
L-models have the highest density and most shallow density profile
outside of the iron core and thus the largest mass accretion rates
after core bounce with the slowest decay in time. Despite their
longest delays until the onset of the explosion, however, they do not
stick out concerning their NS kicks.

The determination of clear trends in dependence of progenitor,
explosion properties, and remnant properties will therefore require a
more extended investigation with a systematic variation
in the involved 
degrees of freedom.  This will need many more 3D models than currently
available, which is computationally extremely demanding, even with the
approximations of the neutrino transport used in our work.  At the
present time we therefore refrain from drawing premature conclusions
or getting entrapped in unfounded speculations.

%
\begin{figure}[!]
\includegraphics[width=\hsize]{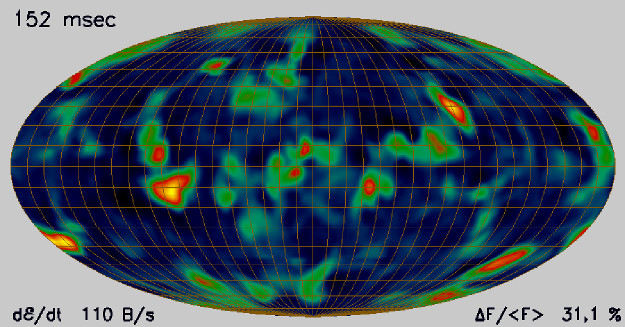}\\
\includegraphics[width=\hsize]{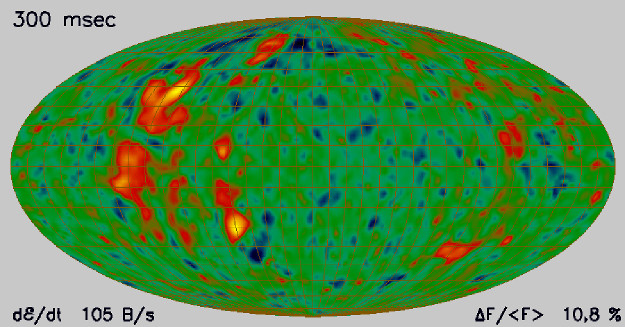}\\
\includegraphics[width=\hsize]{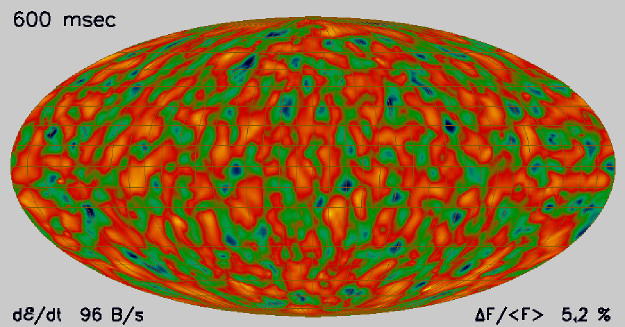}
\caption{Neutrino flux asymmetry at $\sim$150, 300, 600\,ms after
  bounce for model W15-3 as representative case.  The $4\pi$-maps show
  the relative variations $\Delta F/\langle F\rangle$
  in the total 
  (i.e., sum for $\nu$ and $\bar\nu$ of all three neutrino flavors)
  neutrino energy flux in all directions.  Hot spots (yellow and red)
  during the postbounce accretion phase ($t \lesssim 400$\,ms; {\em
    top and middle panels}) are connected to strong convective
  downflows in the neutrino-heated postshock layer. The typical scale
  of the corresponding high-emission areas is 20--30 degrees, and the
  amplitude of variations ranges around 10--30\% during most of the
  time, though rare, single maxima can reach up to $\sim$70\%.  After
  the accretion has ended ($t\sim 400$--500\,ms), the flux variation
  pattern is determined by the cell structure of the convection inside
  the proto-neutron star (i.e., below the neutrinosphere) and the
  variation amplitute decreases to a few percent ({\em bottom
    panel}). The number in the lower left corner of the plots
  ($\mathrm{d}{\cal{E}}/\mathrm{d}t$) gives the total neutrino energy
  loss rate as flux integral over the whole emitting surface.}
\label{fig:neutrinoflux}
\end{figure}
%

\subsection{Contribution by anisotropic neutrino emission}
\label{sec:neutrinokick}

A directional asymmetry of the neutrino flux radiated by the nascent
NS leads to anisotropic energy and momentum loss from the compact
remnant. This can provide a kick of the compact remnant, too.  The
contribution of this neutrino-induced kick to the NS velocity,
$\mathbf{v}_{\mathrm{ns},\nu}$, is equal to the negative of the
momentum carried away by the neutrinos divided by the NS mass,
\begin{equation}
\mathbf{v}_{\mathrm{ns},\nu} = -\,\frac{\mathbf{P}_\nu}{M_\mathrm{ns}}\,.
\label{eq:nukickv1}
\end{equation}
This recoil can be expressed in terms of the total neutrino energy
loss, $E_\nu$, by introducing dimensionless parameters $k_{\nu,i}$
that describe the asymmetry of the neutrino momentum loss in all three
coordinate directions:
\begin{equation}
\mathbf{v}_{\mathrm{ns},\nu} = -\,\mathbf{k}_\nu 
\,\frac{E_\nu}{M_\mathrm{ns}c}\,,
\label{eq:nukickv2}
\end{equation}
where $c$ is the speed of light. The vector components $k_{\nu,i}$ are
computed as
\begin{equation}
k_{\nu,i} \equiv \int_0^\infty \mathrm{d}t
\oint\mathrm{d}S\, n_j P_{ij} \cdot 
\left ( \frac{1}{c}\int_0^\infty \mathrm{d}t
\oint\mathrm{d}S\,\mathbf{\hat n}\, \mathbf{F}_\nu
\right )^{\! -1} \,.
\label{eq:etaasymm}
\end{equation}
Here $P_{ij}$ are the components of the neutrino pressure tensor
$\cal{P}$; $P_{ij}$ defines the flux component $j$ of the neutrino
momentum component $i$. $\mathbf{F}_\nu$ is the neutrino energy flux,
$\mathrm{d}S$ the surface element on the radiating (neutrino)sphere,
$\mathbf{\hat n}$ (with components $n_j$) the unit vector
perpendicular to the surface, and the time and surface integral in the
denominator yields $E_\nu$.  From Eq.~(\ref{eq:nukickv2}) an
order-of-magnitude estimate of the NS velocity is derived as
\begin{equation}
v_{\mathrm{ns},\nu} \approx  3.3\times 10^9\,
\,k_\nu\,\frac{E_\nu}{3\times 10^{53}\,\mathrm{erg}}\,
\frac{1.5\,M_\odot}{M_{\mathrm{ns}}} \,\left\lbrack
\frac{\mathrm{cm}}{\mathrm{s}}\right\rbrack\, .
\label{eq:neutrinokick}
\end{equation}
This yields a recoil velocity of $v_{\mathrm{ns},\nu}\approx
330\,$km\,s$^{-1}$ for $k_\nu = 0.01$.

In principle, ultrastrong magnetic fields could lead to large-scale,
potentially strong dipolar asymmetries of the neutrino transport out
of the hot PNS, e.g.  by changing the neutrino-matter interaction
opacities, coupling to a neutrino magnetic moment, or affecting
neutrino-flavor conversions via the MSW effect
\citep[e.g.,][]{Bisnovatyi-Kogan96,ArrasLai99,Kusenko09,Kishimoto11}.
However, little is known about the strength and structure of the
magnetic field inside a nascent NS. It could well be chaotic due to
convective activity, in which case a large global asymmetry with
dominant low-order multipoles is unlikely.  Not counting on support by
strong dipolar B-field components, a directional bias of the neutrino
emission of even only one percent is extremely difficult to obtain.

%
\begin{figure*}
\includegraphics[width=0.55\hsize]{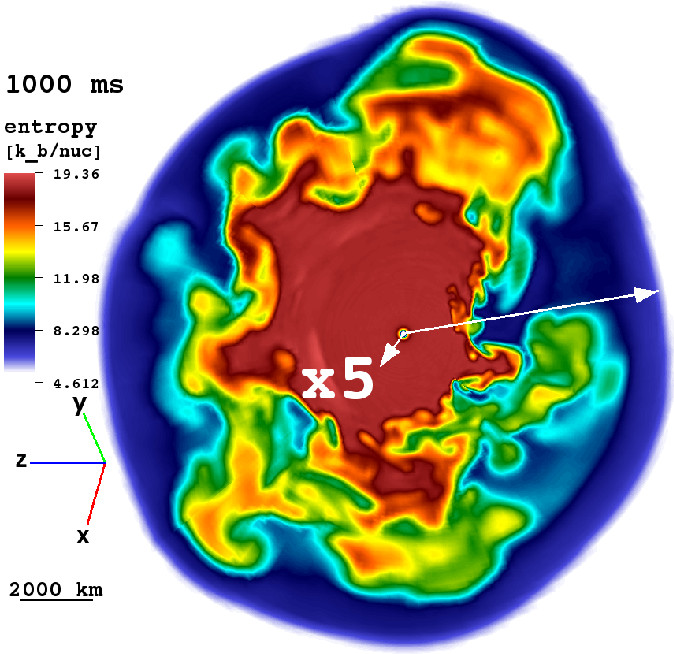}
\includegraphics[width=0.45\hsize]{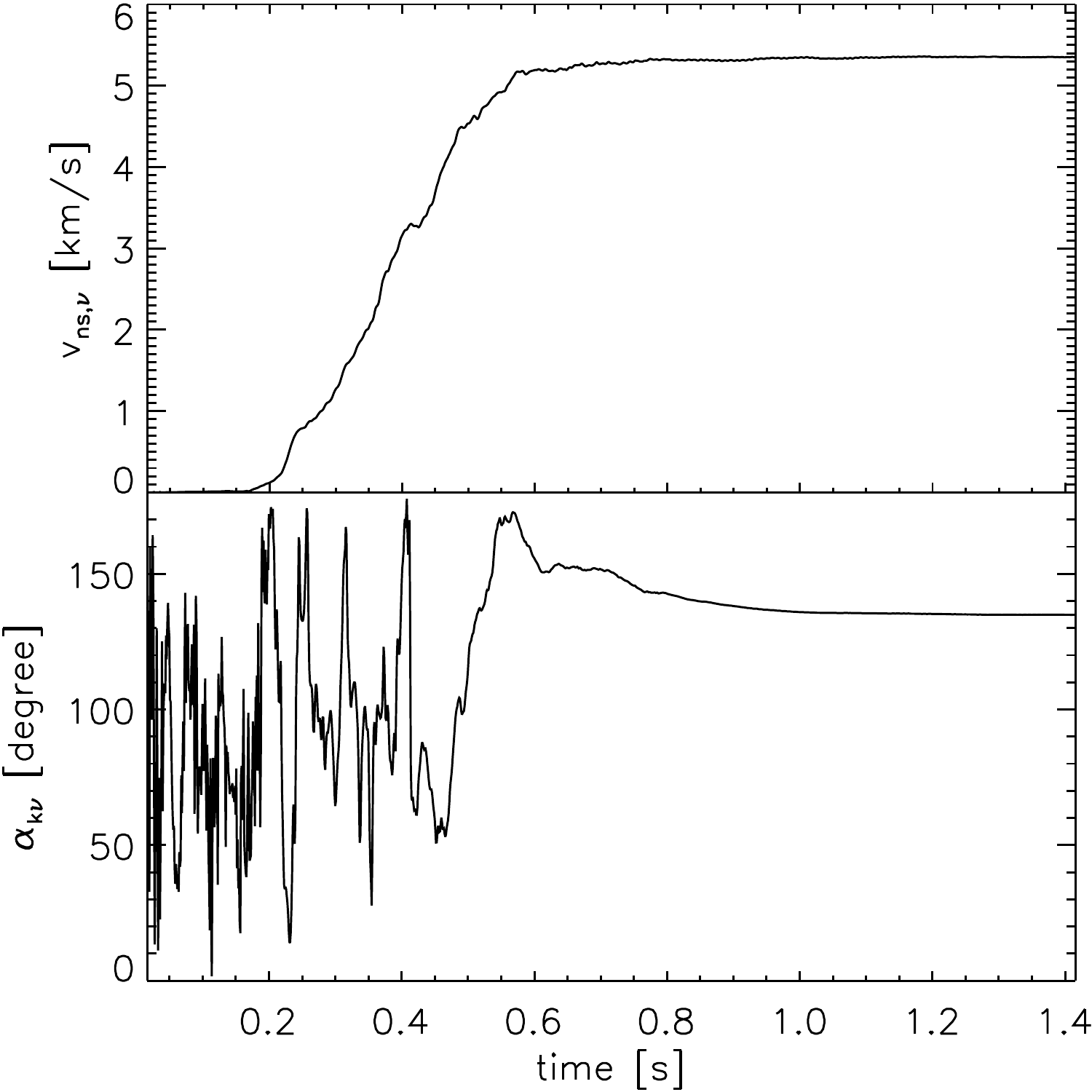}
\caption{{\em Left:} Entropy distribution in a cross-sectional plane
  of model L15-1 at 1000\,ms after bounce. The plane is chosen such
  that it contains the final vectors of the NS kick velocity induced
  by aspherical mass ejection ($\mathbf{v}_\mathrm{ns}$, long white
  arrow pointing to the right) and of the kick velocity caused by
  anisotropic neutrino emission ($\mathbf{v}_{\mathrm{ns},\nu}$, short
  white arrow pointing towards the 7 o'clock position). The relative
  lengths of both arrows correspond to the absolute values of the
  velocities ($v_\mathrm{ns} \approx 160$\,km/s and
  $v_{\mathrm{ns},\nu} \approx 5$\,km/s), but the short arrow for
  $\mathbf{v}_{\mathrm{ns},\nu}$ is stretched by a factor of 5 for
  better visibility.  A yardstick in the lower left corner of the
  figure indicates a spatial length scale of 2000\,km.  The
  contribution of anisotropic neutrino emission to the total NS kick
  velocity is clearly a small effect.  {\em Right:} Time evolution of
  the neutrino-induced kick velocity $v_{\mathrm{ns},\nu}$ and of the
  angle $\alpha_{\mathrm{k}\nu}$ between $v_\mathrm{ns}$ and
  $v_{\mathrm{ns},\nu}$. The final velocities point roughly 
  in opposite directions.}
\label{fig:vnsnu}
\end{figure*}
%

Asymmetric neutrino emission can be caused by convective overturn in
the postshock layer and by convective activity inside the NS, i.e.,
below the neutrinospheric region.  In the postbounce accretion phase,
asymmetric neutrino emission is dominated by low-entropy accretion in
the neutrino-heating layer between gain radius and shock.  Hot spots
of high neutrino emission are connected to the strongest accretion
funnels. They encompass typical angular scales of 20--30 degrees and
are scattered over the whole emitting surface
(Fig.~\ref{fig:neutrinoflux}, top panel).  The low-emission regions,
which surround the hot spots, trace the distribution of high-entropy
bubbles of expanding, neutrino-heated matter and are significantly
more extended. The flux variations between low-emission and
high-emission regions are usually 10--30\%, but in very bright spots
can reach up to about 70\% for short times (typically of order
milliseconds).  These local variations are caused by enhanced fluxes
of electron neutrinos and antineutrinos, both being abundantly
produced in the accretion layer, but the fractional variations are
computed relative to the total neutrino flux. The latter includes
neutrinos of all three flavors, i.e. also heavy-lepton neutrinos and
antineutrinos, whose emission (about 30--50\% of the total flux) comes
from the core and is not strongly affected by accretion emission and
thus more isotropic.  After the explosion has been launched, typically
a fewer accretion downflows survive and the number of hot spots
decreases. The activity is now concentrated in a few regions
(Fig.~\ref{fig:neutrinoflux}, middle panel) instead of showing the
previous high-order multipole pattern that encompasses the whole
sphere. After the accretion has finally ceased, the surface flux
variations are determined by the convectively modified neutrino
transport from below the neutrinosphere. The granular pattern of the
neutrino emission then reflects the smaller convective cells in the
radially less extended convective layer inside the PNS
(Fig.~\ref{fig:neutrinoflux}, lower panel).  In this phase the
variation amplitude of the total neutrino surface flux is usually
around 5\%, in peak activity periods up to 10\%.

During all three stages, however, the flux variation pattern is
extremely time-dependent and non-stationary. The accretion hot spots
appear, fade away again, and grow once more at different locations on
the surface. Only in the phase of shock acceleration and simultaneous
accretion, when the overall structure of the postshock ejecta begins
to freeze out and a nearly self-similar expansion sets in, the
coherence timescale of the emission pattern increases. Nevertheless,
there are still large temporal fluctuations because the accretion
flows along the downflows are not continuous but change in strength.
In the post-accretion phase the cellular pattern is uniform over the
whole surface. The cell sizes show a tendency to increase at later
times because the width of the convective layer grows with time.
Simultaneously, the amplitude of the flux variations as well as the
positions of cooler and hotter areas exhibit rapid temporal
variability on timescales of milliseconds to some 10 milliseconds.

As a consequence of the non-stationary, time-variable, and small-scale
cellular structure of the surface-flux pattern, in particular during
the long-time neutrino emission in the Kelvin-Helmholtz cooling phase
of the PNS when most of the neutrino energy is radiated away, we do
not expect any sizable NS acceleration due to anisotropic neutrino
emission.

Following \citet{Schecketal06}, we calculate the neutrino momentum
from the numerical results as
\begin{equation}
\mathbf{P}_\nu(t)=\int_{R_\mathrm{ib}<r<R_\mathrm{ob}}p_\nu\;
\hat{\mathbf{r}}\;{\mathrm{d}}V+\int_0^t{\mathrm{d}}t'
\oint_{r=R_\mathrm{ob}}p_\nu c\;\hat{\mathbf{r}}\;{\mathrm{d}}S
\, ,
\label{eq:P_nu}
\end{equation}
where $p_\nu=|{\mathbf{F}_\nu}|/c^2$ is the neutrino momentum density,
and $\hat{\mathbf{r}}$ is the unit vector in radial direction, which
in our transport description (``ray-by-ray approximation'') also
defines the direction of the neutrino flux (flux components in the
angular directions are assumed not to exist).  The first term accounts
for the total momentum associated with all neutrino energy on the
computational grid at a given time $t$, while the second term is the
time integral of the neutrino momentum that has left the computational
domain through the outer boundary until this time $t$.

The numbers for the neutrino-kick velocity $v_{\mathrm{ns},\nu}$
listed for all models in Table~\ref{tab:results} confirm our
expectation: The data show that $v_{\mathrm{ns},\nu}$ makes only a
very small, in most cases a negligible, correction to the final NS
kick velocity. It is typically much less than 5\,km\,s$^{-1}$ except
in model N20-1-lr, where $v_{\mathrm{ns},\nu}$ reaches
7\,km\,s$^{-1}$, which is still only $\sim$4\% of the mass-ejection
induced kick velocity $v_\mathrm{ns}$ at 1.3\,s after bounce.  An
entropy slice of model L15-1, which is a better resolved case with a
relatively high neutrino-produced kick, is displayed at 1000\,ms after
core bounce in Fig.~\ref{fig:vnsnu} together with the time evolution
of $v_{\mathrm{ns},\nu}$ and of the angle $\alpha_{\mathrm{k}\nu}$
between $v_{\mathrm{ns},\nu}(t)$ and $v_\mathrm{ns}(t)$.  The chosen
slice contains the final (at $t = 1.3$\,s p.b.) vectors of both
velocities.  The final angle $\alpha_{\mathrm{k}\nu}$ is significantly
larger than 100$^\circ$, a result that is also found in most of our
other 3D models (Table~\ref{tab:results}).

The velocity $v_{\mathrm{ns},\nu}$ increases steeply only after about
200\,ms p.b.\ when a sizable accretion and emission asymmetry begins
to develop. Before this phase the small angular variations
in the
neutrino emission do not allow for the growth of any noticeable value
of $v_{\mathrm{ns},\nu}$. Towards the onset of the explosion (at
$\sim$400\,ms; Table~\ref{tab:results}) a nearly stationary pattern of
accretion downflows and high-entropy bubbles emerges in the postshock
layer and persists during the subsequent expansion of the supernova
shock and neutrino-heated ejecta. The steep rise of
$v_{\mathrm{ns},\nu}$ between 200\,ms and 600\,ms after bounce in
model L15-1 is associated with a massive and extended, long-lasting
downflow funnel that carries fresh gas swept up by the accelerating
shock from the immediate postshock region towards the NS surface. This
leads to a neutrino emission maximum in the right hemisphere (roughly
at the 2:30 o'clock position) and thus to a neutrino-induced kick
pointing in the opposite direction (short white arrow for
$\mathbf{v}_{\mathrm{ns},\nu}$ in Fig.~\ref{fig:vnsnu}).  The relics
of the massive downflow funnel are still visible to the right of the
center in the left image of Fig.~\ref{fig:vnsnu}.  At about 600\,ms
p.b.\ the accretion ends and the neutrino emission pattern adopts the
high-multipole cellular structure characteristic of the PNS convection
during the subsequent Kelvin-Helmholtz cooling phase (cf.\ above
discussion and bottom panel in Fig.~\ref{fig:neutrinoflux}). At this
stage the neutrino-kick velocity has nearly reached its terminal
value. It hardly changes later on because short-wavelength variations
and short-timescale fluctuations diminish the integrated neutrino
emission asymmetry over longer periods of time.  Nevertheless,
$\alpha_{\mathrm{k}\nu}$ continues to exhibit some drift even between
600\,ms and $\sim$1000\,ms, because the gravitational pull by the
anisotropic ejecta still causes an evolution of the associated kick
direction. Only at $t \gtrsim 1000$\,ms $\mathbf{v}_\mathrm{ns}$ has
also found its final orientation about 135$^\circ$ away from
$\mathbf{v}_{\mathrm{ns},\nu}$ (Fig.~\ref{fig:vnsnu}), while the
gravitational acceleration of the NS continues on a significant level
even beyond 1.3\,s after bounce (Table~\ref{tab:results}). The
neutrino kick typically damps the ejecta-induced kick: Large angles
$\alpha_{\mathrm{k}\nu}$ are a natural consequence because
long-lasting downflows determine the accelerating gravitational tug on
the one hand and produce the dominant neutrino emission anisotropy on
the other. While the NS kick associated with the asymmetric ejecta
distribution is directed towards the downflow, the enhanced neutrino
emission on the downflow side carries away momentum that leads to a NS
recoil in the opposite direction.

%
\begin{figure*}
\includegraphics[width=\hsize]{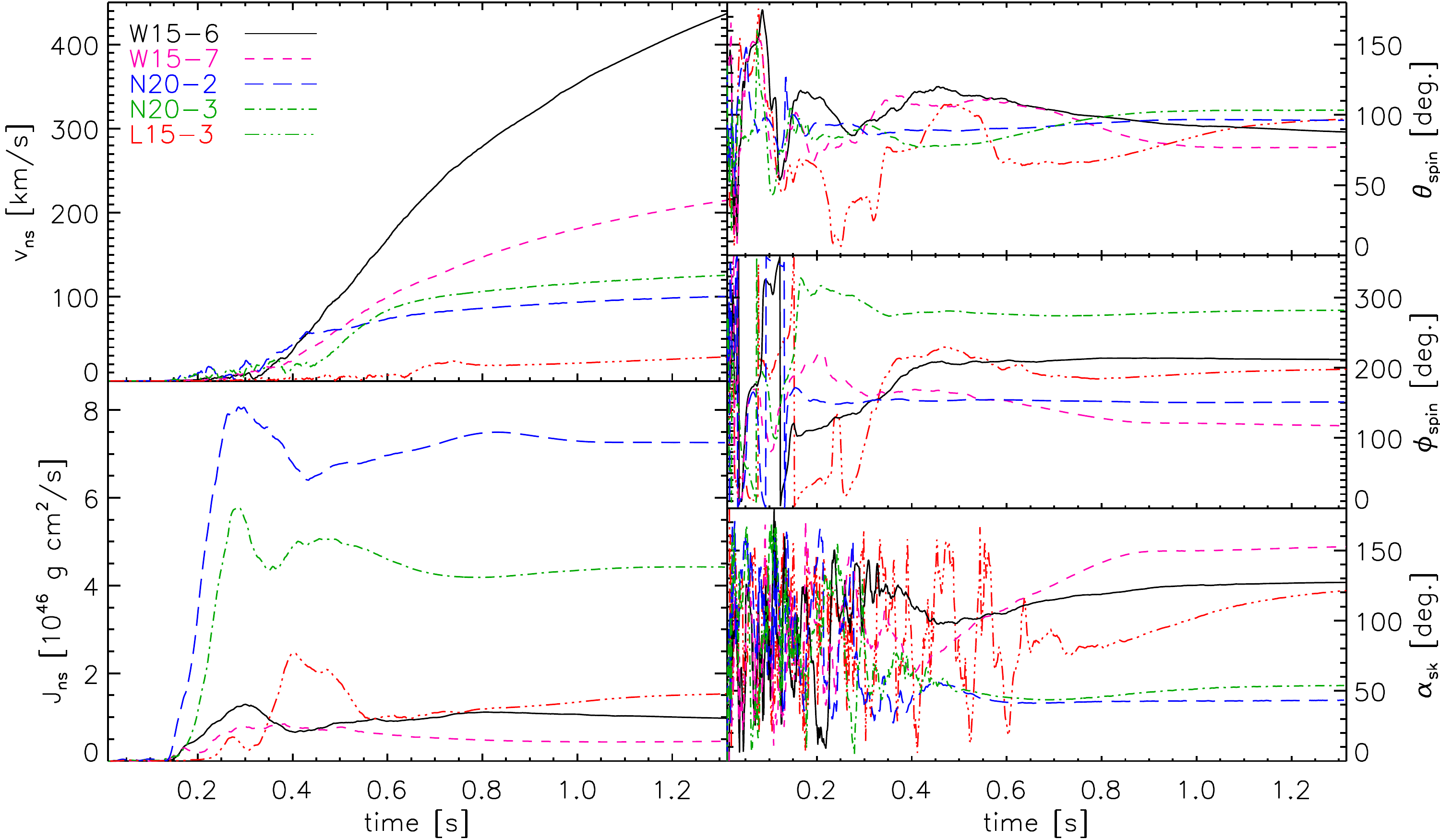}
\caption{Time evolution of the NS kick velocity ({\em upper left
    panel}) compared to the absolute value of the NS angular momentum
  ({\em lower left panel}), the polar and azimuthal angles
  ($\theta_\mathrm{spin}$, $\phi_\mathrm{spin}$) of the NS spin
  direction ({\em right top and middle panels}), and the relative
  angle $\alpha_{\mathrm sk}$ between spin and kick directions ({\em
    right bottom panel}) for five selected models. Two of these models
  represent high-spin cases (N20-2, N20-3), two are low-spin cases
  (W15-6, W15-7), and one is an intermediate case (L15-3) from our
  model set. The build-up of the angular momentum occurs in
  mainly a single impulsive episode during the convectively perturbed
  accretion phase of the NS.  Together with its angles the angular
  momentum reaches its terminal value long before the NS acceleration
  has ended. The clustering of all five plotted models around a
  final $\theta_\mathrm{spin} \approx 90^\circ$ is a mere coincidence,
  and the whole set of computed models exhibits the expected random
  distribution in polar-angle space.  }
\label{fig:jns}
\end{figure*}
%

\section{Neutron star spins}
\label{sec:nsspins}

Mechanisms that impart a natal kick to the NS can also lead to a NS
spin even in the case of nonrotating progenitor stars. This is
expected theoretically
\citep{Burrowsetal95,SpruitPhinney98,FryerYoung07} and suggested by
observations \citep[e.g.,][]{Farretal11}.  NS spin-up may, for
example, be a stochastic process that is connected to random,
nonradial impacts of the fast accretion downflows, by which
low-entropy gas is channelled through the gain region onto the PNS.
Accretion downflows, which hit the NS asymmetrically and not exactly
head-on, exert torques and thus spin up the NS. A mass $\Delta m$
colliding with the NS with impact velocity $v_\mathrm{i} \sim
v_\mathrm{ff} = \sqrt{2GM_\mathrm{ns}/r}$ and impact parameter $d$
transfers an angular momentum of
\begin{eqnarray}
\Delta J_\mathrm{ns} 
&=& \Delta m \,\sqrt{\frac{2GM_\mathrm{ns}}{r}}\,d  \cr
&\sim& 
6\times 10^{46}\,\left[\mathrm{g}\,\frac{\mathrm{cm^2}}{\mathrm{s}}\right]
\,\Delta m_{-3}\, d_{30}\, v_{\mathrm{i},10}
\label{eq:angmom1}
\end{eqnarray}
and for rigid rotation leads to a spin period of
\begin{equation}
T_\mathrm{spin}\, =\,\frac{2\pi I_\mathrm{ns}}{\Delta J_\mathrm{ns}}
\,\sim\, 0.2\,\mathrm{s}\,
(\Delta m_{-3}\, d_{30}\, v_{\mathrm{i},10})^{-1} \,,
\label{eq:tspin1}
\end{equation}
where $\Delta m$ is normalized by $10^{-3}\,M_\odot$, $d$ by 30\,km,
$v_\mathrm{i}$ is given in units of $10^{10}$\,cm\,s$^{-1}$, and
$I_\mathrm{ns}\sim 2\times 10^{45}$\,g\,cm$^2$ is the NS moment of
inertia.

Equations~(\ref{eq:angmom1}) and (\ref{eq:tspin1}) give rough order of
magnitude estimates and show that for moderate values of the
nonaxisymmetrically accreted mass ($\Delta m\sim 10^{-3}\,M_\odot$)
the NS may receive an angular momentum of some
$10^{46}$\,g\,cm$^2$s$^{-1}$, corresponding to a specific angular
momentum of order $10^{13}$\,cm$^2$s$^{-1}$ and spin periods of some
100\,ms. However, neither the number of accretion downflows nor their
impact parameters can be computed analytically. Quantitative answers
require 3D hydrodynamical simulations of the postbounce accretion
phase.

Alternatively, it has been proposed that NS spin-up may not result
from a stochastic process connected to random, nonradial impacts of
accretion downflows. Instead, SASI spiral modes ($m=1,\,2$ spherical
harmonics components) might develop in the postshock accretion flow,
by which a radial redistribution and separation of angular momentum
between the compact remnant and the ejecta could be established: While
NS spin is accumulated by the accretion of rotating gas, the ejecta
carry away counter-rotating matter with an angular momentum of the
same magnitude but with opposite direction
\citep{BlondinMezzacappa07,BlondinShaw07,Iwakamietal09,Fernandez10}.
This phenomenon was seen in idealized numerical models of collapsing
stellar cores, making use of an ideal-gas EoS and ignoring neutrino
heating \citep[except by][]{Iwakamietal09}.  It was also observed in
laboratory experiments of a shallow water analogue of the SASI in two
dimensions \citep{Foglizzoetal12}. However, the timescale for SASI
spiral modes to develop in the core without initial rotation was found
to be close to one second, which is longer than the timescale for the
onset of the explosion and for the saturation of the NS spins in all
of our SN simulations. Correspondingly, we do not observe any clear
signatures of SASI spiral modes in our models.  In contrast, $m=1,\,2$
modes can dominate the asphericity of the shock surface,
$R_\mathrm{s}(\theta,\phi)$, and of the postshock mass distribution,
$\Sigma(\theta,\phi) = \int_{R_\mathrm{ns}}^{R_\mathrm{s}}
\mathrm{d}r\,r^2\rho$ (Eq.~\ref{eq:masspersa}), in models where the NS
acquires only an unspectacular angular momentum.  In all our models,
however, the normalized amplitudes of the $m \neq 0$ spherical
harmonics components of the shock deformation never exceed a few
percent (Fig.~\ref{fig:mode_rotk}). This is much smaller than the
order unity shock displacements obtained by
\citet{BlondinMezzacappa07,BlondinShaw07} and \citet{Fernandez10}.  It
is also possible that neutrino heating and neutrino-driven convection
destroy the coherence of the spiral modes and thus impede their
growth, or it might enhance mode coupling, thus amplifying energy
redistribition between different modes. Moreover, even when shells
with strong differential rotation in opposing directions arise in the
turbulent region between PNS and shock, the question still remains how
much of the induced angular momentum is finally
accreted. \citet{Rantsiouetal11} suspect that most of it is likely to
be ejected in the explosion.

\subsection{Numerical results}

Because of violent convective activity, the matter in the
neutrino-heated layer develops nonradial motion in different
directions. These nonradial flows can be connected with sizable
amounts of angular momentum.  Integral values can indeed reach up to
several $10^{46}$\,g\,cm$^2$\,s$^{-1}$ in different regions of the
layer between PNS surface and accretion shock, compatible with the
order-of-magnitude estimate of Eq.~(\ref{eq:angmom1}).

Since our simulations start from nonrotating collapsing stellar cores,
any angular momentum acquired by the nascent NS must be balanced by
angular momentum in the opposite direction in the gas outside of the
NS.  Different from the linear momentum of the compact remnant, which
is calculated as the negative value of the gas momentum integrated
over the whole computational grid (see Eq.~\ref{eq:vns}), numerical
tests showed that it is advisable to constrain the volume integration
for the angular momentum to the immediate vicinity of the forming
NS. This helps minimizing numerical errors connected to the imperfect
conservation of angular momentum of gas that expands outwards over
radial distances extending over many orders of magnitude on a
spherical grid with varied resolution.
As in \citet{kickletter} we therefore determine an estimate of the
angular momentum $\mathbf{J}_\mathrm{ns}$ transferred to the NS by
hydrodynamic forces associated with accreted or ejected gas from the
following equation:
\begin{equation}
\mathbf{J}_\mathrm{ns}(t)=-\left(\int_{R_\mathrm{ns}}^{r_\mathrm{o}}
\mathrm{d}V \,\rho \mathbf{j}(t) + \int_0^t
\mathrm{d}t' r_\mathrm{o}^2 \oint_{4\pi}\mathrm{d}\Omega
\, (\rho \mathbf{j}v_r)\mid_{r_\mathrm{o}}\right) \, ,
\label{eq:Jns}
\end{equation}
where $\mathbf{j}$ is the specific angular momentum, $\Omega$ denotes
the solid angle, and $A_\mathrm{o} =4\pi r_\mathrm{o}^2$ is the area
of a sphere of chosen radius $r_\mathrm{o}$. The first term in the
brackets on the rhs of Eq.~(\ref{eq:Jns}) accounts for the angular
momentum contained by the spherical shell bounded by the NS
``surface'' at $R_\mathrm{ns}$ (defined as the radius where the
density is $10^{11}$\,g\,cm$^{-3}$) on the one side and bounded by
$r_\mathrm{o}$ on the other. The second term gives the angular
momentum that is carried by mass leaving the shell through surface
$A_\mathrm{o}$ until time $t$.  The introduction of the surface
integral is possible in the context of NS spin-up, because angular
momentum can be transferred to the central object only when the latter
exchanges mass with its environment. Long-range gravitational forces
cannot exert a torque on the compact remnant, because the NS (in a
very good approximation) is described in our simulations as a
gravitating point mass plus an essentially spherical near-surface
layer of matter surrounding the inner grid boundary.  The numerical
evaluation of Eq.~(\ref{eq:Jns}) yields results that do not depend on
the exact location of $r_\mathrm{o}$ between 500\,km and 1000\,km.

In Table~\ref{tab:results} the results for the angular momentum
$\mathbf{J}_\mathrm{ns}$, normalized by
$10^{46}$\,g\,cm$^2$\,s$^{-1}$, the relative angle between spin and
kick directions, $\alpha_{\mathrm{sk}}$, and the estimated final NS
spin period, $T_{\mathrm{spin}}$, are listed for all our model runs at
the end of most of the simulations (1.1--1.4 seconds after bounce). We
compute $T_{\mathrm{spin}}$ from the given numbers for
$\mathbf{J}_\mathrm{ns}$ and the NS moment of inertia,
$I_\mathrm{ns}$, as $T_\mathrm{spin} = 2\pi
I_\mathrm{ns}/|\mathbf{J}_\mathrm{ns}|$, considering the NS as a rigid
rotator and making the rough approximation that $I_\mathrm{ns}$ is
given by the Newtonian expression for a homogeneous sphere:
$I_\mathrm{ns}\approx\frac{2}{5} M_\mathrm{ns}R_\mathrm{ns}^2$.  The
(baryonic) NS mass $M_\mathrm{ns}$ is also listed in
Table~\ref{tab:results} and we adopt a final NS radius of
$R_\mathrm{ns}=12$\,km. It is important to note that we assume that
the angular momentum of the NS is conserved when the remnant contracts
to its final radius during the neutrino-cooling evolution.  This
ignores a possible $\sim$40\% reduction associated with neutrino
losses \citep[see][]{Janka04} and with mass ejection in the
neutrino-driven wind.

The angular momentum values listed in Table~\ref{tab:results} span a
range from $\sim$0.1$\times 10^{46}$\,g\,cm$^2$\,s$^{-1}$ to more than
$7\times 10^{46}$\,g\,cm$^2$\,s$^{-1}$, while the corresponding spin
periods vary by over a factor of 60, ranging from close to 0.1\,s to
nearly 8\,s. Higher angular momenta usually lead to the smaller
$T_\mathrm{spin}$; this order is reversed by differences of the NS
masses only in a few cases.

While there is no obvious systematic differences of the NS angular
momenta between the W15 and L15 models, although the L
models clearly have the longest delays to explosion, three of the four
members of the N15 series stick out having the highest $J_\mathrm{ns}$
values of the whole model set. In addition, two of the three B models
are among those acquiring the lowest NS angular momentum. These
tendencies, cautiously interpreted in view of the limited number of
models in each series, seem to be linked to the convectively stirred
ejecta mass in the neutrino-heating region. Assuming similar values
for the specific angular momentum $\mathbf{j}$ in all cases, the
integration of Eq.~(\ref{eq:Jns}) depends crucially on the mass
involved. In particular model B15 is special by its steep density
decline outside the iron core in addition to the short evolution
timescale to explosion, both of which tend to give a small angular
momentum transfer to the accreting remnant.

The values of the final angle between spin and kick vectors,
$\alpha_{\mathrm{sk}}$, in Table~\ref{tab:results} do not reveal any
clear correlations beween spin and kick directions, in particular no
strong tendency towards alignment or counter-alignment.  The
distribution is nearly symmetric for $\alpha_{\mathrm{sk}} < 90^\circ$
and $\alpha_{\mathrm{sk}} > 90^\circ$ (9:11 cases) and essentially
flat when sorted into 60$^\circ$ intervals (7:6:7 cases), although a
statistically equal distribution in all directions would require twice
as many events in the equatorial belt than in each polar wedge. Our
model sample, however, is still too small to draw firm conclusions in
a statistical sense.

\subsection{Neutron star spin-up mechanism}

It is important to note that kicks and spins are attained by the NS
through different mechanisms in our simulations, although both
phenomena are connected to the development of large-scale nonradial
flows in the collapsing stellar core, which lead to momentum and
angular momentum exchange between the accreting compact remnant and
the asymmetrically expanding ejecta gas. In order to understand the
differences, we present in Fig.~\ref{fig:jns} the time evolution of
$J_\mathrm{ns}$ compared to $v_\mathrm{ns}$ for a selection of
representative models, namely for the two high-resolution N15-runs
with the largest values of $J_\mathrm{ns}$, for the two W15-cases with
the lowest $J_\mathrm{ns}$, and for model L15-3 as an intermediate
case. The left two panels show that the angular momentum increases
much faster and reaches a value close to its asymptotic one much
earlier than the recoil velocity does. Different from the NS kick most
of the NS spin is attained in an impulsive event within a time
interval of 100--200\,ms relatively early after core bounce. No
significant $J_\mathrm{ns}$ evolution takes place beyond $t \gtrsim
500$\,ms, i.e. after the onset of the explosion (measured by
$t_\mathrm{exp}$). In contrast, the NS velocity is very small at the
time when the PNS receives its angular momentum, but the kick can grow
monotonically with considerable rate for several seconds.

The reason for this difference is the fact that angular momentum can
be transferred to the NS only by nonspherical mass flows and thus
during a few hundred milliseconds of accretion. This takes place
before the onset of the explosion and can go on afterwards only as
long as accretion still continues simultaneously with the outward
expansion of the SN shock\footnote{The neutrino-driven wind
  that develops after the onset of the explosion is essentially
  spherically symmetric in our simulations. Possible angular
  variations caused by PNS convection have low amplitudes and short
  wavelengths and thus do not carry any significant amount of
  momentum. Also the rate of angular momentum loss in the dilute
  outflow is typically very low.}.  
In contrast, the recoil of the newly formed NS is mainly established
by the long-range anisotropic gravitational attraction between the
compact remnant and the asymmetric ejecta as described in detail in
Sect.~\ref{sec:kicks}.  In Fig.~\ref{fig:jns} this difference
manifests itself in the long-lasting, monotonic rise of
$v_\mathrm{ns}(t)$ during the whole displayed time and even much
longer (cf.\ Fig.~\ref{fig:longvns}), whereas most of the angular
momentum is accumulated in the NS during a period of massive accretion
shortly after the convective and SASI instabilities have grown to
their full strength. When the accretion to the NS is diminished by the
outward acceleration of the gas behind the explosion shock, sporadic
downdrafts reaching the NS can still lead to modest variations
in 
$J_\mathrm{ns}$ for some 100\,ms.  The duration of this phase depends
on the progenitor structure and the detailed dynamics of the
explosion, but even for relatively low explosion energies it does not
last any longer than $\sim$500\,ms after the revival of the SN shock.

%
\begin{figure}[!]
\includegraphics[width=\hsize]{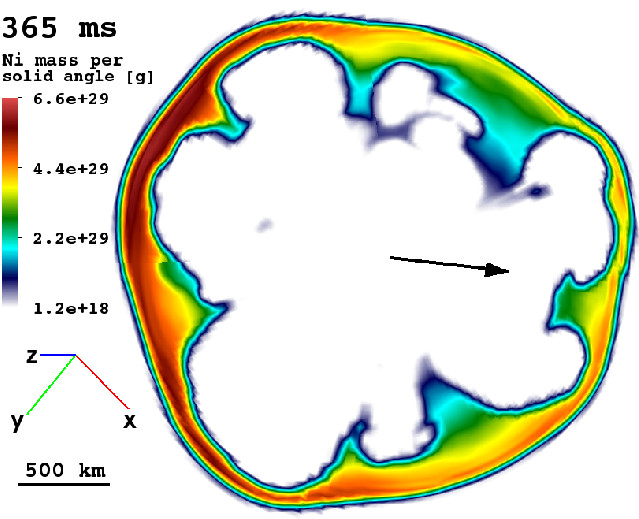}\\
\includegraphics[width=\hsize]{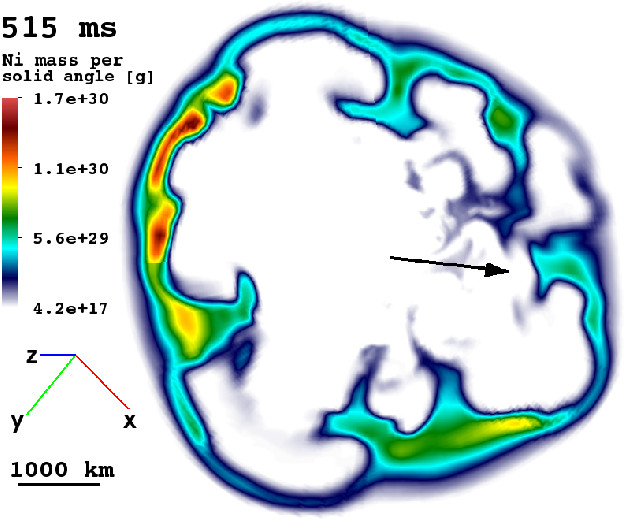}
\caption{$^{56}$Ni distribution for model W15-2 at 365\,ms and 515\,ms
  after bounce as produced by explosive burning in
    shock-heated ejecta. The integrated nickel mass per solid angle
  is plotted 
  in the cross-sectional plane that contains the NS kick (indicated by
  the black arrows pointing to the right) and spin vectors. Red
  colored regions show that more nickel is formed in the direction
  {\em opposite} to the NS kick direction. The size of the displayed
  volume can be estimated from the yardsticks given in the lower left
  corner of each image.}
\label{fig:niprod}
\end{figure}
%

The differences between the kick and spin mechanisms are reflected not
only by the evolution of the NS kick and spin magnitudes but also by
the evolution of the kick and spin directions.  During the initial
steep growth of $J_\mathrm{ns}$ the spin direction
($\theta_\mathrm{spin}$, $\phi_\mathrm{spin}$) mostly settles to its
final direction, in particular in the two high-spin N20-models.
Long-time variations in the NS angular momentum then correlate with
only smaller changes in the spin direction, which
signal the impact of 
new accretion downdrafts. Such changes become less frequent at later
times and finally die away when the accretion ends. In the L15-models,
which have the highest densities in the shells surrounding the iron
core and therefore explode with the longest delays (a good example is
model L15-3, where the shock is revived as late as $\sim$500\,ms after
bounce), the period of large $J_\mathrm{ns}$ variations is more
extended.  Slow fluctuations of the spin-vector angles and a slight
drift of $J_\mathrm{ns}$ occur even until more than one second post
bounce because of the angular momentum transported away with the more
massive neutrino-driven wind in the L models, in contrast to the W and
N models. The long-time variation in the spin differs from that of the
kick, whose direction freezes out as soon as the bulk of the mass
carrying the ejecta asymmetries expands essentially self-similarly,
although $v_\mathrm{ns}$ continues to increase.  The bottom right
panel of Fig.~\ref{fig:jns} provides the time evolution of the
relative angle between kick and spin vectors. The high-frequency
variations in $\alpha_\mathrm{sk}$ are caused by rapid changes
in the 
kick direction ($\theta_\mathrm{kick}$, $\phi_\mathrm{kick}$) until
about 200\,ms after the onset of the explosion, while a slow drift of
$\alpha_\mathrm{sk}$ on longer timescales (up to $\sim 1$\,s in the
most extreme case) is connected to an ongoing
reorientation of the spin direction.

Interestingly, the highest NS kicks and spins are not obtained for the
same models or progenitor. While the remnants of the N20 progenitor
exhibit the tendency to develop the highest spins, those of the W15
series are found to end up with the largest kicks.  This can be
considered as another manifestation that spin and kick mechanisms are
linked to different aspects of anisotropic mass flows in the SN core,
namely the former to asymmetries of the accretion flow and the latter
mainly to asphericities of the ejecta distribution.

It is unclear whether NS spin periods that are more extreme than those
found in some of our N20 models, in particular periods below 100\,ms,
can be obtained in nonrotating stars. It is possible that faster NS
rotation requires the progenitor core to rotate
\citep[e.g.,][]{Hegeretal05,Ottetal06}. In a rotating environment,
however, the growth of SASI spiral modes will be severely altered and
potentially fostered
\citep{BlondinMezzacappa07,YamasakiFoglizzo08,Iwakamietal09}, and
three-dimensional simulations are therefore indispensable to make
predictions for the connection between progenitor and NS rotation.


\begin{table*}
\caption{Hemispheric ejecta yields of helium and metals, and total
  nickel mass for the high-kick models W15-1 and W15-2, and the
  moderate-kick models L15-1 and L15-2. The numbers are integral
  values of the elemental abundances in the whole stellar matter that
  gets expelled by the supernova explosion. The tracer column provides
  the yield of Fe-group nuclei in neutrino-processed ejected material,
  some undetermined fraction of which may be $^{56}$Ni. The
  north-polar direction is defined to coincide with the NS kick
  direction.}
\begin{center}
\begin{tabular}{lccccccccccc}
\hline
\hline
\multirow{2}{*}{Model} &
\multicolumn{2}{c}{$^4$He $[M_\odot]$} &
\multicolumn{2}{c}{$^{12}$C  $[10^{-1}\,M_\odot]$} &
\multicolumn{2}{c}{$^{16}$O  $[10^{-1}\,M_\odot]$} &
\multicolumn{2}{c}{$^{20}$Ne $[10^{-2}\,M_\odot]$} &
\multicolumn{2}{c}{$^{24}$Mg $[10^{-2}\,M_\odot]$} \\
& North & South & North & South & North & South & North & South &
North & South\\
\hline
\hline
W15-1 & 2.78 & 2.66 & 1.18 & 1.10 & 3.68 & 3.75 & 8.90 & 8.49 & 2.41 &
2.85 \\
W15-2 & 2.78 & 2.65 & 1.16 & 1.12 & 3.43 & 3.84 & 8.67 & 8.49 & 2.16 &
2.86 \\
L15-1 & 2.39 & 2.34 & 0.90 & 0.87 & 2.77 & 2.89 & 5.00 & 5.06 & 2.12 &
2.49 \\
L15-2 & 2.40 & 2.39 & 0.89 & 0.87 & 2.85 & 2.79 & 5.21 & 4.88 & 2.47 &
2.42 \\
\hline\\
\hline
\hline
\multirow{2}{*}{Model} &
\multicolumn{2}{c}{$^{28}$Si $[10^{-2}\,M_\odot]$} &
\multicolumn{2}{c}{$^{40}$Ca $[10^{-2}\,M_\odot]$} &
\multicolumn{2}{c}{$^{44}$Ti $[10^{-3}\,M_\odot]$} &
\multicolumn{2}{c}{$^{56}$Ni $[10^{-2}\,M_\odot]$} &
\multicolumn{2}{c}{Tracer $[10^{-2}\,M_\odot]$} &
\multirow{2}{*}{Total $^{56}$Ni Mass $[M_\odot]$} \\
& North & South & North & South & North & South & North & South &
North & South\\
\hline
W15-1 & 1.88 & 2.92 & 1.33 & 4.81 & 0.68 & 2.43 & 1.26 & 4.28 & 2.23 &
6.08 & 0.055--0.139 \\
W15-2 & 1.74 & 2.83 & 1.27 & 4.66 & 0.81 & 2.17 & 1.37 & 4.09 & 2.22 &
6.27 & 0.055--0.139 \\
L15-1 & 1.75 & 2.33 & 1.76 & 2.47 & 1.49 & 2.40 & 1.34 & 1.87 & 4.78 &
7.20 & 0.032--0.152 \\
L15-2 & 2.13 & 2.15 & 2.54 & 2.74 & 2.32 & 2.55 & 1.81 & 1.89 & 8.68 &
9.74 & 0.037--0.221 \\
\hline
\end{tabular}
\end{center}
\label{tab:niNS}
\end{table*}

\section{Heavy-element production and explosion asymmetries}
\label{sec:heavyelements}

\citet{FryerKusenko06} pointed out that ejecta asymmetries could be
used to distinguish between different theoretical suggestions for the
kick mechanism. Ejecta-driven kicks like the gravitational tug-boat
mechanism discussed here must be expected to be associated with
stronger ejecta motion, i.e. with higher ejecta momentum, in the
direction opposite to the kick. However, diagnosing momentum
asymmetries of the ejecta may be difficult except in cases where the
center of mass of the gaseous SN remnant is clearly displaced relative
to the NS position or where the gas cloud exhibits extreme deformation
without gradients of the environmental conditions being responsible
for the asymmetric ejecta expansion. The SN remnants Cassiopeia~A
\citep{Isenseeetal10, Delaneyetal10, Restetal11, HwangLaming12} and
Puppis~A \citep{Petreetal96, WinklerPetre07, Katsudaetal08,
  Katsudaetal10, Beckeretal12} may be such lucky cases, because in the
former all the SN ejecta in the remnant, whose mass is dominated by
oxygen, while in the latter fast-moving, oxygen-rich optical filaments
and knots seem to have been recoiled opposite to the direction of the
high-velocity compact stellar remnant. In general, however, a major
problem for observing momenta asymmetries linked to an asymmetric
beginning of the explosion may result from the fact that the outgoing
SN shock transfers the bulk of the explosion energy and ejecta
momentum to the He and H shells, which dominate the ejecta mass in
events other than stripped SNe. An initially asymmetric explosion thus
becomes much more spherical when the SN shock sweeps up massive
stellar He and H layers. In this context it is important to note that
the NS kinetic energy and the NS momentum are very small compared to
the kinetic energy of the SN explosion, $E_\mathrm{k,SN}$, and the
corresponding measurable momentum, $P_\mathrm{k,SN} \equiv
\sqrt{2M_\mathrm{ej}E_\mathrm{k,SN}}$, associated with the radial
motion of the ejecta mass $M_\mathrm{ej}$. To determine the
(relatively small) linear momentum of the ejecta, which balances the
momentum of the NS, it is necessary to sum up all components of the
momentum vectors of the whole ejecta mass. Because of projection
effects this may be difficult and error-prone even if all of the
ejecta can be observed.

For this reason we propose here to look for correlations between NS
kicks and asymmetries in the distribution of heavy elements. These are
explosively produced in the innermost, shock-heated (and
neutrino-heated) SN matter during the first seconds of the blast, and
their globally asymmetric distribution will not be destroyed
lateron. In particular iron-group nuclei, most of which are
radioactive $^{56}$Ni decaying to stable $^{56}$Fe, are good tracers
of such early explosion asymmetries. This is connected to the fact
that the initial Fe-group material in the progenitor core is
completely photo-disintegrated during collapse and then buried in the
forming NS. All expelled Fe-group nuclei are freshly assembled in the
neutrino-heated SN ejecta and in the silicon and oxygen layers that
are heated to sufficiently high temperatures by the accelerating SN
shock.

Figure~\ref{fig:niprod} visualizes the corresponding directional
asymmetry of the $^{56}$Ni production in the plane containing the NS
velocity and spin vectors in the high-kick model W15-2. The images
show the nickel mass per unit solid angle at post-bounce times of
365\,ms and 515\,ms (the explosion in this model sets in about 250\,ms
after bounce; Table~\ref{tab:results}), which is obtained by
integrating the nickel-mass density over radius within each zone:
\begin{equation}
\left ( \frac{{\mathrm d}M_\mathrm{Ni}}{{\mathrm d}\Omega} 
\right )_i(\theta,\phi)  = 
\int_{r_{i-1/2}}^{r_{i+1/2}}{\mathrm d}r\,r^2 
\rho({\mathbf r}) X_\mathrm{Ni}({\mathbf r}) \, ,
\label{eq:nickelcell}
\end{equation}
where $X_\mathrm{Ni}({\mathbf r})$ is the mass fraction of nickel and
$\rho({\mathbf r})$ the mass density at location ${\mathbf r}$. The
color coding clearly shows more nickel production (intense red) in the
direction {\em opposite} to the kick vector.

At around 450--500\,ms after bounce the nickel fusion in the
shock-heated layers comes to an end. While most of the nickel expands
in a thin and dense shell behind the shock, some of it is channeled
into massive clump-like ``pockets'' between the more dilute
high-entropy bubbles of rising and outward pushing neutrino-heated
matter.  Also these bubbles may contain a fair amount of nickel from
the $\alpha$-rich freeze-out that occurs during the expansion cooling
of the bubbles. However, because these regions have a slight neutron
excess in our models, the nuclear flow is assumed to lead to the
production of the ``tracer'' species included in our network as
described in Sect.~\ref{sec:nucleosynthesis}. Since the actual
composition of this tracer material strongly depends on the
inaccurately determined value of $Y_e$ set by $\nu_e$ and $\bar\nu_e$
absorption and emission in the neutrino-processed ejecta, the amount
of $^{56}$Ni in this material is uncertain and we have not depicted it
in Fig.~\ref{fig:niprod}. The tracer material, however, enhances the
hemispheric asymmetry of the nickel distribution. This is visible from
the data listed in Table~\ref{tab:niNS}, which provides abundance
yields of the nuclear species in our burning network after all
nucleosynthesis processes have ended (depending on the progenitor and
explosion energy, this is the case typically 100--300\,s after the
onset of the explosion). Higher values of $^{56}$Ni in the southern
hemisphere correlate with a similarly large excess of tracer material
on the same side. We define the northern hemisphere as the half-sphere
around the NS kick direction, which implies that more $^{56}$Ni and
tracer material are ejected opposite to the NS kick direction. This
effect is especially pronounced (with hemispheric contrast up to a
factor 3.5) in high-kick models like W15-1 and W15-2.  The last column
of Table~\ref{tab:niNS} provides our determination of the total
$^{56}$Ni yields of the listed models. The uncertainty range is
defined by the assumption that in the lower limit the tracer material
contains no $^{56}$Ni and in the upper limit all tracer material is
$^{56}$Ni.

%
\begin{figure*}
\includegraphics[width=0.5\hsize]{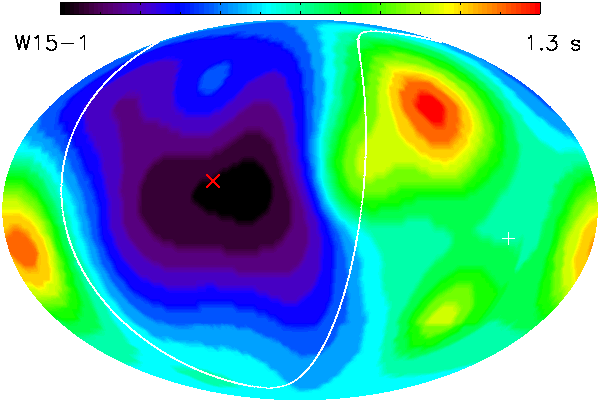}
\includegraphics[width=0.5\hsize]{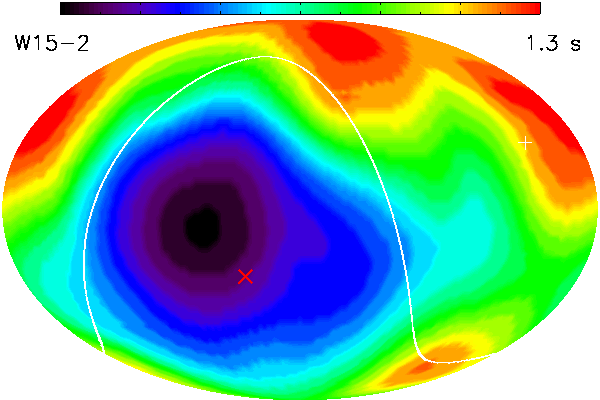}\\
\includegraphics[width=0.5\hsize]{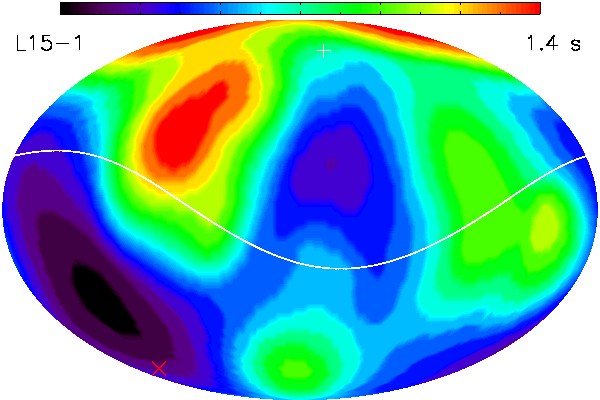}
\includegraphics[width=0.5\hsize]{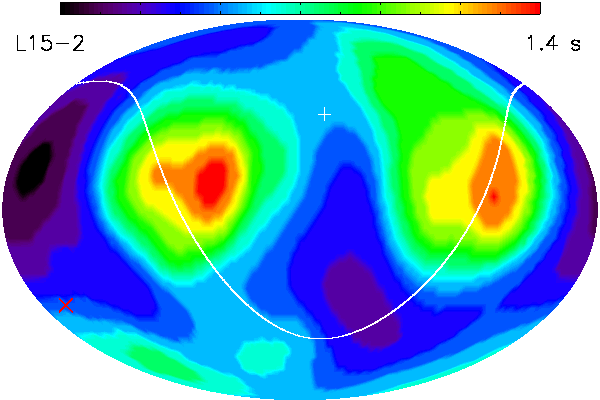}\\
\caption{Shock deformation for the high-kick models W15-1 and W15-2
  and for the moderate-kick models L15-1 and L15-2 at 1.3\,s or 1.4\,s
  after bounce, as given in the upper right corner of each map. These
  times are close to the evolution phases when the nickel production
  in the shock-heated ejecta takes place. The color scale is linear
  with black indicating minima and red maxima. The differences between
  maxima and minima relative to the average shock radii are 30\%,
  35\%, 34\%, and 29\%, respectively.  Red crosses indicate the NS
  kick directions while white plus signs mark the directions {\em
    opposite} to the kicks.  The white lines separate the
  corresponding two hemispheres.  The northern hemisphere is defined
  as centered around the NS kick vector.}
\label{fig:shockmap}
\end{figure*}
%

%
\begin{figure*}
\includegraphics[width=0.5\hsize]{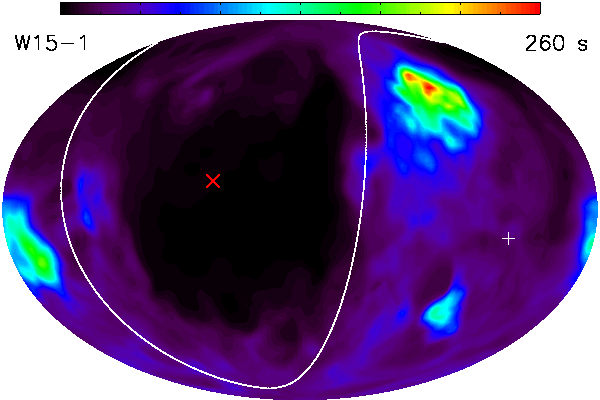}
\includegraphics[width=0.5\hsize]{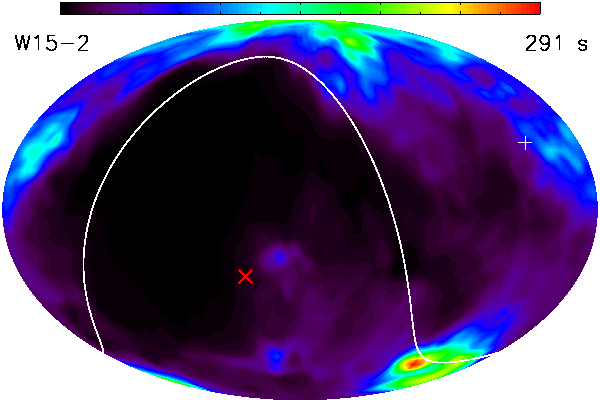}\\
\includegraphics[width=0.5\hsize]{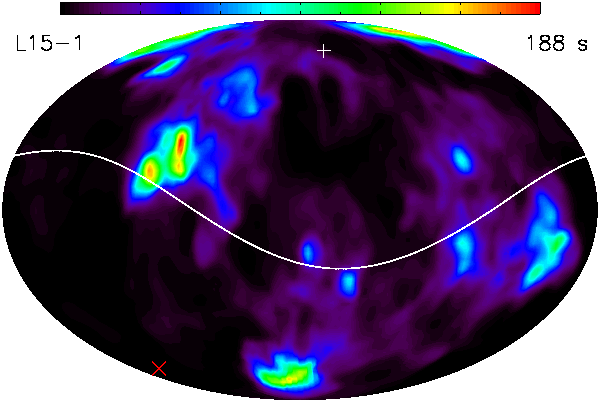}
\includegraphics[width=0.5\hsize]{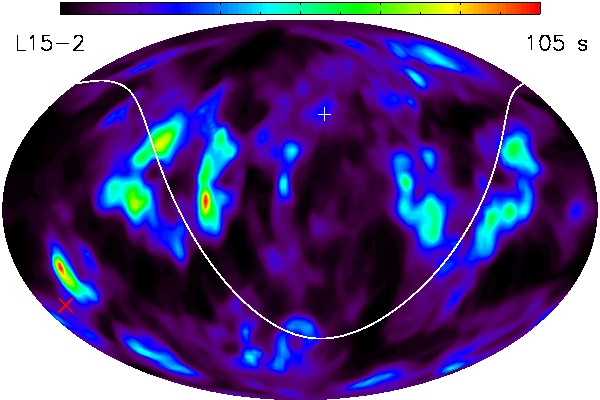}\\
\caption{Integrated nickel masses per solid angle in all directions
  displayed in $4\pi$ maps for models W15-1, W15-2, L15-1, and L15-2
  at postbounce times well after all nucleosynthesis processes have
  seized in our simulations.  Red crosses indicate the NS kick
  direction while white plus signs mark the direction opposite to the
  kick. The corresponding hemispheres are separated by a white equator
  line.  Note the differences between the high-kick W15 models and the
  moderate-kick L15 models. The former show a clear enhancement of the
  nickel production in the hemisphere opposite to the kick, whereas
  the latter eject nickel significantly more isotropically.}
\label{fig:niNSmap}
\end{figure*}
%

%
\begin{figure*}
\includegraphics[width=0.5\hsize]{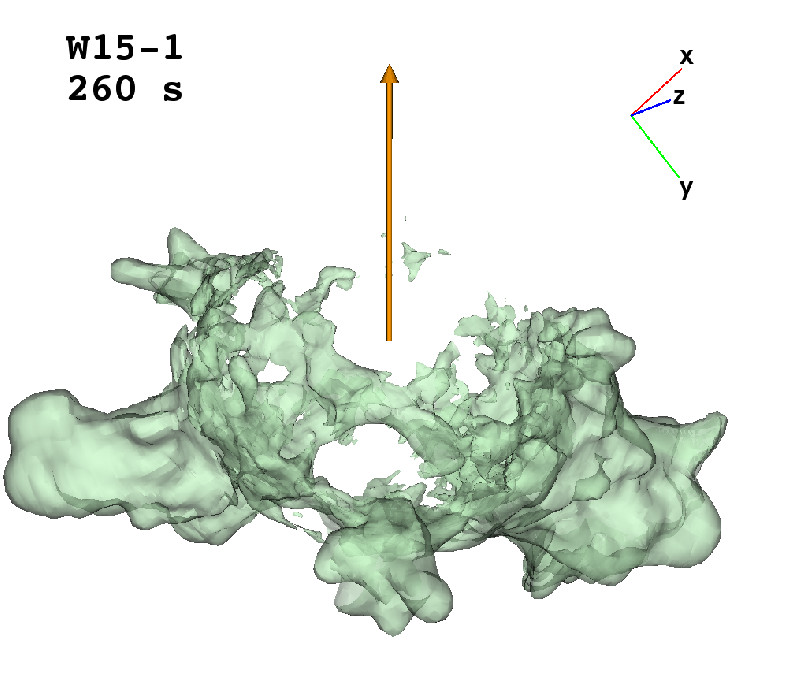}
\includegraphics[width=0.5\hsize]{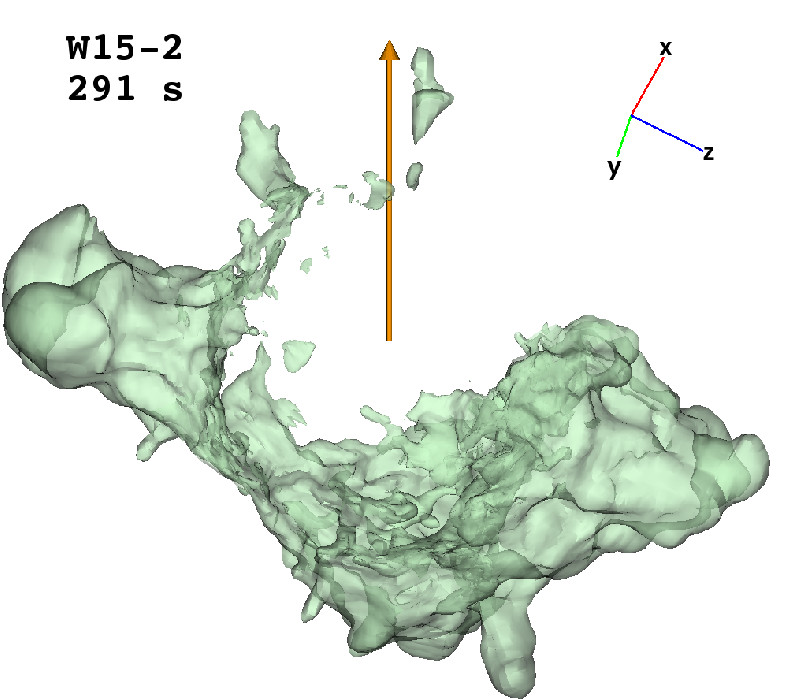}\\
\includegraphics[width=0.5\hsize]{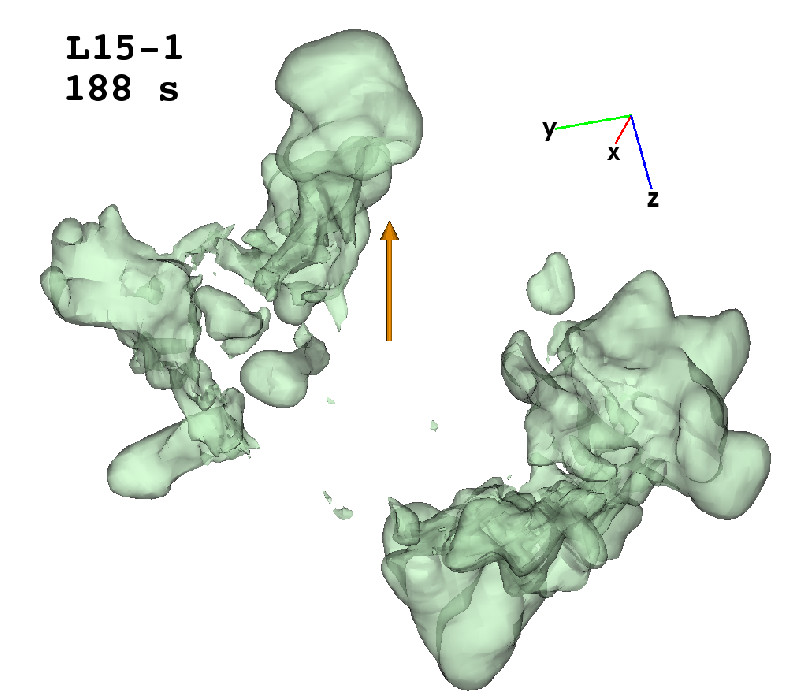}
\includegraphics[width=0.5\hsize]{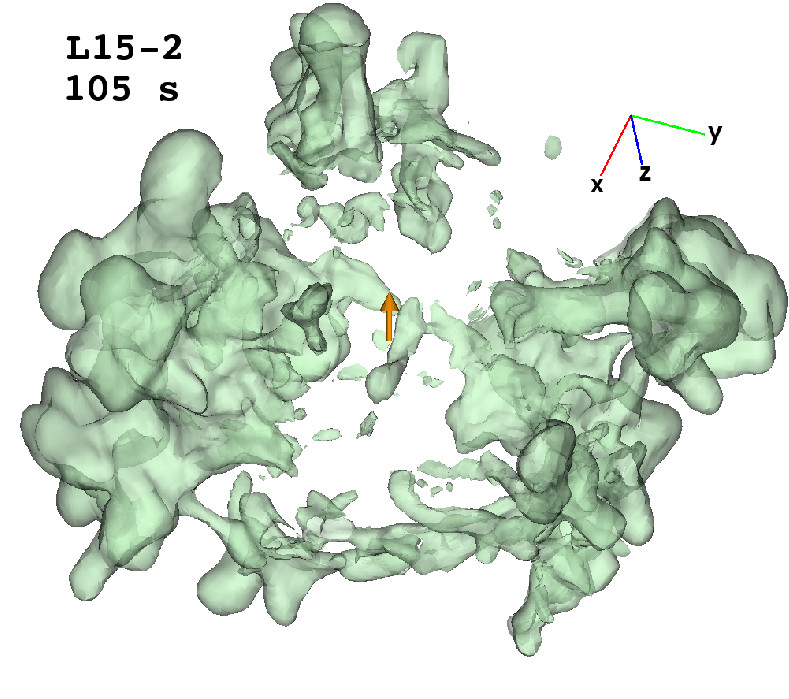}\\
\caption{Volumetric three-dimensional visualization of the nickel
  distribution for models W15-1, W15-2, L15-1, and L15-2 at the
  postbounce time given in the top left corner of each panel along
  with the model name. The semi-transparent isosurfaces correspond to
  a chosen value of the nickel mass per grid cell of
  $3\times10^{26}$\,g and are displayed at a stage well after all
  nucleosynthesis processes have seized in our simulations. The orange
  vectors represent the NS kick directions and are scaled by the
  corresponding NS velocities (96\,km\,s$^{-1}$ for the shortest arrow
  and 575\,km\,s$^{-1}$ for the longest one). The high-kick models
  W15-1 and W15-2 in the upper two panels exhibit a clear asymmetry
  with much more nickel being ejected in the hemisphere opposite to
  the kick direction. In contrast, the moderate-kick models L15-1 and
  L15-2 in the lower two panels exhibit a more isotropic distribution
  of the nickel, in particular no obvious hemispheric asymmetry
  between kick and anti-kick directions. While the radial
  distribution of the nickel may be strongly affected and changed by
  subsequent mixing instabilities that develop after the outgoing
  shock has passed the composition-shell interfaces of the progenitor
  star, the hemispheric differences in the nickel ejection will not be
  destroyed during the later supernova explosion.} 
\label{fig:niNS3D}
\end{figure*}
%

The comparison of the high-kick models W15-1 and W15-2, both of which
have NS recoil velocities in excess of 500\,km\,s$^{-1}$
(Table~\ref{tab:results}), with the moderate-kick model L15-1
($v_\mathrm{ns} \sim 250$\,km\,s$^{-1}$) and model L15-2, whose NS
receives a recoil velocity of only $\sim$100\,km\,s$^{-1}$, is
enlightening. The data in Table~\ref{tab:niNS} demonstrate that the
hemispheric differences of nickel and tracer ejection are largest for
W15-1 and W15-2, still considerable for L15-1, and nearly disappear
for L15-2. Similarly large hemispheric differences as for $^{56}$Ni
and our tracer species show up in the masses channeled by the network
into $^{28}$Si, $^{40}$Ca, and $^{44}$Ti. All of these nuclear
species\footnote{It should be noted that the small number of species in the
  employed network is likely to be responsible for a considerable
  overestimation of the $^{44}$Ti production in our simulations.  A
  better determination of the $^{44}$Ti yields will require more
  sophisticated network calculations.}  
emerge from the innermost ejecta regions, which carry the explosion
asymmetries and the mass-distribution inhomogeneities that are
responsible for the gravitational acceleration of the newly formed
NS. While some degree of hemispheric differences correlating with the
NS kick magnitude is still visible for $^{24}$Mg, the effect in the
case of $^{16}$O is very small. Although oxygen is more abundantly
synthesized behind the stronger shock in the southern hemisphere, an
asymmetry of the distribution of this element is largely obscured by
the dominance of the spherically symmetric ejecta mass that originates
from the unperturbed oxygen shell of the progenitor star. The other
lighter species ($^{20}$Ne, $^{12}$C, and $^4$He) do not exhibit the
southern excess.  In contrast, their yields are slightly reduced in
the southern hemisphere ---most pronouncedly in the case of
$^{20}$Ne--- because of the explosive production of the higher-mass
nuclei at the expense of these lighter abundances.

The dependence of the explosion strength on direction is visualized in
Fig.~\ref{fig:shockmap} in terms of the variations
in the shock radius 
and thus shock velocity at around the time when the nickel
nucleosynthesis has just taken place. Figure~\ref{fig:niNSmap} shows
the corresponding radially integrated nickel masses per solid angle,
\begin{equation}
\frac{{\mathrm d}M_\mathrm{Ni}}{{\mathrm d}\Omega}(\theta,\phi) 
= \int_{R_\mathrm{ib}}^{R_\mathrm{ob}}{\mathrm d}r\,r^2 
\rho({\mathbf r}) X_\mathrm{Ni}({\mathbf r}) \, ,
\label{eq:nickelint}
\end{equation}
where the radial integration is performed between the inner and outer
grid boundaries, $R_\mathrm{ib}$ and $R_\mathrm{ob}$, respectively.
In Fig.~\ref{fig:niNS3D} we provide volumetric impressions of the
nickel distribution in the ejecta by plotting isosurfaces of a
suitably chosen constant value of the nickel mass per grid
cell\footnote{The displayed quantity has the advantage that the bulk of
  the produced nickel becomes visible.  Both the nickel density, which
  essentially follows the radial gradient of the mass density, and the
  nickel mass fraction, which does not necessarily trace regions with
  high integral nickel mass, are less suitable for this purpose.}.
The coordinate system in these images has been rotated such that the
NS kick vector points to the 12 o'clock position.

In Fig.~\ref{fig:shockmap} one can see clear differences of the shock
deformation between low-kick and high-kick models.  In the high-kick
cases of W15-1 and W15-2, the northern side, which is defined as the
hemisphere around the NS kick direction (marked by the red cross),
contains a huge, deep trough (blue and black) whose minimum roughly
coincides with the kick direction.  In contrast, the southern
hemisphere is filled by a wide bump with several local maxima.
Consequently, there is a clear dipolar shock deformation along the
kick axis (cf.\ also Fig.~\ref{fig:mode_rotk}).  The moderate-kick
model L15-1 still exhibits a tendency for the same asymmetry pattern
with a shock minimum close to the red cross and a maximum near the
white plus sign in the opposite direction. Now, however, shallow
minima and maxima also appear in the southern and northern hemisphere,
respectively, indicating the growth of a quadrupolar deformation
component. The low-kick model L15-2 violates the north-south
imbalance. Both shock maxima and minima cross the equatorial line and
are obviously not correlated with the locations of the cross and plus
signs. Instead of the dipolar shock deformation of models W15-1 and
W15-2 we now see a quadrupolar pattern, consistent with the left lower
panel of Fig.~\ref{fig:mode_rotk}.  We note that the relative
variation in the shock radius in all four displayed cases is
similar. The shock deformation in high-kick and low-kick models
therefore does not need to differ in the variation amplitude, but
instead the type of dominant spherical harmonics mode can be the more
important factor.

In Fig.~\ref{fig:niNSmap} the highest concentrations of ejected nickel
are found exactly at the locations of the shock maxima in
Fig.~\ref{fig:shockmap}, while big ``holes'' of the nickel
distribution appear in the regions of shock minima. Again, the angular
pattern of the $^{56}$Ni distribution of the low-kick model L15-2 is
clearly dominated by higher spherical harmonics modes than in the
high-kick models W15-1 and W15-2. Model L15-1 with a moderate kick
defines an intermediate case. The directional asymmetries depicted in
Fig.~\ref{fig:niNSmap} are complemented by the volumetric information
of Fig.~\ref{fig:niNS3D}. The black areas seen in
Fig.~\ref{fig:niNSmap} correspond to directions where holes exist in
the spatial nickel distribution, while the green-yellow-red maxima in
Fig.~\ref{fig:niNSmap} appear as big, clump-like, coherent structures
of nickel in Fig.~\ref{fig:niNS3D}. The hemispheric nickel imbalance
in models W15-1 and W15-2 is obvious, whereas L15-1 shows a bipolar
structure that is misaligned with the NS kick vector. Different from
the three others, L15-2 exhibits a more isotropic nickel distribution
and considerably more power in structures on smaller angular scales.

In summary, Figs.~\ref{fig:niprod}--\ref{fig:niNS3D} confirm that the
production of nickel and of adjacent products of explosive burning
(cf.\ Table~\ref{tab:niNS}) are excellent tracers of explosion
asymmetries.  An observational determination of major anisotropies of
the motion of heavy elements between silicon and nickel in SN remnants
and an expansion of these metals mainly opposite to the direction of
the NS motion would lend strong support to the NS acceleration by the
described gravitational tug-boat mechanism.


\section{Summary and conclusions}
\label{sec:summary}

We have performed a set of 20 SN simulations from shortly after core
bounce to more than one second after the onset of the explosion for
four different 15\,$M_\odot$ and 20\,$M_\odot$ nonrotating progenitor
stars, making use of an axis-free Yin-Yang grid in three
dimensions. The grid helped us to avoid numerical artifacts as well as
severe time-step constraints near the polar axis of a usual spherical
grid. A further gain in performance was achieved by excising the
high-density core of the PNS (at densities significantly above the
neutrinosphere) and replacing it by an inner boundary condition, where
the contraction of the PNS core was imposed and the neutrino
luminosities were chosen such that neutrino-energy deposition behind
the shock led to SN explosions of defined energy. The neutrino
transport on the grid was treated by a computationally efficient,
analytic characteristics integrator, in which a grey approximation by
was used averaging over Fermi-Dirac spectra and flux variations with
latitude and longitude were represented by employing a ray-by-ray
description \citep[see][]{Schecketal06}.

A subset of our explosion models was evolved beyond three seconds
postbounce to follow the approach of the NS recoil velocity to the
saturation level.  Some of these simulations were carried on for
several 100 seconds to determine the explosive nucleosynthesis of
heavy elements and the spatial distribution of the burning products.

Our results demonstrate the viability of the gravitational tug-boat
mechanism for NS acceleration in three dimensions. In this scenario
the newly formed NS is accelerated over a timescale of many seconds
mainly by the anisotropic gravitational attraction of the
asymmetrically expanding neutrino-heated ejecta gas, in which
hydrodynamic instabilities had created large-scale asphericities prior
to the onset of the explosion. Before the accretion flows to the NS
are quenched by the accelerating explosion, hydrodynamic forces on the
NS play a nonnegligible role, too. We emphasize, however, that
anisotropies of the neutrino emission, which our ray-by-ray transport
approximation tends to overestimate, contribute to the NS kick
velocity only on a minor level of percents or less.  The reason for
this small effect is simple: Anisotropies of the neutrino emission
exhibit short-timescale intermittency and are characterized by a
higher-order multipole structure in angular space. As a consequence,
the neutrino-induced momentum transfer is diminished by statistical
averaging and is therefore unable to cause any significant net kick in
a certain direction.  Moreover, only a small fraction of the total
neutrino energy loss is radiated anisotropically during the simulation
time. Instead, the far majority of the escaping neutrinos contribute
to the isotropic background flow that diffuses out of the essentially
spherical, hot accretion layer around the nascent NS.

NS acceleration by the gravitational tug-boat mechanism was first
discussed on the basis of axisymmetric (2D) simulations by
\citet{Schecketal04,Schecketal06} and was confirmed in 3D with a small
set of explosion models by \citet{kickletter}. Moreover, support for
this mechanism was recently also provided by the 2D simulations of
\citet{Nordhausetal10,Nordhausetal12}, in which the NS was allowed to
move self-consistently out of the grid center.

While the artificial constraint to axisymmetry enforces a collimation
of large-scale flows parallel to the polar grid axis and thus tends to
favor a pronounced, dipolar deformation of the explosion, the 3D
asphericities appear less extreme and seemingly less promising for
high NS kicks.  Nevertheless, the inhomogeneities and anisotropies in
the massive postshock shell are sizable also in 3D. In fact, low
spherical harmonics modes of dipolar and quadrupolar character can
dominate the asymmetry of the shock and postshock ejecta also in 3D,
although with somewhat smaller amplitude than in extreme 2D situations
(Fig.~\ref{fig:mode_rotk}).  Correspondingly, also in our set of 20
three-dimensional models several cases were found that produced NS
kick velocities well above of 500\,km\,s$^{-1}$, and our current
record holder developed a NS recoil of more than 700\,km\,s$^{-1}$
after 3.3\,s (Table~\ref{tab:results} and
Fig.~\ref{fig:longvns}). This is compatible with the 2D results of
\citet{Schecketal06}, who obtained velocities up to 800\,km\,s$^{-1}$
after one second and estimated the final NS speed in one out of more
than 70 models to be nearly 1000\,km\,s$^{-1}$ and in another one to
be more than 1200\,km\,s$^{-1}$.  Simple analytic estimates of the
long-time influence of the gravitational forces by the expanding SN
ejecta confirm that dense clumps of some $10^{-3}\,M_\odot$
(accounting for mass distribution asymmetries of around 10\% in the
postshock ejecta shell) could well accelerate the NS to velocities of
1000--2000\,km\,s$^{-1}$ as measured for the fastest young pulsars
(cf.\ Eq.~\ref{eq:kickest1}). With an ejecta-shell mass of a few
$10^{-2}\,M_\odot$, a similar accelerating effect on the NS can be
achieved by hemispheric asymmetries of the expansion velocity of
several 10\% (cf.\ Eq.~\ref{eq:kickest2}), which is in the ballpark of
shock expansion differences seen in some of our models
(Fig.~\ref{fig:shockmap}).

The long-time acceleration of the NS is therefore connected to
slow-moving, massive ejecta ``clumps'', whose expansion lags behind
the rest of the ejected material of the SN core. For this reason the
NS gains momentum opposite to the direction of fastest expansion of
the explosion shock and postshock shell.  Correspondingly, the
explosive nucleosynthesis of heavy elements from $^{28}$Si to the iron
group, especially also of radioactive $^{56}$Ni, occurs
primarily 
in the hemisphere pointing away from the NS velocity vector. This
effect clearly correlates in strength with the size of the kick. We
therefore propose that an observational confirmation of opposite
momentum directions of the NS and of the bulk of elements heavier than
silicon in supernova remnants would lend support to the gravitational
kick mechanism discussed in this paper. Such an analysis for the
Cassiopeia~A (Cas A) SN remnant with its central point source could be
enlightening.

Nonradial (off-center) accretion downflows in the postshock layer can
also lead to a spin-up of the nascent NS in nonrotating progenitor
stars. This requires angular momentum separation in the postshock flow
during the postbounce accretion phase of the PNS.  The corresponding
angular momentum can accumulate to values of several
$10^{46}$\,g\,cm$^2$\,s$^{-1}$ in our simulations, in agreement with
an order-of-magnitude estimate on grounds of simple considerations
(Eq.~\ref{eq:angmom1}). Assuming angular momentum conservation during
the later evolution, we derive spin periods between $\sim$100\,ms and
several seconds for the NS in our models. We note, however, that
angular momentum conservation is difficult to assure in the numerical
models, and thus these numbers should be considered only as a rough
range of possibilities.

While the NS acceleration is a long-time (continuing over seconds)
phenomenon mostly achieved by the gravitational pull of the postshock
ejecta shell, the spin-up of the nascent NS is connected to
anisotropic mass flows that transfer angular momentum to the
accretor. The relevant postbounce accretion phase of the PNS may last
for several 100\,ms beyond the onset of the SN blast.  This limits the
time interval for the spin-up of the NS to a much shorter period than
the NS acceleration.  Kick and spin of the NS in this scenario are
therefore not obtained by the same process: The former is linked to
mass ejection asymmetries, the latter to anisotropic, nonradial
accretion flows; they are mediated by different forces
---gravitational versus hydrodynamical---, and they also build up on
different timescales. It is therefore not astonishing that we do not
find any correlation of spin and kick directions in our explosion
models of nonrotating progenitor stars, and a possible spin-kick
alignment suggested by observations (e.g.,
\citealt{Johnstonetal05,Johnstonetal07,Kaplanetal08}; also see
\citealt{Wangetal06,Wangetal07}) remains unexplained.

In this context, the question of the consequences of (even a modest
amount of) angular momentum in the progenitor core is interesting and
deserves future exploration. In the described kick scenario, however,
rotational averaging of the kick, which could cause an alignment of
pulsar velocity and spin as suggested by \citet{Laietal01} and
\citet{Wangetal07}, is unlikely to be efficient despite the long
timescale of the NS acceleration. Rotational motions in the expelled
gas, whose clumpy asymmetries accelerate the NS by gravitational
forces, are diminished during expansion, because under the constraint
of angular momentum conservation the angular velocity decays with
growing radius like $r^{-2}$.  It is therefore not possible to destroy
the randomness of the kick direction just by the rotation of the
accelerating ejecta asymmetries.  An asphericity that
tends to accelerate the NS along the spin direction
would have to be imprinted 
in the ejecta gas already at the onset of the explosion. It is unclear
whether core rotation can produce such an asymmetry in the case of
reasonable and realistic assumptions about the spin rate of the
progenitor core prior to collapse. With predicted typical Fe-core
rotation periods of ${\cal O}(100\,\mathrm{s})$ \citep{Hegeretal05},
which lead to NS spin periods of ${\cal O}(10\,\mathrm{ms})$, the
neutrino-heated postshock layer has rotation periods of about one
second. This is much too slow to have any dynamical influence. On the
other hand, if sufficiently fast rotation allowed for a rapid growth
of SASI spiral modes or triaxial dynamical or secular instabilities,
kicks perpendicular to the NS spin appear more likely than the
alignment of both.  In our simulations we do not find any evidence
of 
the large-amplitude SASI spiral modes that were observed in idealized
setups (without neutrino effects) by \citep{BlondinMezzacappa07} and
\citet{Fernandez10} and experimentally by \citet{Foglizzoetal12}. The
growth of such large-scale, coherent patterns in the postshock
accretion flow might be inhibited by the presence of neutrino heating,
or their growth timescale might be longer than the evolution period
until the onset of the explosion in our models.  Numerical studies
\citep{BlondinMezzacappa07,Iwakamietal09} and analytical treatment
\citep{YamasakiFoglizzo08}, however, predict faster growth of SASI
spiral modes in rotating environments.

Our still limited set of simulations does not allow us to draw any
firm conclusions on progenitor or explosion dependences of the spin
and kick magnitudes. One of the employed 15\,$M_\odot$ stars (the W15
progenitor) seems to exhibit a tendency to favor higher recoil
velocities on average than the other three investigated progenitors,
but the exact reason for this trend is not determined.

\citet{Podsiadlowskietal04} hypothesized that the collapse and
explosion of low-mass SN progenitors with ONeMg cores leads to NS with
low-velocity kicks. They argue that in the case of ONeMg cores in
contrast to more massive Fe-core progenitors the explosion sets in
more quickly and does not provide the time for the growth of
large-amplitude asymmetries of low-order multipole type in the
neutrino-heated accretion layer between NS and stalled shock. Indeed,
the bounce shock accelerates quickly in the steep density profile at
the edge of collapsing ONeMg cores and the explosion sets in on a
timescale of only $\sim$100\,ms after core bounce
\citep{Kitauraetal06,Jankaetal08}. This leaves time for only one
convective turnover. Therefore, the angular size of the convective
plumes remains rather small. Because of the extremely fast, spherical
expansion of the shock, the convective structures are also rather
symmetric in all directions \citep[cf.\ Fig.~1 in][]{Wanajoetal11}.
These facts together with the extremely small mass swept up from the
dilute environment of the ONeMg core by the outgoing shock disfavor
high NS kicks by the gravitational tug-boat mechanism\footnote{We note
  in passing that \citet{Podsiadlowskietal04} as well 
  as \citet{vandenHeuvel10} erroneously attributed the acceleration of
  the NS to anisotropic neutrino emission instead of asymmetric mass
  ejection.}.  
In our scenario of NS acceleration the differences of low-kick and
high-kick cases are connected to differences in the explosion
dynamics, manifesting themselves in a different growth of hydrodynamic
(convective and SASI) instabilities during the onset of the
explosion. The growth of these instabilities can be seeded by small
random perturbations (which should be present in any convectively
perturbed stellar environment and are imposed by us artificially on
spherical pre- or post-collapse progenitor models) and leads to
anisotropic mass ejection. Large-scale asymmetries that might develop
in the progenitor core during convective shell-burning phases prior to
collapse as advocated by \citet{BurrowsHayes96,vandenHeuvel10}, and
\citet{ArnettMeakin11} are therefore an interesting possibility but do
not seem to be necessary for an explanation. Of course, if such
large-scale asymmetries of the progenitor structure occurred, they
could have important consequences for the SN explosion and the NS kick
mechanism. Unquestionably, the availability of more realistic,
three-dimensional progenitor models for core-collapse simulations
would be highly desirable.


\begin{acknowledgements}
We thank Rodrigo Fern{\'a}ndez, Thierry Foglizzo, Jerome Guilet, and
Jeremiah Murphy for useful discussions, and the latter in particular
for suggesting the term ``gravitational tug-boat mechanism''. This
work was supported by the Deutsche Forschungsgemeinschaft through the
Transregional Collaborative Research Center SFB/TR~7 ``Gravitational
Wave Astronomy'' and the Cluster of Excellence EXC~153 ``Origin and
Structure of the Universe'' (http://www.universe-cluster.de). Computer
time at the Rechenzentrum Garching (RZG) is acknowledged.
\end{acknowledgements}


\def\eprinttmppp@#1arXiv:@{#1}
\providecommand{\arxivlink[1]}{\href{http://arxiv.org/abs/#1}{arXiv:#1}}
\def\eprinttmp@#1arXiv:#2 [#3]#4@{\ifthenelse{\equal{#3}{x}}{\ifthenelse{
\equal{#1}{}}{\arxivlink{\eprinttmppp@#2@}}{\arxivlink{#1}}}{\arxivlink{#2}
  [#3]}}
\providecommand{\eprintlink}[1]{\eprinttmp@#1arXiv: [x]@}
\providecommand{\eprint}[1]{\eprintlink{#1}}

\end{document}